\documentclass{aa}  

\usepackage{graphicx}
\usepackage{txfonts}
\usepackage{amsmath}
\usepackage{xcolor}
\usepackage[normalem]{ulem}
\usepackage[colorlinks=true, linkcolor=blue, citecolor=blue, urlcolor=blue]{hyperref}
\usepackage{indentfirst}

\begin{document}

   \title{Chemistry and IR emission of acetylene in planet-forming {regions of T Tauri} disks}

   \subtitle{Impact of elemental abundances and dust properties}

   \author{P. Estève
          \inst{1}
          \and
          B. Tabone\inst{1}
          \and
          E. Habart\inst{1}
          \and
          E. F. van Dishoeck\inst{2,3}
          \and
          M. Vlasblom\inst{2}
          \and
          I. Kamp\inst{4}
          \and
          A. M. Arabhavi\inst{4}
          \and
          S. Bruderer\inst{3}
          }

   \institute{Université Paris-Saclay, CNRS, 
   Institut d’Astrophysique Spatiale, 91405 Orsay, France\\
              \email{pacome.esteve@universite-paris-saclay.fr}
        \and
        Leiden Observatory, Leiden University, 2300 RA Leiden, Netherlands
        \and
        Max-Planck Institut für Extraterrestrische Physik (MPE), Giessenbachstr. 1, 85748, Garching, Germany
        \and
        Kapteyn Astronomical Institute, Rijksuniversiteit Groningen, Postbus 800, 9700AV Groningen, The Netherlands
             }

   \date{Received **}
   \abstract 
   {JWST has revealed a broad diversity of spectra, pointing toward
   a wide range of physical or chemical conditions in the planet-forming regions of disks (<10 au) around T Tauri stars. In particular, acetylene (C$_2$H$_2$) is widely detected, but it remains unclear whether the spread in observed line flux is due to variation in the gas-phase C/O ratio or caused by other factors.
   }
   {Our goal is to explore the parameters that influence the mid-infrared (MIR) emission of C$_2$H$_2$ and H$_2$O, and if the spread observed in $F\rm{_{C_2H_2}}$/$F\rm{_{H_2O}}$ is tracing a variation in the C/O ratio. }
   {Our work is based on the DALI 2D thermochemical model, which self-consistently computes the thermal and chemical structure of the disk, and predicts spectra readily comparable to JWST/MIRI observations. To robustly model organics in inner disks, several improvements have been made: (1) carbon chemistry adapted for warm environments, (2) updated UV shielding treatment, and (3) mutual line overlap in the ray-tracing.
   }
   {With the model improvements, we are able to reproduce the observed C$_2$H$_2$ fluxes {of T Tauri disks} with a realistic disk geometry and solar C/O ratio. Our models show that C$_2$H$_2$ is bright and detectable by JWST even with a solar C/O. Its abundance is primarily set by a balance between formation initiated by CO dissociation by X-rays and destruction of carbon chains by atomic oxygen, the latter being generated by X-ray-induced destruction of H$_2$O and CO. The water UV shielding and hot temperatures (500-1000 K) of the inner disk also favor acetylene formation, as they prevent the destruction of carbon chains and allow the activation barriers of reactions with H$_2$ to be overcome. C$_2$H$_2$ and H$_2$O emissions are not only sensitive to the C/O ratio but also to the total O/H elemental abundance, supporting recent claims. In particular, we find that {enhanced} {O/H} reduces acetylene emission due to an excess of atomic oxygen. $F_{\rm{C_2H_2}}$/$F_{\rm{H_2O}}$ is thus a promising tracer of the elemental composition of inner disks. Still, the dust size distribution also plays a key role in this line flux ratio, reflecting the intrinsic link between the gas and dust components. We find that increasing the abundance of small grains relative to large grains favors C$_2$H$_2$ flux over H$_2$O flux.  Grain depletion does not affect the $F_{\rm{C_2H_2}}$/$F_{\rm{H_2O}}$ ratio as previously suggested by observational works. {A preliminary comparison with published JWST observations indicates a gas-phase C/O ratio below unity and suggests that enhanced O/H ratios may be common in T Tauri disks.}} 
   {Thermochemical models are essential for interpreting JWST spectra of inner disks, which ultimately provide information on the composition of exoplanet atmospheres. Yet, robustly estimating the C/O ratio {through $F_{\rm{C_2H_2}}$/$F_{\rm{H_2O}}$} requires one to constrain dust properties from dust emission and better characterize the carbon chemistry.}


   \maketitle
   
\section{Introduction}

   \noindent It has been nearly three decades since the first exoplanet was discovered {\citep{Mayor_1995Natur.378..355M}}. The detection of thousands of such objects since then has revealed the remarkable diversity of planetary systems. In the next decade, this diversity will be scrutinized from a new vantage point: their atmospheric compositions thanks to cutting-edge space-borne and ground-based observatories (JWST, ELT, and ARIEL).
   
   However, understanding the starting point of planet formation is essential to make the most of these future missions and link the properties of exoplanets to their formation history. Indeed, the atmospheric composition of gas giant planets can give valuable information on their formation history {\citep{Bitsch_2015bA&A...582A.112B,madhusudhan_exoplanetary_2016, Bitsch_2022A&A...665A.138B}}. The composition of gas and dust that is accreted by the planet is thought to vary spatially in the protoplanetary disk, following a hierarchical structure from the different sublimation temperatures of the volatile species in the gas \citep{oberg_effects_2011, ligterink_mind_2024}. The elemental composition of planet atmosphere, in particular the C/O ratio, may indicate when and where the planet forms by being directly linked to the composition of the disk at that location {\citep{cridland_connecting_2019b, oberg_astrochemistry_2021, schneider_how_2021, O'Donovan_2026A&A...706A..30O}}. Recently, other processes that shape the elemental composition of disks have emerged: the drift of icy pebbles from the outer disk that enriches the inner disk in oxygen {\citep{banzatti_hints_2020,Kalyaan_2021ApJ...921...84K,banzatti_jwst_2023, Kalyaan_2023ApJ...954...66K, Sellek_2025A&A...701A.239S,Houge_2025aMNRAS.537..691H, Williams_2025MNRAS.544.3562W}} and sublimation or chemisputtering of carbon grains that would enrich the gaseous disk in carbon {\citep{booth_chemical_2017, Lenzuni_1995ApJ...447..848L,Borderies_2025A&A...694A..89B, Houge_2025bA&A...699A.227H}}. All of these processes will thus have a significant impact on the elemental C/O ratio, ranging from subsolar (<0.5) to super-solar (>0.5) or even C/O > 1. 
   
   These scenarios have been confronted by new observations taken by ALMA, revealing that evolved disks harbor ubiquitous substructures, showing that the radial distribution of species is far from smooth \citep{ALma_HLTau_2015ApJ...808L...3A,Andrews_2020ARA&A..58..483A,Oberg_2021ApJS..257....1O,Booth_2024AJ....167..164B}. Indeed, pressure bumps can block the pebbles that drift inward, preventing the supply of oxygen in the inner disk {\citep{Zhu_2012ApJ...755....6Z,Pinilla_2012A&A...538A.114P,banzatti_water_2025,gasman_minds_2025-1, Temmink_2025A&A...699A.134T}}, which therefore challenges the scenarios mentioned above. {However, recent theoretical works suggest that these gaps might be leaky, in particular to submicron-sized grains \citep{Weber_2018ApJ...854..153W, Drazkowska_2019ApJ...885...91D, Stammler_2023A&A...670L...5S}, in line with recent observations combining ALMA and\textit{ James Webb} Space Telescope (JWST) data \citep{gasman_minds_2025-1}. Regarding the gas content of outer disks ($r$ > 10 au)}, ALMA observations show that they are surprisingly faint in CO, which is interpreted as a chemical conversion of CO into less volatile species, resulting in C/O$>$1 \citep{miotello_lupus_2017,bergin_hydrocarbon_2016,bosman_co_2018, Miotello_2019A&A...631A..69M}, as well as locking up CO ice in larger bodies \citep{Krijt_2020ApJ...899..134K}. Together with thermochemical models, CO and C$_2$H proved to be good tracers of the C/O ratio in the outer disk by being widely detected \citep{Bosman_2021ApJS..257....7B,Sturm_2022A&A...660A.126S}. 
   
   Inner disks ($r$<1-5 au) have been less studied, but \textit{Spitzer} along with ground-based instruments (CRIRES, iSHELL, and TEXES) gave interesting first results in the mid-infrared (MIR). Unlike outer disks, inner disks of T Tauri stars are oxygen-rich, with bright H$_2$O and OH emissions \citep{salyk_h_2008,carr_organic_2008}, revealing a potentially low C/O. Regarding the main C, O, and N carriers, HCN, C$_2$H$_2$, and CO$_2$ (and CO) have also been widely detected \citep{pontoppidan_spitzer_2010,carr_organic_2011}. JWST is now revolutionizing the field with increased sensitivity and spectral resolution, allowing us to detect new species and isotopologs, and thus better constrain inner disks conditions. Slab models indicate that IR emission from 5 to 27 $\mu$m is emitted from hot, upper layers at $T \sim 300$ - $800$ K, with column densities typically of around $N \sim 10^{16}$ - $10^{19}$ cm$^{-2}$ \citep{Salyk_2011ApJ...731..130S,kamp_chemical_2023,van_dishoeck_diverse_2023,Grant_2023ApJ...947L...6G}. 
   
   Interestingly, even if T Tauri disks are dominated by water emission, C$_2$H$_2$ is systematically detected with a detection rate of 90\% \citep{Arulanantham_2025AJ....170...67A, Grant_2025A&A...702A.126G}, suggesting that this molecule might have a special place among hydrocarbons. Moreover, JWST observations of very low-mass stars (VLMSs) show a very rich carbon chemistry, with new detections of C$_4$H$_2$, C$_6$H$_6$, C$_3$H$_4$, and CH$_4$, and very little water {\citep{Tabone_2023NatAs...7..805T,arabhavi_abundant_2024,Kanwar_2024A&A...689A.231K, Arabhavi_2025A&A...699A.194A}}. In particular, C$_2$H$_2$ is the most prominent feature with a pseudo-continuum reaching $N \sim 10^{22}$ cm$^{-2}$, reinforcing the fact that C$_2$H$_2$ is central in carbon chemistry. These results suggest that, as opposed to T Tauri stars, VLMSs might have a high C/O ratio {\citep{Kanwar_2026A&A...705A.222K}}.
   
   Nevertheless, \cite{Grant_2025A&A...702A.126G} found that the emissions of C$_2$H$_2$ and H$_2$O vary by 2 orders of magnitude among T Tauri disks, showing that the chemistry in inner disks is not entirely dictated by the mass of the central star. This {spread} possibly traces the {diversity of} C/O ratio in the inner disk, with C$_2$H$_2$ tracing the carbon and H$_2$O the oxygen. We now need thermochemical models to better understand the physical and chemical processes at stake in these regions. 
   
   Previous works on thermochemical models show that MIR emission comes from the surface of the disk atmosphere, but the layer depends on the species, with CO and OH at the very top, then H$_2$O and CO$_2$, and C$_2$H$_2$ and HCN deeper down \citep{woitke_modelling_2018}. However, numerous studies have shown that MIR molecular emission depends on the atmospheric conditions of the disk, including dust properties. \cite{Maijereink_2009ApJ...704.1471M} and \cite{Glassgold_2009ApJ...701..142G} revealed that water emission can be significantly increased by depleting grains in the atmosphere or by flattening the dust size distribution. \cite{Antonellini_2015A&A...582A.105A} confirmed that dust opacity sets the water emission, expanding the dependency to the UV field and the disk geometry. More recently, \cite{antonellini_mid-infrared_2023} have claimed that the trend of HCN/H$_2$O with dust disk mass seen in \cite{Najita_2013ApJ...766..134N} can be explained by dust evolution, following the prescription of \cite{Greenwood_dustevo_2019A&A...626A...6G} that includes dust settling, growth, and radial drift, resulting in the gas-to-dust ratio and grain sizes increasing over time.
   
   Results concerning organic molecules, especially acetylene, are still debated, probably due to the organic chemistry being much more complex than the oxygen chemistry. \cite{Agundez_2008A&A...483..831A} predicted that the steady-state abundance of C$_2$H$_2$ and organic molecules should be higher in X-ray dominated regions, but they should decrease with more X-ray irradiation, later confirmed by \cite{Najita_2011ApJ...743..147N}. In contrast, \cite{woitke_2d_2024} predicted the opposite and \cite{Greenwood_2019A&A...631A..81G} found no dependence of C$_2$H$_2$ on X-rays. However, most of these studies significantly underproduced acteylene \citep{Najita_2011ApJ...743..147N, woitke_modelling_2018, Greenwood_2019A&A...631A..81G}. \cite{kanwar_hydrocarbon_2024} extended the ProDiMo chemical network DIANA \citep{kamp_consistent_2017} to include hydrocarbons until eight atoms of carbon, but it did not increase the emission of C$_2$H$_2$. \cite{woitke_2d_2024} solved the underproduction of acetylene by adopting a smooth inner rim, {which enables the formation of acetylene above the optically thick dust layer in a very hot region.} The column densities derived by \cite{walsh_molecular_2015} and \cite{anderson_observing_2021} were consistent with \textit{Spitzer}, but no detailed radiative transfer was performed.  Nevertheless, they agree that the emission of water and acetylene are sensitive to the elemental abundances, and that the C/O ratio is a driver of the brightness of hydrocarbons. The line flux ratios HCN/H$_2$O and C$_2$H$_2$/H$_2$O were proposed to be promising tracers of the C/O ratio, allowing us to constrain the chemical composition of inner disks \citep{Najita_2011ApJ...743..147N, anderson_observing_2021}. {Using the same model as \cite{woitke_2d_2024}, \cite{Arabhavi_2026A&A...708A..82A} also did the synthetic prediction of the elemental abundances with detailed radiative transfer, but did not include water UV shielding. Indeed,} \cite{bosman_water_2022-1} and \cite{duval_water_2022} recently showed that water can efficiently shield the molecular layers by absorbing UV photons that photodissociate molecules. This process significantly boosts the abundance of organic molecules, which means that models do not require a high C/O ratio to produce abundant C$_2$H$_2$.
   
   In this paper we explore the formation of C$_2$H$_2$ and the physical processes that can enhance its emission to match the observations, {including water UV shielding}. We also self-consistently predict IR emission of water and acetylene to better understand the diversity of JWST spectra of T Tauri disks {and put first constraints on elemental abundances.} 
   
   This paper is organized as follows. Section \ref{sec: General model} details the thermochemical model DALI, with several improvements for the study of inner regions of protoplanetary disks. Section \ref{sec: Results General} first presents the results of a fiducial model to highlight the processes shaping the water and acetylene emission, and then describes the results of a grid of models covering elemental abundances, dust properties, and disk geometry. A detailed discussion of different modeling considerations as well as a comparison with observations is proposed in Sect. \ref{sec: Discussion}. The main results are summarized in Sect. \ref{sec: conclusion}. 

\section{Model}
\label{sec: General model}
\noindent This section describes the DALI model, with a focus on the processes added to the code to better model inner disks. The setup and parameters explored in this work are then summarized.
   \subsection{DALI model}
   
   \noindent This work is based on the 2D thermo-chemical model DALI \citep{bruderer_warm_2012,bruderer_survival_2013, bruderer_ro-vibrational_2015} to self-consistently compute $\rm{H_2O}$ and $\rm{C_2H_2}$ emission. The code is divided in several steps. After creating a grid with the input dust and gas density structure, the first step computes the dust temperature and the dust mean intensity for each cell using a 2D Monte Carlo method. Then, the thermo-chemistry module computes iteratively the abundance of each species from a given chemical network (see Sect. \ref{sec: Networks}), the atomic and molecular excitation, and the thermal balance of the gas to get a self-consistent gas temperature, $T_{\rm{gas}}$, and chemical abundances. This specific step has been recently improved by including the UV shielding from molecules, especially $\rm{H_2O}$ \citep{bosman_water_2022-1}, which has a prominent impact on molecular layers of inner disks. In this work, we further improved this step by including an extended chemical network (see Sect. \ref{sec: Networks}) and a refined treatment of the UV shielding (see Sect. \ref{sec: UV_shielding, method}). The last step of DALI builds synthetic spectra from the radiative transfer equation. Since we want to reconstruct IR spectra including thousands of lines, we use an improved fast ray-tracer \citep{bosman_co2_2017}, based on the fact that the velocity gradient along the line of sight is approximately linear \citep{Horne_1986MNRAS.218..761H}. We upgraded it by including so-called line overlap, an effect that can efficiently reduce the optically thick emission of molecules, particularly $\rm{C_2H_2}$ (see \citealt{Tabone_2023NatAs...7..805T}). 

\subsection{Improvements in DALI}
    \subsubsection{UV shielding}
   \label{sec: UV_shielding, method}
   \noindent Atoms or molecules absorb the UV field irradiated from the star. This absorption changes the spectral shape of the radiation field and reduces its strength. As a result, a species not only protects itself by reducing its photodissociation rate (self-shielding), but also protects the other species (mutual shielding). To properly treat UV shielding, the absorption of the radiation field by all species at all wavelengths should be computed. However, cross sections, which describe the interaction of atoms and molecules with the radiative field, exhibit steep variations with very narrow features. This would require a very fine wavelength grid to sample cross sections, which is computationally prohibitive. 
   
   To reduce the computational time, we evaluate the rates of photoprocesses by using the “computational efficiency” format of the Leiden Database\footnote{\url{https://home.strw.leidenuniv.nl/~ewine/photo/cross_sections.html}} (see Appendix \ref{appendix:UV shielding} for technical details). This format splits each cross section into two parts: narrow features (called “lines'';$\Delta \lambda < 1$ nm) and smooth variations (called the “continuum”). With this approach, we can compute photodissociation rates with a limited wavelength grid of $\sim$100 points, because the “lines” are integrated thanks to their respective oscillator strength. The lines are particularly efficient to self-shield because the peaks can be very high (around $\sigma_{\rm{peak}} \sim 10^{-16}$ - $10^{-15}$ $\rm{cm^2}$) meaning that a column density of $N = 10^{15}$ $\rm{cm^{-2}}$ is sufficient to shield the gas (reached relatively easily in inner regions of protoplanetary disks for molecules such as CO, H$_2$, and N$_2$). However, since the lines are very narrow, the impact on the radiation field is limited. Thus, we only consider these lines for the self-shielding. In contrast, the continuum {part of the cross section} affects a broad wavelength range, which allows the species not only to self-shield but also to protect other molecules. Water is a good example of efficient mutual shielding \citep{Bethell_2009Sci...326.1675B, Adamkovics_2014ApJ...786..135A,duval_water_2022}. We note that high column densities are required for mutual shielding ($N \sim 10^{18}$ - $10^{19}$ $\rm{cm^{-2}}$), so this process is only relevant for abundant molecules in protoplanetary disks. 
   
   In the previous version of DALI, self-shielding was taken into account for well-known molecules (H$_2$, C, N$_2$, CO, and isotopologs; \citealt{Miotello_2014A&A...572A..96M, visser_nitrogen_2018}), and the mutual shielding is usually considered for water, although the user can choose which molecules can play a role in the attenuation of the UV flux.
   Our new implementation of the UV shielding extends the self-shielding to all species in the chemical network. In case the UV cross section is not available for a species, we arbitrarily choose the {cross section of $c-\rm C_3H_2$} for the photoreactions of this species {(except for C$_4$H$_2$ for which we take C$_4$H cross section)}. {This cross section should more accurately reflect the typical photodissociation rate of hydrocarbons than that of water, chosen in \cite{bosman_co_2018}}\footnote{List of hydrocarbons with  $c-\rm C_3H_2$ cross section in Appendix \ref{appendix: water cross section}}. The calculation of the photorates in DALI relies on cross-section data from the Leiden Database \citep{heays_photodissociation_2017,hrodmarsson_photodissociation_2023}. We added other abundant molecules in the mutual shielding, for a total of 12 species: S, Fe, H$_2$O, OH, CO$_2$, HCN, CN, C$_2$H$_2$, C$_3$, C$_2$H$_4$, CH$_4$, and C$_2$H$_6$. {We find that H$_2$O is the dominant species to mutual shield with a C/O < 1 (see Sect. \ref{Sec: C2H2chem}). For C/O > 1, C$_3$ takes over the role of UV attenuation, followed to a lesser extent by C$_2$H$_2$, since they are both abundant and have a relatively broad UV cross section.} We do not expect other molecules (for which a cross section is available) to be abundant enough to play a significant role in mutual shielding. The mutual shielding efficiency of 6 over these 12 species are shown in Appendix \ref{Appendix: mutual shielding}.

\subsubsection{Chemical networks}
\label{sec: Networks}
   \label{Sec: constr_net}
   \noindent The original chemical network of \cite{bruderer_warm_2012} was first extended to include CO isotopolog chemistry \citep{Miotello_2014A&A...572A..96M}, and then refined to add key molecules in the chemistry of protoplanetary disks, such as HCN or $\rm{C_2H}$ \citep{visser_nitrogen_2018,Miotello_2019A&A...631A..69M}. However, the simple hydrocarbon chemistry in \cite{Miotello_2019A&A...631A..69M} stopped at $\rm{C_2H_3}$, and was tailored for outer regions of disks. {\cite{bosman_water_2022-1} built a network for inner regions, based on the \texttt{RATE12} network of UMIST \citep{mcelroy_umist_2013} and added three-body reactions from \cite{walsh_molecular_2015}. This very large network (674 species and 9441 reactions) extends the carbon chemistry and has already been used for the study of hydrocarbons and organic molecules in the inner disk \citep{duval_water_2022, colmenares_jwstmiri_2024}. Nevertheless, this network does not particularly focus on carbon chemistry, and significantly slows down the code.} The aim of this new network is {twofold: have an accurate description of the carbon chemistry while having a simple network to avoid runtime issues\footnote{This new chemical network is 10x faster ($\sim$31h CPU for the chemistry of the fiducial model) than the network of \cite{bosman_water_2022-1}}.} For its construction, we start from the network of \cite{Miotello_2019A&A...631A..69M} and we add new species and reactions. We mainly focus on gas-phase reactions, since the emitting layers in inner disks are warm, typically $T_{\rm{gas}} \sim 300 -800$ K \citep{Gamsan_2023A&A...679A.117G,Grant_2023ApJ...947L...6G,Temmink_2024A&A...686A.117T,Schwarz_2024ApJ...962....8S}. This new chemical network includes {223} species and {3523} reactions, with {89} new species and {1653} new reactions. The construction follows three steps:
   \begin{enumerate}
       \item Select the pure hydrocarbons (C$_x$H$_y$) with five atoms of carbon maximum.
       \item Add gas-phase reactions between these new hydrocarbons and any other species already in the network, following the types of reactions in DALI detailed in \cite{bruderer_warm_2012}. Photoreactions are included, with the updated UV cross sections from \cite{hrodmarsson_photodissociation_2023} as well as the X-ray and cosmic ray induced reactions. We also add the channel of X-ray/cosmic ray induced photodissociation of H$_2$O leading to O which was missing in the network from \cite{Miotello_2019A&A...631A..69M} and in the UMIST database. 
       \item Add charge exchange with polycyclic aromatic hydrocarbons following the prescription of \cite{2003ApJ...587..278W} as done in \cite{bruderer_warm_2012}.
       \item {Add adsorption and desorption for new neutral species. The binding energies are adopted from \cite{mcelroy_umist_2013} and \cite{bosman_water_2022-1}. When the energy is not available, we take that of the corresponding isomer. If there is no isomer, we assume a desorption energy of 3000 K (only for C$_5$H$_5$ and C$_5$H$_6$).}
       \item {Add three-body reactions from \cite{bosman_water_2022-1}, exported from the network of \cite{walsh_molecular_2015}.}
   \end{enumerate} 

   We identified several missing reactions in UMIST {(\texttt{RATE22} release \citealt{millar_umist_2023})} compared to the KIDA database, especially reactions with high activation barriers. Many abstractions of H are missing, in particular, the key reaction with C$_2$:
   \begin{equation}
       \mathrm{ C_2 + H_2 \xrightarrow{} C_2H + H}~~~~~~~~~~~~~E_a = 1420~\mathrm{K}.\\
   \\ \end{equation}
   UMIST is tailored for cold interstellar medium (ISM) conditions, typically 10-100 K, so reactions with activation barriers of the order of 1000 K or more, are not likely to happen in dense molecular clouds. However, these reactions may have a prominent impact on the chemistry in inner regions of protoplanetary disks, with the example of water formation with $E_a = 3150$ K. 
   
   To include the reactions with high activation barriers, we combined UMIST \citep{millar_umist_2023} with reactions in the KIDA Database \citep{wakelam_2024_2024} \footnote{We exported the online KIDA database \url{https://kida.astrochem-tools.org/export/}. For this network, the exportation was done in January 2025. We included only gas-phase reactions and removed duplicated reactions, as well as reactions with rate coefficient of 0. }.
   This KIDA exportation and “cleaning” is essentially composed of the cold ISM network from \citet[kida.uva.2024]{wakelam_2024_2024} and extra reactions from planetology, mainly from the work of \cite{hebrard_how_2009} who modeled the atmospheric chemistry of Titan. The online KIDA database also contains several interesting hydrocarbon reactions from the high-temperature network of \cite{harada_new_2010}. In addition, we included the revised endothermicities calculated by \cite{tinacci_gretobape_2023}. We removed all reactions with $\rm\Delta H > 100$ kJ.mol$^{-1}$ ($\sim$ 12 000 K). Because of the large uncertainties of the calculated enthalpies (1200 K),  we amended a reaction rate only when the reaction was endothermic by more than 1200 K and we adopted an activation energy equal to the enthalpy change at $T=0$K. 
   Following this procedure, only seven reactions were corrected (see Appendix \ref{appendix:endo_tinacci}), including the key reaction C$_3$+H$_2$. Finally, we added this KIDA network to our network by following exactly the same procedure mentioned above for UMIST.

   {Additionally, we included the reaction $\rm C+H_2O \xrightarrow[]{} HCO + H$ as it can be crucial for the total carbon reservoir for hydrocarbons, but not referenced in UMIST. \cite{woitke_2d_2024} noted that this reaction is included in the online KIDA database with an unreasonable high rate, inconsistent with the upper limit in NIST \citep{husain_reactions_1971} and the low temperature measurements of \cite{Hickson_2016arXiv160808877H}. Further details on the chosen rate are provided in Appendix \ref{Appendix : C+H2O}. This reaction reduces the acetylene emission by $\sim30$\%.}
   
    We do not include grain surface reactions in the network, meaning that the ices are only formed via thermal adsorption, and destroyed via thermal and photo-desorption. IR active layers from which $\rm{H_2O}$ and $\rm{C_2H_2}$ emit, are expected to be at temperatures above 300 K, so the ices do not play a major role in our study. This chemical network is therefore not suitable for studying cold regions, particularly near the midplane where ices are abundant, and surface reactions important. Finally, we do not consider the UV photolysis of carbon grains \citep{Alata_2014A&A...569A.119A}.

       \subsubsection{Line overlap}
    \noindent Gas-phase molecular emission lines are usually spectrally narrow. The line overlap is therefore marginal between species and even in $Q$ branches where thousands of lines are clustered together. However, once the lines start to be highly optically thick, they broaden and can “screen” each other significantly. In planet-forming regions, column densities of abundant species can reach $N \geq 10^{20}$ cm$^{-2}$, producing a forest of optically thick lines. As a result, neglecting line overlap  can strongly overestimate the fluxes, especially in $Q$ branches of species such as C$_2$H$_2$ \citep{Tabone_2023NatAs...7..805T}. To better predict acetylene emission, we update the fast ray-tracer implemented in \cite{bosman_co2_2017} to include line overlap. It reduces the $Q$-branch emission of C$_2$H$_2$ by a factor of 1.5 for a model with a solar C/O ratio (shown in Appendix \ref{appendix:line_overlap}). \\ 
    
   \subsection{Model setup}
   \begin{table}[]
       \centering
       \caption{Input parameters in DALI.}

       \begin{tabular}{c c c c}
        \hline
        \hline
           Parameter & Notation & Fiducial & Range \\
        \hline
           Mass      & $M_*[M_\odot]$    & 1.0      & - \\
           Luminosity      & $L_*[L_\odot]$    & 1.0      & - \\
           Effective Temp      & $T_{\rm{}eff}[\rm K]$    & 4250      & - \\
           Acrretion lum.      & $L_{\rm{}acc}[L_\odot]$    & 0.12      & - \\
           \textbf{X-ray lum.}      & $L_X[\rm erg.s^{-1}]$    & $10^{30}$   & $10^{27}$-$10^{33}$ \\
           X-ray temp.      & $T_X[\rm K]$    & $4.6\times10^7$      & - \\
           Ly-$\alpha$ contrib. & $Ly\alpha$ [$L_{\rm{}acc}$]    & 0.15   & - \\
           \hline
           Disk mass      & $M_D[M_\odot]$    & 0.03      & - \\
           Disk size      & $R_{\rm{}out}[\rm{au}]$    & 10      & - \\
           Inner radius   & $R_{\rm{}in}[\rm{au}]$    & 0.1      & - \\
           Critical radius      & $R_{c}[\rm{au}]$    & 46      & - \\
           \textbf{Disk aspect ratio\tablefootmark{a}}     & $h_c[\rm{rad}]$    & 0.09      & 0.05-0.25 \\
           \textbf{Flaring angle\tablefootmark{a}}      & $\psi$    & 0.15      & 0.05-0.20 \\
           
           \hline
           \textbf{C/O}      & -    & 0.47      & 0.2-1.5 \\
           \textbf{O/H}      & -    & 2.9e-4     & 2.9e-5-2.9e-3 \\
           \hline
           \textbf{Gas-to-dust ratio}\tablefootmark{b}     & $gd$    & $10^3$     & $10^2$-$10^4$ \\
           \textbf{Min. grain size}     & $a_{min}[\rm{nm}]$    & 5     & 5-100 \\
           \textbf{Max. grain size }     & $a_{max}[\rm{mm}]$    & 1     & 1-100 \\
           \textbf{Power law index}\tablefootmark{c}     & $q$    & 3.5     & 3.0-4.0 \\
           \hline
       \end{tabular}
       \tablefoot{\tablefoottext{a} The disk aspect ratio here is actually a scale height angle. The flaring angle is therefore $\psi = \beta - 1$ where $\beta$ is the usual flaring index defined from the scale height (in units of distance).
       \tablefoottext{b}The gas-to-dust ratio value is vertically averaged.
       \tablefoottext{c}Dust opacities adopted from \cite{facchini_different_2017}.}
       \label{tab:inputs_dali}
   \end{table}
   \noindent The input parameters used in this work are listed in Table \ref{tab:inputs_dali}.
   The 2D density structure follows a viscous accretion disk ansatz given by \citep{lynden-bell_evolution_1974}   \begin{equation}
       \Sigma_{gas}(R) = \frac{M_D}{2\pi R_C^2}(2-\gamma)\left(\frac{R}{R_C}\right)^{-\gamma}\exp{\left[-\left(\frac{R}{R_C}\right)^{2-\gamma}\right]},
       \label{eq:rad_struc}
   \end{equation}
   with a characteristic disk radius of $R_C=46$ au, a surface density index of $\gamma = 1.0$, and a disk mass of $M_D = 0.03$ $M_\odot$. 
   The vertical distribution follows an isothermal profile:
   \begin{equation}
       \rho_{gas}(R,\theta) = \frac{\Sigma_{gas}(R)}{\sqrt{2\pi} Rh(R)}\exp{\left[-\frac{1}{2}\left(\frac{\pi/2-\theta}{h(R)}\right)^2\right]}
   ,\end{equation}
   where the disk aspect ratio is given by $h(R) = h_c\left(\frac{R}{R_C}\right)^{\psi}$. For the fiducial model, we adopt $h_c=0.09$ (disk aspect ratio at $R_C$) and a flaring angle of $\psi=0.15$\footnote{The disk aspect ratio is defined here as a scale height angle. The corresponding flaring angle is thus defined from this scale height angle, and differs from the usual flaring index for which the scale height is in units of distance.}. {This setup does not include a smooth inner rim, which could increase molecular emission, as shown in the case of EX Lupi by \cite{woitke_2d_2024}. }

   For this work, we used the DALI module developed by \cite{facchini_different_2017} to self-consistently compute the size-dependent dust settling, following \cite{RiolsLesur2018} and assuming a mixing parameter $\alpha = 10^{-3}$. This corresponds to the prescription used by \cite{woitke_2d_2024}{, which includes the vertical dependence of the Stokes number with the gas density}. For our fiducial model, {the maximum grain size is $a_{max}=1$ mm, set by fragmentation \citep{birnstiel_dust_2011}, which is the typical regime in inner disks \citep{Birnstiel_2012A&A...539A.148B, birnstiel_dust_2015}. The minimum grain size was set to $a_{min} = 5$ nm, consistent with \cite{facchini_different_2017}. Before settling, we considered an MRN dust-size distribution \citep{Mathis_1977ApJ...217..425M} with a power-law index $q=3.5$ ($f(a) \propto a^{-q}$), consistent with the fragmentation-limited regime \citep{Birnsitel_2024ARA&A..62..157B}. Finally, a fiducial global gas-to-dust ratio was set to 1000, which is the typical gas-to-dust ratio achieved after 2-3 Myr in the inner disk \citep{Bitsch_Mah_2023A&A...679A..11B}}. This gas-to-dust ratio is vertically averaged. {Our models do not include the vertical mixing of species, although it could change the IR emission of molecules by a factor of up to 3 \citep{Woitke_2022A&A...668A.164W}.}

   The fiducial abundances are adopted from \citet[our Table \ref{tab:init_ab}]{bruderer_warm_2012} and are consistent with those of \cite{bruderer_survival_2013} and \cite{bosman_water_2022}. They correspond to the volatile ISM abundances based on \cite{Jonkheid_2006A&A...453..163J}. The C/O ratio in the fiducial model is C/O = 0.47. Hereafter, we refer to this value as the “solar C/O” throughout the remainder of the paper, rather than “volatile ISM C/O.” The abundances are obtained by the time-dependent solver LIMEX \citep{ehrig_1999} with an end-time of 3 Myr, which is enough to reach steady-state abundances for oxygen and carbon chemistry \citep{Agundez_2008A&A...483..831A,kamp_consistent_2017, Kanwar_2025A&A...698A.294K}. 

   Regarding the radiation field, the disk orbits a T Tauri star with an effective temperature of $T_{\rm{eff}}=4250$ K, a photospheric luminosity of $L_*=1$ $L_\odot$, and a mass of $M_*= 1~M_\odot$. The {shape of the }far-ultraviolet (FUV) excess corresponds to a blackbody at $T=20~000$ K {following the prescription of \cite{tabone_oh_2024}}. {We include a Lyman-$\alpha$ line with a full width at half maximum (FWHM) of 200 km.s$^{-1}$ and a total luminosity of 0.15 $L_{\rm{}acc}$ \citep{tabone_oh_2024}, which contributes to $\sim 80$\% of the FUV luminosity \citep{Schindhelm_2012ApJ...756L..23S, France_2014ApJ...784..127F}. We neglect the scattering by H atoms.} The X-ray spectrum is given by a blackbody at $T_X=4.6\times 10^7$ K between $10^3$ and $10^5$ eV with a luminosity of $L_X = 10^{30}$ erg.s$^{-1}$.

   Throughout this work, the excitation of H$_2$O was performed in nonlocal thermal equilibrium (non-LTE; \citealt{Faure_Josselin_2008A&A...492..257F}), whereas the population levels of C$_2$H$_2$ was calculated in LTE (due to a lack of collisional rate coefficients). According to \cite{bruderer_ro-vibrational_2015}, the excitation of HCN and C$_2$H$_2$ should be similar. We observe a minor difference ($\sim25$\% reduction) in our models between LTE and non-LTE calculations for HCN thanks to infrared pumping \citep{bruderer_ro-vibrational_2015}, so non-LTE effects should not dramatically change acetylene emission.
   
   \subsection{Model grid}
    \noindent Throughout this work, we explore a set of input parameters to highlight the possible drivers of the acetylene emission in T Tauri disks, and its relationship with water emission. These parameters are listed in bold in Table \ref{tab:inputs_dali}.
    Regarding the elemental abundances, we vary the C/O ratio between 0.2 and 1.5, by keeping O/H constant and varying C/H accordingly. We also run models with 10 times more {and less} oxygen than the ISM {to quantify the impact of an oxygen enrichment due to the drift of icy pebbles and depletion due to advection of water vapor onto the star, respectively.} As with the standard grid, we vary the C/O ratio by varying C/H and keeping O/H constant. These models will be termed {“enhanced O/H” and “depleted O/H”} hereafter.
    For the disk structure, we changed the disk aspect ratio from 0.05 to 0.25. The flaring angle varies from 0.05 to 0.20, but we adapted the disk aspect ratio, $h_c$, at $R_C=46$ au to keep the same disk aspect ratio of the other models at 0.5 au. By doing so, we could better isolate the effect of the flaring angle in the inner disk.
    
    Regarding the properties of the dust, we consider a range of gas-to-dust mass ratios, ranging from 10$^2$ to 10$^4$. {The dust size distribution is known to vary in protoplanetary disks due to vertical settling, radial drift, growth and fragmentation \citep{Weidenschilling_1977MNRAS.180...57W, Brauer_2008A&A...480..859B,Zsom_2010A&A...513A..57Z,Birnsitel_2024ARA&A..62..157B}. The minimum grain size in the inner disk is not well constrained, so we explore a large range, from 5 nm to 100 nm. This range is consistent with observation in scattered light, showing that submicron-sized grains are present in the atmosphere of outer disks, with an upper limit of $a_{min}<400$ nm \citep{Tazaki_2022A&A...663A..57T}. The large uncertainty on the fragmentation velocity ($v_{frag} \sim 1-10$ m.s$^{-1}$ \citealt{Blum_2008ARA&A..46...21B, Gutler_2010A&A...513A..56G, Birnstiel_2012A&A...539A.148B}) leads to a variation in $a_{max}$ of a factor of 100, which is covered in this work. We also explore the impact of a different dust size distribution, since dust size distributions limited by radial drift tend to be top-heavy \citep{birnstiel_dust_2015, Birnsitel_2024ARA&A..62..157B}, and spectral energy distribution fitting typically retrieved power-law indexes between 3 and 4 \citep{Ribas_2020A&A...642A.171R, Kaufer_2023A&A...672A..30K}. }
    
    Finally, we explore a range of X-ray luminosity since its impact on the emission of C$_2$H$_2$ is still debated in the literature \citep{Najita_2011ApJ...743..147N, Greenwood_2019A&A...631A..81G, Notsu_2021A&A...650A.180N,woitke_2d_2024}. {This grid does not cover the effect of varying the inner disk radius since we focus only on full disks, but it could also change molecular emission as shown in \cite{Vlasblom_2024A&A...682A..91V}.}

\section{Results}
\label{sec: Results General}
\noindent This section first describes the results obtained by the fiducial model (see Table \ref{tab:inputs_dali}) and explains the key processes setting the abundance of acetylene. Then, the results of the model grid are analyzed.
\subsection{Fiducial model}
\begin{figure*}[ht]
\centering
\resizebox{\hsize}{!}{
\includegraphics[scale=0.4, trim={0 0 0 0cm}, clip]{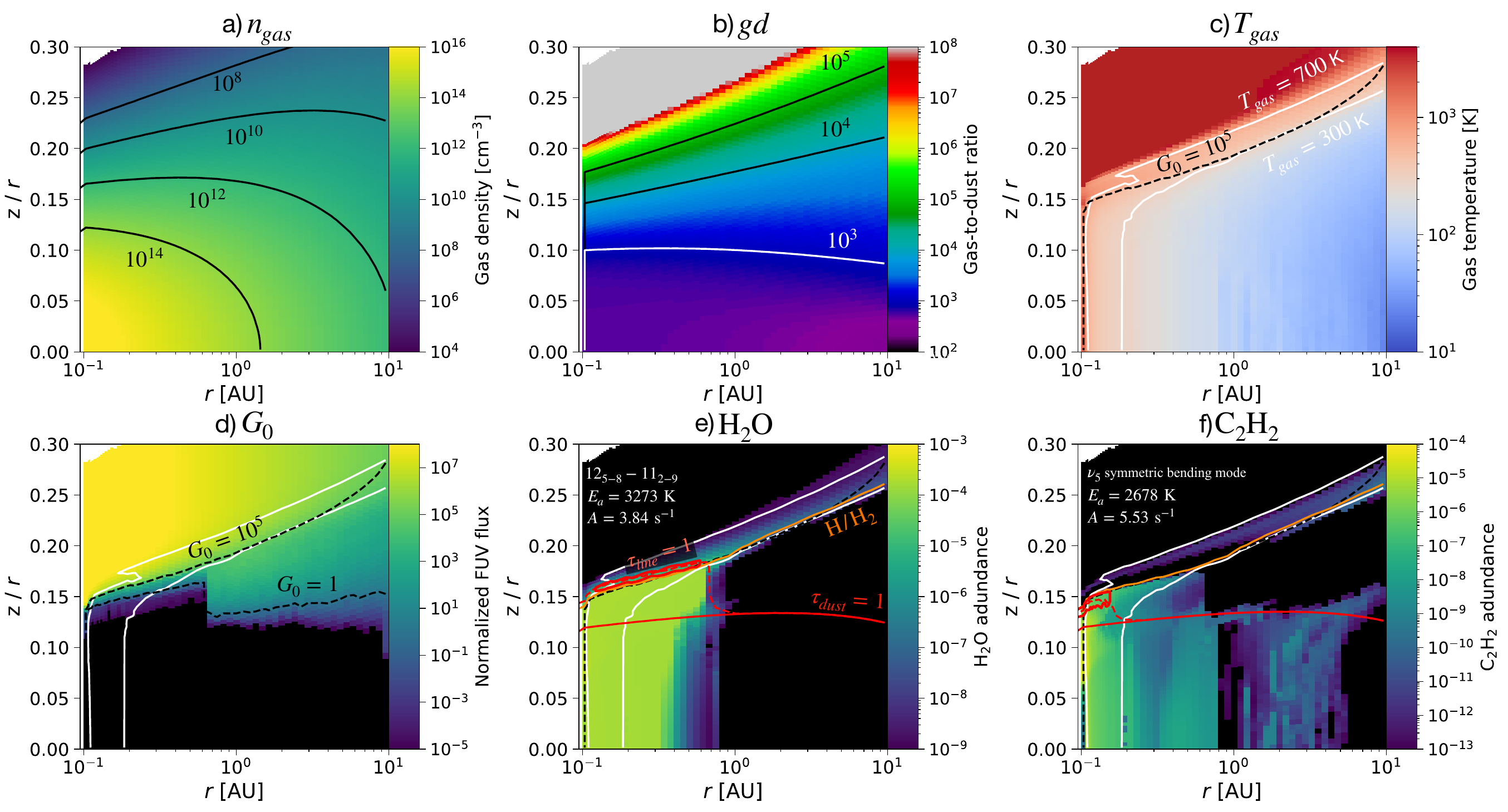}}
\caption{Disk structure of the fiducial model. Top panels: Gas density, local gas-to-dust ratio, and gas temperature. Bottom panels: Normalized UV field $G_0$ (Habing units) and the abundance of H$_2$O and C$_2$H$_2$. The white lines indicate the 300 K and 700 K gas temperature contours. The bottom solid red line shows the dust optically thick surface ($\tau_{\rm{dust}}=1$ at 14 $\mu$m), while the dashed red line represents the surface where $\tau_{\rm{line}}=1$. The red contours correspond to 80\% of the total emitting flux. The dashed orange line in the bottom panels indicate the H/H$_2$ transition, which is sometimes difficult to distinguish from the dashed black line, $G_0=10^5$. }
\label{fig: Fid_model}
\end{figure*}
    
       \subsubsection{Abundances and synthetic spectrum}
       \label{sec: Fid_results}
       \noindent The results of our fiducial model (C/O = 0.47, $gd=10^3$, O/H = $2.88\times 10^{-4}$) are presented in Fig. \ref{fig: Fid_model}. Figure \ref{fig: Fid_model}a shows the gas density structure, while Fig. \ref{fig: Fid_model}b reveals that the dust structure is significantly different, with a strong depletion above $z/r=0.15$ due to the dust settling. Figure \ref{fig: Fid_model}e shows that water is very abundant in the inner disk, in both the shielded regions inside of $r\simeq 0.6~$au and in the warm upper atmosphere extending to large radii. The formation of water is indeed greatly enhanced at high temperature to overcome the barrier to form OH \citep{vanDishoeck_2013ChRv..113.9043V}. Water emits in this warm ($300~\rm{K}$ < $T_{\rm{gas}} < 700~\rm{K} $), upper atmosphere ($z/r > 0.15$) indicated by the red contours, which represent 80\% of the emission of H$_2$O (water line $12_{5-8}  - 11_{2-9}$, $E_u = 3273$ K, $A = 3.84$ s$^{-1}$, and $\lambda=17.10$ $\mu$m). According to Fig. \ref{fig: Fid_model}b, dust is significantly depleted where water emits, with a gas-to-dust ratio $gd\sim5\times 10^4$. The normalized FUV flux $G_0$ (Fig. \ref{fig: Fid_model}d) highlights the important role of H$_2$O in the absorption of the UV flux in inner disks. Indeed, the water distribution perfectly matches the structure of the UV field in the inner disk: when water becomes abundant, $G_0$ drops by orders of magnitude in a couple of cells (see Figs. \ref{fig: Fid_model}d and \ref{fig: Shielding}). It allows more complex, organic molecules such as acetylene to form underneath (Fig. \ref{fig: Fid_model}f and Sect.  \ref{Sec: C2H2chem}).

       Figure \ref{fig: Fid_model}f shows that C$_2$H$_2$ is present three reservoirs:  the inner disk ($r$ < 1 au), the upper layer of the outer disk ($r$ > 1 au, $z/r > 0.20$), and near the midplane in the outer disk ($r$ > 1 au, $z/r < 0.10$), consistent with other thermochemical models \citep{kanwar_hydrocarbon_2024}. Interestingly, even though C/O = 0.47, acetylene is particularly abundant in the first reservoir (inner disk), which will be the main focus of this paper since the IR emission seen by JWST originates from this region. Indeed, the red contours indicate that 70\% of the emission of C$_2$H$_2$ (line with $E_u = 2678$ K, $A = 5.527$ s$^{-1}$ and $\lambda=13.6872$ $\mu$m) is concentrated in the innermost regions of the disk, where $450~\rm{K}$ < $T_{gas} < 1000~\rm{K} $, and $n_{\rm{H}} \sim 10^{13}$-$10^{14}$ cm$^{-3}$. These temperatures are in the typical range derived from slab model fits of the JWST spectra of T Tauri disks\footnote{It is further discussed in Sect. \ref{Sec: excited_cond}} \citep{Gamsan_2023A&A...679A.117G,Grant_2023ApJ...947L...6G,Vlasbom_CXtau_2025A&A...693A.278V,van_dishoeck_diverse_2023,Arulanantham_2025AJ....170...67A}. The emitting region of C$_2$H$_2$ is therefore slightly lower and closer to the star than water {for lines with similar $E_u$}, in between the UV shielded region by water and the optically thick IR layer of the dust. As acetylene emits deeper than water, the dust is less depleted where acetylene emits ($gd\sim9\times 10^3$) than where water emits. The H/H$_2$ transition perfectly matches the layer where C$_2$H$_2$ becomes abundant, showing the crucial role of warm H$_2$ in its formation (see Sect. \ref{Sec: C2H2chem}).

       \begin{figure*}[ht]
       \resizebox{\hsize}{!}{
        \includegraphics[scale=0.2, trim={0 0 0 0cm},clip]{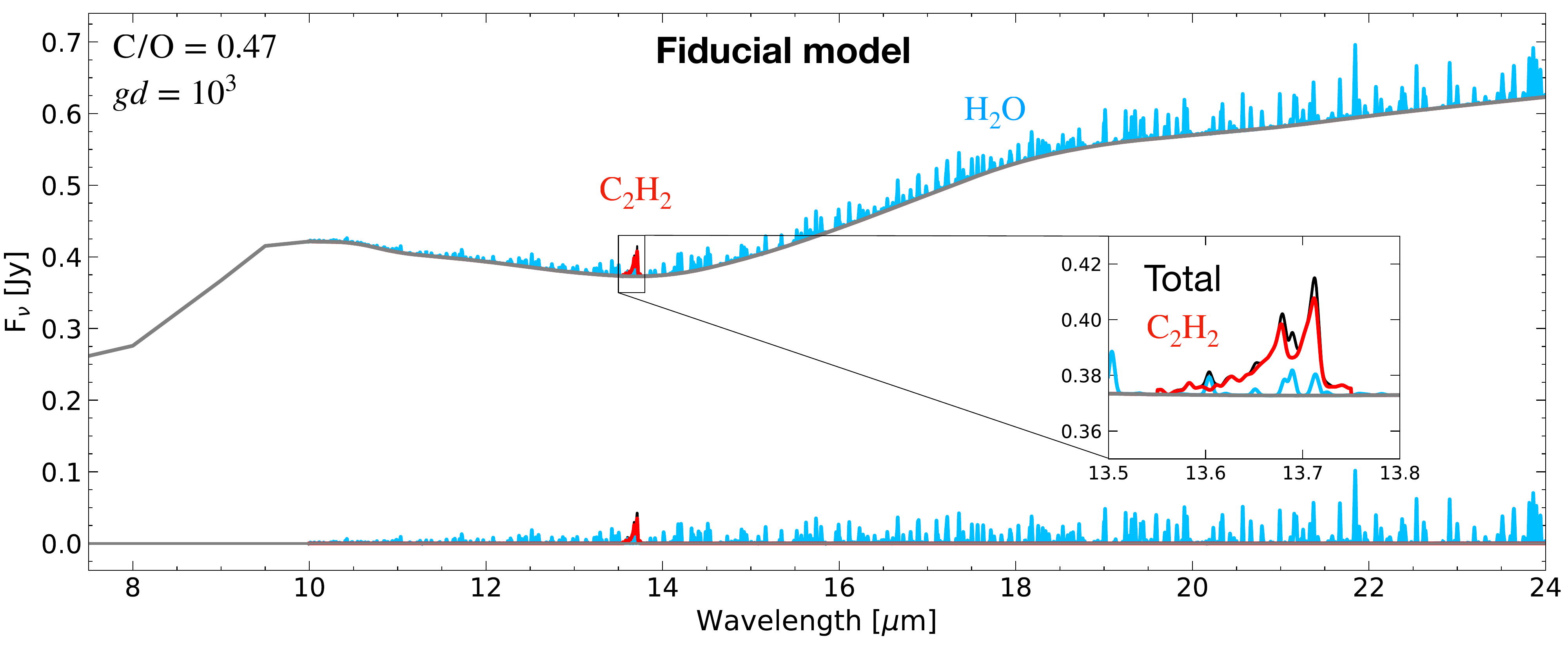}}
        \caption{DALI synthetic spectrum of $\rm{C_2H_2}$ (red) and $\rm{H_2O}$ (blue) for the fiducial model. The total spectrum is shown in black. The spectral resolution is $\lambda/\Delta \lambda = 2000$ to mimic a JWST/MIRI spectrum. Although the fiducial model is richer in oxygen, the $\rm{C_2H_2}$ feature stands out clearly from the forest of water lines.}
        \label{FigVibStab}
       \end{figure*}
    
       The synthetic spectrum of the fiducial model is presented in Fig. \ref{FigVibStab} between 10 and 20 $\mu$m. The continuum is typical for a T Tauri disk, with the silicate bump at 10 $\mu$m. The forest of blue lines corresponds to the water rotational lines, which is expected since the emitting region in Fig. \ref{fig: Fid_model}e is much higher than the dust continuum (surface $\tau_{\rm{dust}}(17.10~\mu m) = 1$). Moreover, the $Q$-branch feature of acetylene at 13.7 $\mu$m clearly stands out, reflecting the high abundance of acetylene seen in Fig. \ref{fig: Fid_model}f, even though the fiducial model has a solar C/O. This is consistent with \textit{Spitzer} and JWST observations of T Tauri disks, systematically detecting H$_2$O and C$_2$H$_2$ \citep{carr_organic_2008,carr_organic_2011,Grant_2025A&A...702A.126G} but in contrast with the very weak emission predicted by modeling works that neglect water UV shielding \citep{Greenwood_dustevo_2019A&A...626A...6G,kanwar_hydrocarbon_2024}.
       
        \subsubsection{What sets the abundance of carbon chains?}
        \label{Sec: C2H2chem}
        \noindent The oxygen chemistry and, in particular, the formation and destruction pathways of H$_2$O have already been largely studied and understood, the main result being that water is abundant in hot, irradiated layers \citep[$T$ > 300 K]{Glassgold_2009ApJ...701..142G,woitke_radiation_2009,vanDishoeck_2013ChRv..113.9043V,walsh_molecular_2015}. On the other hand, the carbon chemistry is much more complex and is not yet fully understood, with it being unclear why C$_2$H$_2$ is so bright in T Tauri disks with a solar C/O. This section focuses on carbon chemistry to address this question. 

\begin{figure}[ht]
        \centering
        \includegraphics[scale=0.25, trim={0cm 0 0 0cm}, clip]{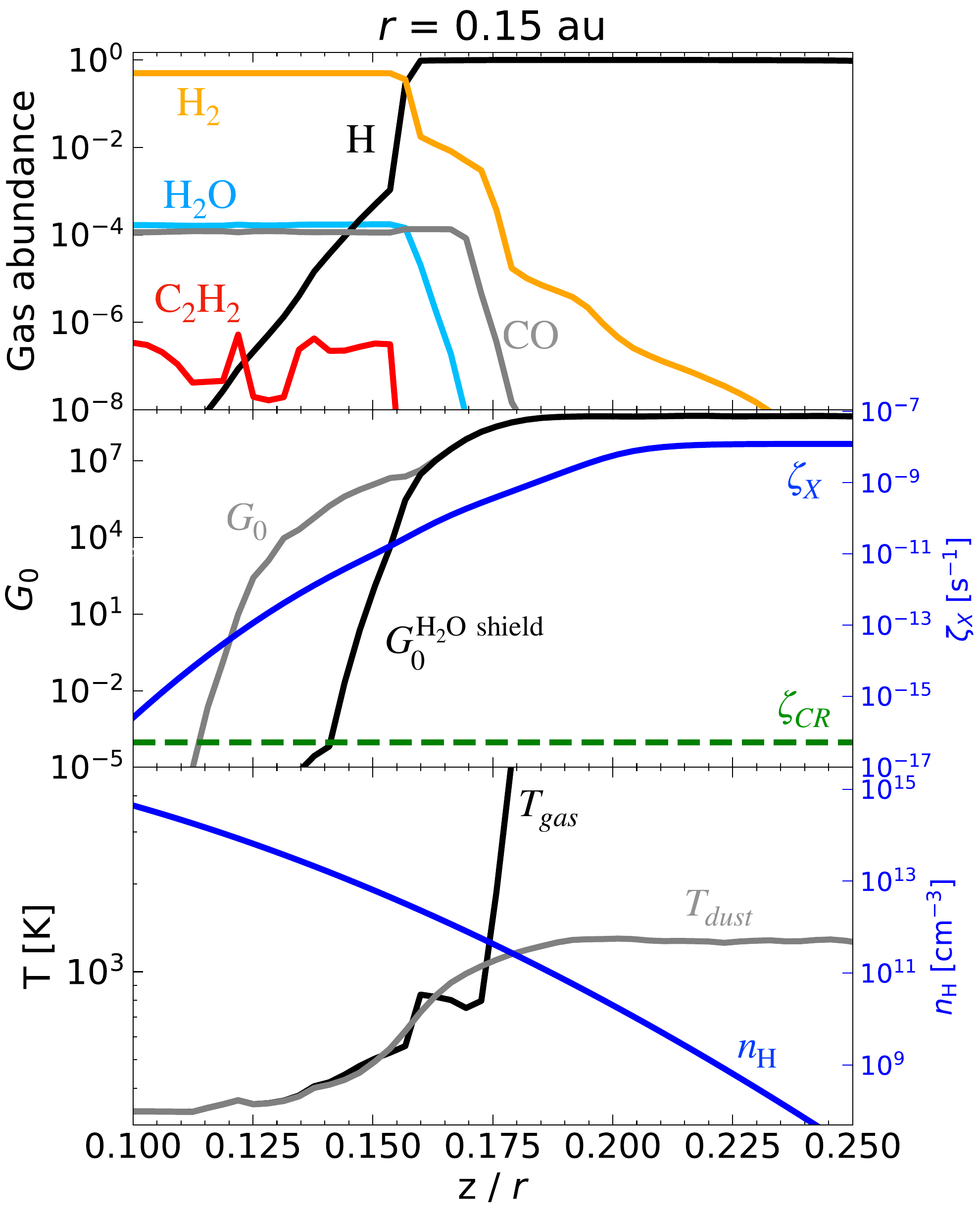}
        \caption{Top: Vertical cut at $r=0.15$ au showing the abundances of several key atoms and molecules. Middle: Irradiation conditions in this vertical cut, with $G_0^{\rm{H_2O~shield}}$ showing the UV field attenuated by water absorption (water shielding). The gray line $G_0$ indicates what the UV field would be without the water UV shielding. Bottom: Gas and dust temperature, with the gas density $n_{\rm{H}}$ in blue.  Water UV shielding allows C$_2$H$_2$ to be abundant much higher in the disk by suppressing UV photodissociation.}
        \label{fig: Shielding}
        \end{figure}

        The formation of hydrocarbons starts from C or C$^+$.
        The first molecule to form is carbon monoxide, CO, from C$^+$ and O (Fig. \ref{fig: Shielding}).
        However, with a solar C/O, there is more oxygen than carbon, so CO locks up almost all the gas-phase carbon. The oxygen left is available to form H$_2$O.
        As has been mentioned in previous studies discussing the formation of organic molecules, the “free” carbon available to form hydrocarbons is released by the dissociation of CO by UV or X-rays \citep{Bast_2013A&A...551A.118B, walsh_molecular_2015, woitke_2d_2024, kanwar_hydrocarbon_2024}. However, our models include the water UV shielding which suppresses the penetration of UV photons in the layer where organic molecules are abundant. This layer is actually X-ray active according to Fig. \ref{fig: Shielding}. Therefore, the main production of carbon comes from the indirect dissociation of CO by X-rays, either by secondary ionization or reaction with He$^+$:
        \begin{equation}
            \mathrm{ CO + \gamma_{FUV,~secondary} \xrightarrow{} C + O}
            \label{eq: Direct CO diss}
        ,\end{equation}
        \begin{equation}
            \mathrm{ CO + He^+ \xrightarrow{} C^+ + O + He}
            \label{eq: indirect CO diss}
        .\end{equation}
        Then, from this carbon released, carbon chains are built mainly from the addition of C (or C$^+$) and H$_2$, and dissociative recombinations with electrons \citep{Bast_2013A&A...551A.118B,kanwar_hydrocarbon_2024}. The H-abstraction reactions often involve large activation barriers, so hydrocarbon formation is favored at high temperatures and in regions where H$_2$ is abundant. This explains why C$_2$H$_2$ is particularly abundant in the inner disk, just below the H/H$_2$ transition, which is deeper than the C/CO transition (Fig. \ref{fig: Shielding}) because the warm chemistry consumes H$_2$ to form OH. However, X-ray induced photodissociation of CO and H$_2$O also releases oxygen. The latter plays an important role in the carbon chemistry by destroying carbon chains to form back CO. This channel of destruction has been overlooked so far in the literature, although \cite{Bast_2013A&A...551A.118B} mentioned that OH can impede the formation of carbon chains at low temperatures by the reaction $\mathrm{C + OH \xrightarrow{} CO + H}$. Therefore, the abundance of carbon chains and especially C$_2$H$_2$ is set by a balance between formation with the carbon released by CO dissociation, and destruction by the oxygen. This balance reveals that C$_2$H$_2$ is not expected to be sensitive to a variation in the X-ray luminosity since X-rays are involved in both its formation and destruction (confirmed by Fig. \ref{fig: diag_plot_negligible_effect}, {with only an increase of a factor of 3 for 6 orders of magnitude in $L_X$}). It is also crucial not to miss any reactions involving H$_2$ or O or it will skew the estimation of carbon chain abundances.

        Finally, \cite{bosman_water_2022-1,bosman_water_2022} and \cite{duval_water_2022} strikingly showed that water UV shielding significantly enhances the abundance of organic molecules. The vertical cut at $r=0.15$ au (Fig. \ref{fig: Shielding}) confirms that C$_2$H$_2$ is abundant when the water UV shielding becomes effective (represented as $G_0^{\rm{H_20~shield}}$). In this layer, water attenuates the UV field by seven orders of magnitude, which suppresses the photodissociation by the UV field {for species with photodissociation threshold at $\lambda \lesssim 180$ nm.}, helping molecules to survive. Because H$_2$O is no longer photodissociated, the abundance of atomic oxygen drops, quenching the destruction route of carbon chain molecules \citep{duval_water_2022}. It also quenches the destruction of H$_2$ by atomic oxygen, shifting the H/H$_2$ transition to higher altitudes. This leads to a higher C$_2$H$_2$ abundance in upper layers as this molecule is particularly sensitive to this transition as well \citep{kanwar_hydrocarbon_2024}. As a result, water UV shielding enhances the acetylene abundance by more than 4 orders of magnitude in the upper atmosphere (consistent with \citealt{duval_water_2022}), and it increases the emission of C$_2$H$_2$ by a factor of 5 (shown in Appendix \ref{appendix: effect water uv shielding on c2H2}). It is interesting to note that when C/O > 1, H$_2$O is much less abundant (Sect. \ref{Sec: modelgrid results}) and no longer attenuates the UV field. Instead, C$_2$H$_2$ and C$_3$ are the main species that shield the gas (see Appendix \ref{appendix:model_co1.5}). 

        In short, the abundance of acetylene is determined by two distinct processes. First, the chemistry shows that it is set by a balance between formation from atomic carbon thanks to CO dissociation by X-rays, and destruction by the atomic oxygen. Second, C$_2$H$_2$ is known to be sensitive to the UV field \citep{walsh_molecular_2015}, so suppressing UV photons with water UV shielding allows organic molecules to be abundant much higher in the disk, revealing C$_2$H$_2$. Our model predicts an emission of C$_2$H$_2$ much brighter than other thermochemical modeling works \citep{woitke_modelling_2018, Greenwood_2019A&A...631A..81G, kanwar_hydrocarbon_2024} and consistent with JWST observations because we enhance C$_2$H$_2$ formation by including more reactions with H$_2$ (see Sects. \ref{Sec: constr_net} and \ref{sec: Discussion}) and we take into account water UV shielding. 
 
    \subsection{Model grid}
    \label{Sec: modelgrid results}
    \begin{figure*}[ht]
    \centering
    \resizebox{\hsize}{!}{
    \includegraphics[scale=0.15, trim={0 0 0 0 cm}, clip]{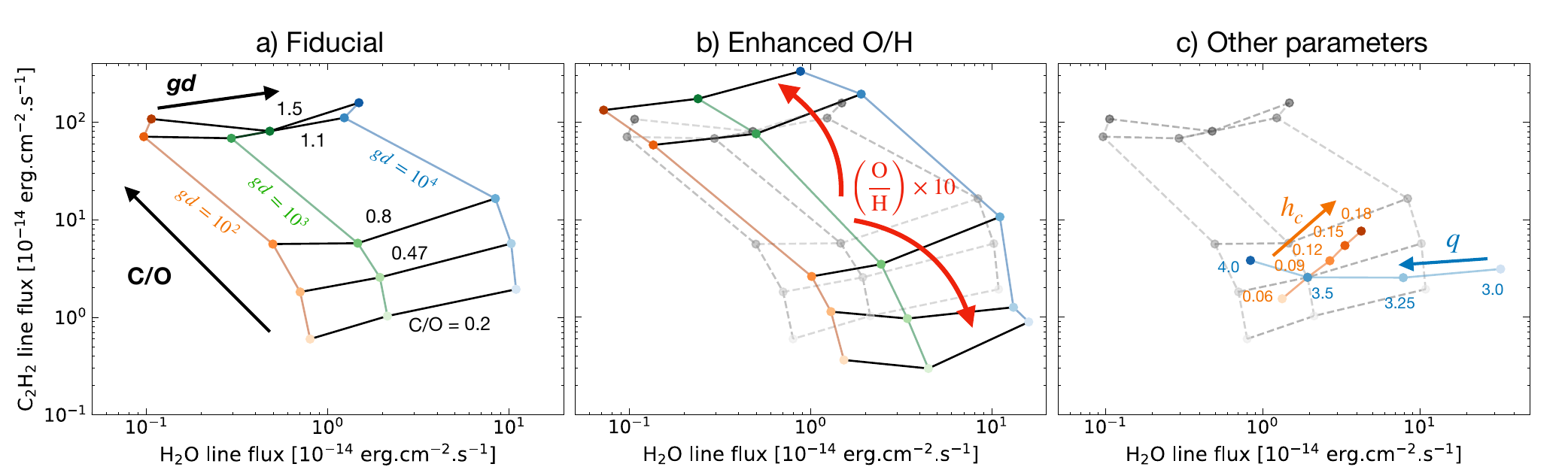}}
    \caption{Left: Result from the fiducial grid. The orange, green, and blue points correspond to $gd=10^2$, $10^3$, and $10^4$, respectively. The black lines highlight a constant C/O ratio. Middle: Result obtained for the {“enhanced O/H”} grid: O/H$\times10$ (C/H is scaled with the C/O ratio). The fiducial grid is overlaid in gray for reference. Right: Results for the disk aspect ratio, $h_C$ (orange), power law index of the dust distribution, $q$ (in blue) with $gd=10^3$. }
    \label{fig: diagnostic_plot}
    \end{figure*}

\subsubsection{Elemental abundances}
\begin{figure}[ht]
\centering
\resizebox{\hsize}{!}{
\includegraphics[scale=0.50, trim={0cm 0 0 0cm}, clip]{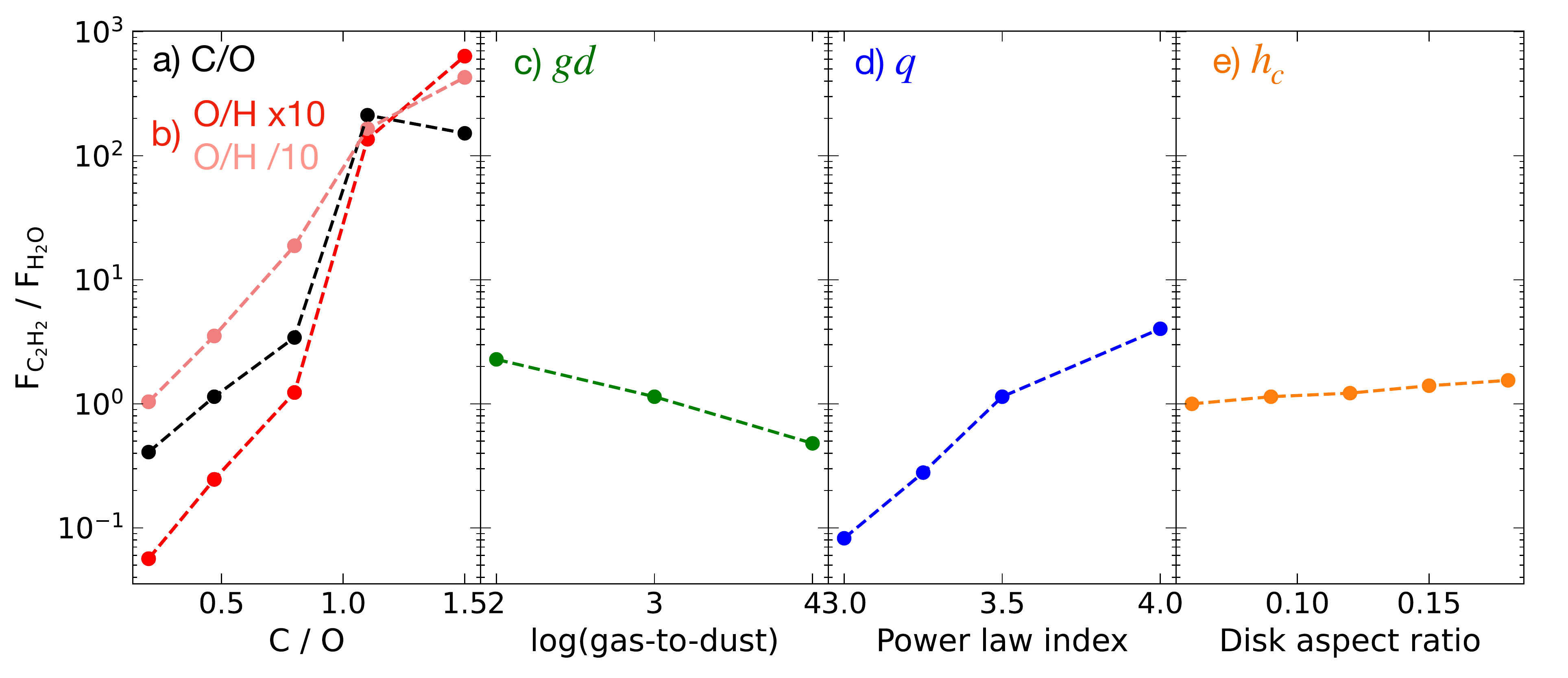}}
\caption{Evolution  of the line flux ratio C$_2$H$_2$/H$_2$O with the elemental abundances (C/O and O/H), gas-to-dust mass ratio ($gd$), power law index ($q$), and disk aspect ratio ($h_C$), from left to right. The fiducial model with C/O=0.47 is adopted in panels (c), (d), and (e), except for the parameters that are varied.}
\label{fig: flux ratio dep}
\end{figure}
 This section focuses on the impact of the elemental composition of disks on the emission of acetylene and water.
 To compare their emission, we followed the integration windows of \cite{Grant_2025A&A...702A.126G}. The emission of C$_2$H$_2$ is integrated over its main $Q$ branch (13.6-13.72 $\mu$m) and the water emission is integrated over three windows: 17.09-17.15 $\mu$m , 17.2-17.245 $\mu$m, and 17.3-17.42 $\mu$m. Figure \ref{fig: diagnostic_plot} presents the flux of water and acetylene obtained with the grid, covering two parameters: C/O ratio and gas-to-dust ratio $gd$. Figure \ref{fig: diagnostic_plot}a (left) shows the fiducial grid.
 
Increasing the C/O ratio {(by increasing C/H)}, more carbon {is available to form molecules, in particular hydrocarbons and CO. In addition, less oxygen is available to form H$_2$O since CO captures more oxygen as well. As a result,} the water emission drops while the acetylene emission sharply increases. The C$_2$H$_2$ emission is more sensitive to the C/O ratio than H$_2$O, varying by a factor of 120 versus 7, respectively, between a C/O ratio of 0.2 and 1.5. Moreover, the jump seen for both molecules around C/O=1 corresponds to a switch in the chemistry. Once C/O > 1, the limiting factor becomes the oxygen. In this case, CO locks up most of the oxygen while free carbon remains: hydrocarbons and organic molecules do not need to form via the dissociation of CO, making them much more abundant (see Appendix \ref{appendix:model_co1.5}). This result is consistent with recent thermochemical model studies \citep{woitke_modelling_2018, Arabhavi_2026A&A...708A..82A}. 
The effect of the C/O ratio is opposite between acetylene and water: one is brighter when the other is fainter. Consequently, the line flux ratio $F_{\rm{C_2H_2}}$/$F_{\rm{H_2O}}$ is strongly dependent on the C/O ratio (Fig. \ref{fig: flux ratio dep}a), which varies by almost three orders of magnitude between C/O = 0.2 and C/O = 1.5. The clear jump at C/O = 1 would also allow two populations of disks to be separated and may thus serve as an important observational signature.

\begin{figure}[ht]
\centering
\includegraphics[scale=0.13, trim={0cm 0 0cm 0cm}, clip]{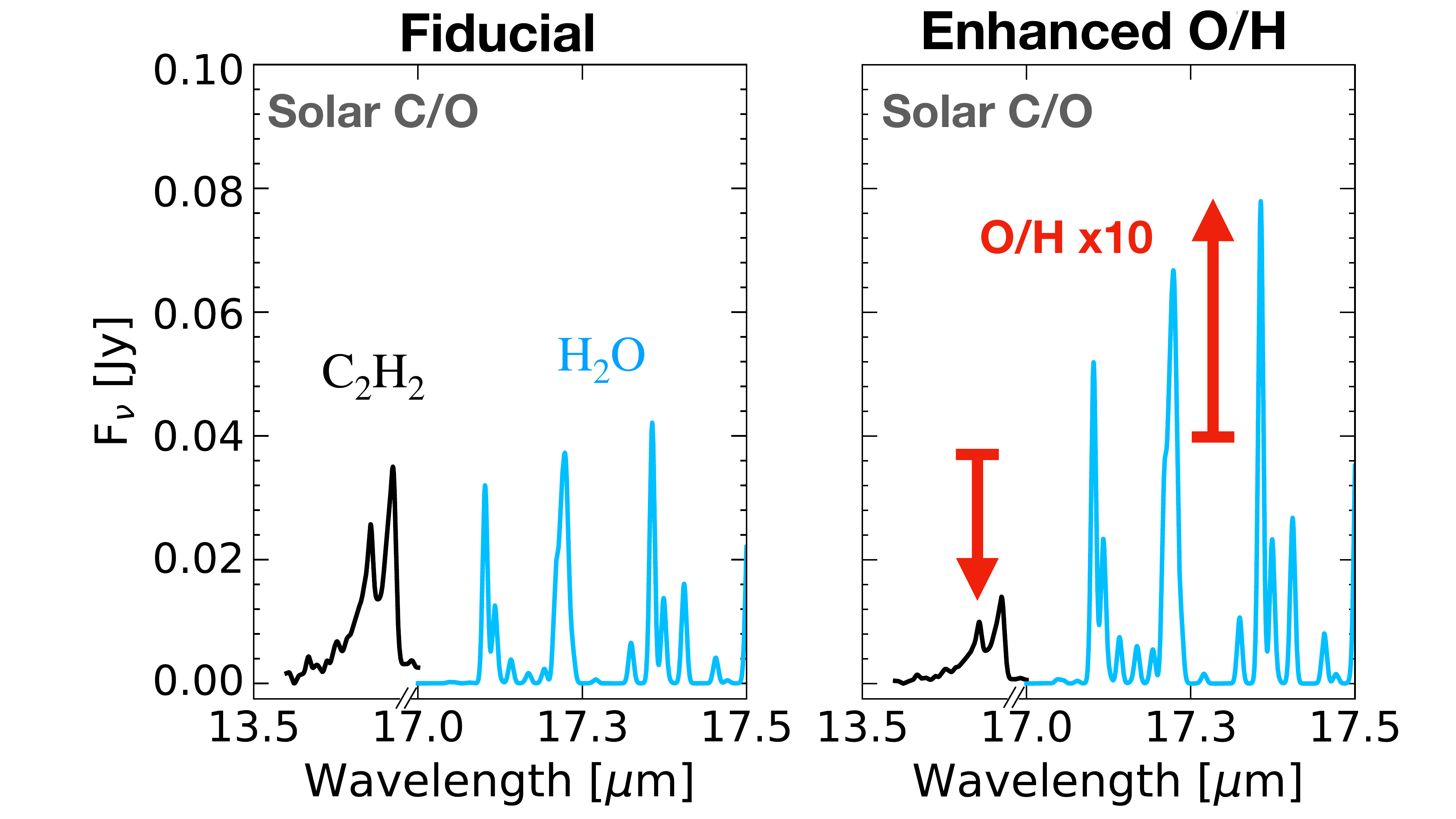}
\caption{DALI synthetic spectra of the fiducial model (left) and a model with O/H x10 (right). The $Q$ branch of C$_2$H$_2$ (black) and the water lines (blue) used for the line flux around 17.25 $\mu$m are shown together for clarity. The carbon and oxygen budget itself acts similarly to a decrease in the C/O ratio: the flux of water increases while acetylene decreases. }
\label{fig: spec_metal}
\end{figure}

As seen in \cite{Arabhavi_2026A&A...708A..82A}, not only the C/O can play a role in the molecular emission, but also the total oxygen and carbon budget (O/H and C/H). With this model grid, we revisited this aspect with DALI by increasing O/H by a factor of ten compared to the standard ISM values ({"enhanced O/H"} models), as could be appropriate for icy grains delivering oxygen in the inner disk (see Sect. \ref{Sec: radial drift}). When comparing two models with exactly the same C/O, Fig. \ref{fig: spec_metal} reveals how different the two spectra are. 
Indeed, the water lines are increased by a factor of $\sim$2 whereas the $Q$ branch of C$_2$H$_2$ drops by about the same factor. The increase in water emission is simply due to the overabundance of oxygen. As water lines are optically thick, line fluxes do not scale linearly with O/H even though its abundance is enhanced by an order of magnitude. The drop in C$_2$H$_2$ emission is, however, due to the destruction of carbon chains by oxygen. Indeed, there is more carbon released by the dissociation of CO, but also more oxygen released by CO and H$_2$O. The excess of oxygen released by water is then converted to CO, by depriving hydrocarbons of the available carbon. Therefore, an increase in elemental abundances pushes the balance of formation/destruction (mentioned in Sect. \ref{Sec: C2H2chem}) toward more destruction of carbon chains, creating more oxygen-rich spectra without changing the C/O ratio. {Similarly, depleting the oxygen abundance pushes this balance toward more formation of acetylene, although the effect is less pronounced due to a weaker water shielding (see the depleted O/H grid in Appendix \ref{appendix:subsolargrid})}. Our results confirm the claim of \cite{Arabhavi_2026A&A...708A..82A} with a different thermochemical model, reinforcing the robustness of these results. 

Figure \ref{fig: diagnostic_plot}b presents the results of the {"enhanced O/H"} grid with an increased elemental abundance of carbon and oxygen (C/H is scaled accordingly to the C/O ratio). Interestingly, the effect discussed above is only true when C/O < 1. When C/O > 1, C$_2$H$_2$ is actually brighter than with a solar O/H because the formation of acetylene is not limited by CO. Thus, increasing the amount of carbon with a C/O > 1 results in more carbon chains. The flux of water decreases slightly because its formation is in competition with CO. {The "depleted O/H" grid (given in Appendix \ref{appendix:subsolargrid}) follows the same trend but in the opposite directions.}

The carbon and oxygen budget itself acts as a variation in the C/O ratio: it stretches the grid in the same direction as the C/O ratio, revealing a {partial} degeneracy between these two chemical parameters. {According to Fig. \ref{fig: flux ratio dep}, it is still possible to differentiate between C/O < 1 and C/O > 1, but determining the exact C/O ratio requires knowledge of the O/H ratio, which can be constrained using other molecular emission lines \citep{Arabhavi_2026A&A...708A..82A}.}

\subsubsection{Dust properties}
\label{sec: dust_prop}
\noindent Our model grid also explores the impact of dust properties on C$_2$H$_2$ and H$_2$O emission. This section first describes the effect of the gas-to-dust ratio, then the parameters that fix the dust size distribution: the power law index, $q$ ($f(a)\propto a^{-q}$), and the minimum and maximum grain size, $a_{min}$ and $a_{max}$. 

An increase in the gas-to-dust ratio (red, green and blue in Fig. \ref{fig: diagnostic_plot}a) enhances all the fluxes, moving the predictions to the top right of $F_{\rm{H_2O}}$-$F_{\rm{C_2H_2}}$ plane. In fact, decreasing the $gd$ ratio (so increasing the amount of dust) strengthens the dust continuum by increasing the total dust cross section per hydrogen atom \citep{facchini_different_2017}. 
The increased dust opacity cools the medium and pushes the line-emitting region into less dense regions, effectively reducing the emission of water and acetylene. 
This opacity effect is in agreement with previous modeling works on H$_2$O emission \citep{Maijereink_2009ApJ...704.1471M, Antonellini_2015A&A...582A.105A}. However, unlike the C/O ratio, the effect of the $gd$ ratio is the same for both water and acetylene; therefore, grain depletion does not affect $F_{\rm{C_2H_2}}$/$F_{\rm{H_2O}}$ (Fig. \ref{fig: flux ratio dep}b) as previously suggested by observational works  \citep{Tabone_2023NatAs...7..805T, arabhavi_abundant_2024, Grant_2025A&A...702A.126G}.

{Dust properties in the inner disks are expected to vary according to spectral energy distribution fitting \citep{Ribas_2020A&A...642A.171R, Kaufer_2023A&A...672A..30K} and theoretical works \citep{birnstiel_dust_2015, Birnsitel_2024ARA&A..62..157B}.
Here, we explore how the power law index $q$ of the dust size distribution ($f(a)\propto a^{-q}$) affects molecular features in the MIR (blue in Fig. \ref{fig: diagnostic_plot}c ).} Unlike the gas-to-dust ratio, the power-law index strongly changes the relative strength of water and acetylene emission. A flatter dust size distribution (smaller $q$, more large grains with regard to small grains) significantly boosts water while acetylene emission decreases slightly. This effect comes from the dust settling. Indeed, small grains are well coupled to the gas as opposed to large grains. Adopting a flatter dust distribution means that the vertical gradient of the gas-to-dust ratio would be significantly more pronounced  because there are only small grains in the upper layers. As water emits vertically higher than acetylene, the contrast between water and acetylene emission is increased because the region where water emits would be much more depleted. 

The minimum and maximum grain size do not appear in Fig. \ref{fig: diagnostic_plot}c as they do not change the flux ratios as significantly {as the power-law index} (see Appendix \ref{appendix:paramwithnegligibleimpact}) . By increasing the minimum grain size from 5 nm to 100 nm, UV photons penetrate deeper in the disk. Molecules are photodissociated more efficiently, {pushing their emitting layers deeper. However, }
the IR continuum remains similar because small grains (below 100 nm) do not change the MIR opacity significantly. {Consequently, C$_2$H$_2$ emission is somewhat reduced as it emits closer to the optically thick dust region. Water emission is less affected by dust continuum because it emits from slightly higher up and further out layers. Pushing its emitting region into denser layers increases its emission, similar to an increase in the gas-to-dust ratio.} This effect may be different if the minimum grain size becomes larger than 0.5 $\mu$m \citep{woitke_consistent_2016}, but this is beyond the scope of this paper. Increasing the maximum grain size for a fixed gas-to-dust ratio is equivalent to increasing the gas-to-dust ratio in the upper atmosphere, because more mass is then carried by large and well-settled grains. The results shown in Appendix \ref{appendix:paramwithnegligibleimpact} confirm this behavior, with limited increase in the resulting molecular emission {and a minor change in the C$_2$H$_2$/H$_2$O flux ratio}.

\subsubsection{Geometrical parameters}
\noindent We finally explored the effect of the geometry on molecular emission by changing the disk aspect ratio, $h_C$, and the flaring angle, $\psi$. When increasing the disk aspect ratio (orange in Fig. \ref{fig: diagnostic_plot}c), the disk can intercept more photons from the star (the solid angle is larger), making it warmer. This extends the emitting layer of molecules, increasing their emission. In particular, C$_2$H$_2$ is more sensitive to the increase in temperature than H$_2$O. As a result, a thicker disk would extend the emitting layer of acetylene more than water, which is why the disk aspect ratio is more effective on acetylene than on water, with a slight increase in the line flux ratio (Fig. \ref{fig: flux ratio dep}).

The effect of the flaring angle is negligible on water and acetylene emission (therefore not displayed in Fig. \ref{fig: diagnostic_plot}c). This result is in apparent contradiction with \cite{Antonellini_2015A&A...582A.105A, Greenwood_2019A&A...631A..81G} who found that both H$_2$O and C$_2$H$_2$ are brighter when the disk is more flared. However, their reference points for the disk height is 0.1 and 100 au, respectively, meaning that their disk aspect ratios at 0.5 or 1 au change as well. Therefore, they likely traced the effect of disk aspect ratio rather than the flaring angle. Nevertheless, it is interesting to note that the dust continuum is significantly flatter for $\lambda\geq15$ $\mu$m when we decrease the flaring angle, even though it is not the focus of this paper.

\section{Discussion}
\noindent In this section we discuss the impact of the chemical networks on C$_2$H$_2$ abundance, and explore whether other carbon chains are expected to be abundant in this region. Then, we confront our results with JWST observations to provide first constraints on C/O and O/H in inner disks around T Tauri stars.
\label{sec: Discussion}
    \subsection{Chemical networks and C$_2$H$_2$}
        \begin{figure}[ht]
        \centering
        \includegraphics[scale=0.18, trim={10cm 0 15 0cm}, clip]{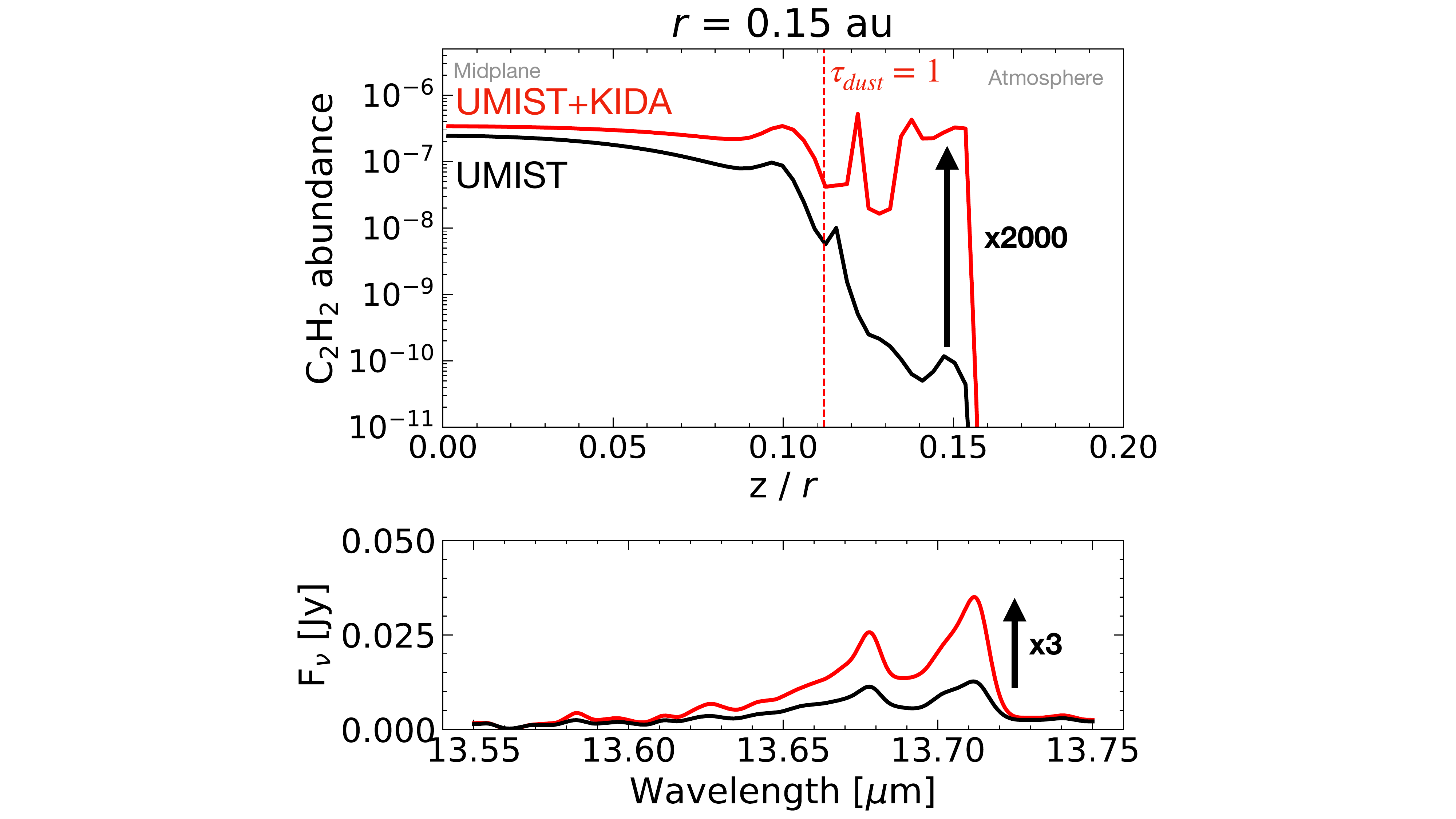}
        \caption{Vertical cut at $r=0.15$ au showing the abundance of C$_2$H$_2$ (top) and the corresponding spectrum (bottom) for two chemical networks: the fiducial (in red) and the one based on UMIST only {(including also three-body reactions and $\rm C+H_2O \xrightarrow{} HCO + H$)}.}
        \label{fig: Umistvskida}
        \end{figure}
    Our models are based on an extended chemical network, combining reactions from the latest version of UMIST \citep[RATE22]{millar_umist_2023} and KIDA \citep{wakelam_2024_2024}. This section highlights the sensitivity of C$_2$H$_2$ to chemical networks, based only on UMIST (as previous DALI networks) or on UMIST+KIDA. {For this comparison, we build a network following the method described in Sect. \ref{Sec: constr_net} without adding reactions from KIDA (so this network also includes three-body reactions and $\rm C+H_2O \xrightarrow{} HCO + H$)}. Figure \ref{fig: Umistvskida} presents the C$_2$H$_2$ abundance as a function of the vertical height at $r = 0.15$ AU obtained with these two chemical networks. 
    When comparing the emitting layers (right side of the vertical dashed red line), there are more than 3 orders of magnitude difference in C$_2$H$_2$ abundance, resulting in a factor $\sim4$ in C$_2$H$_2$ emission (bottom panel). This difference is mainly explained by reactions between H$_2$ and hydrocarbons (Appendix \ref{appendix:key_reactiionkida}), which are present in KIDA from the high temperature network of \cite{harada_new_2010}, but missing in UMIST. Interestingly, \cite{anderson_observing_2021} also used reactions from this network, and C$_2$H$_2$ was abundant in the inner disk of their heated model, which probably comes from the reactions of \cite{harada_new_2010}. However, these reactions between carbon chains and H$_2$ are poorly characterized and little studied. A good example is the reaction $\mathrm{C_2+H_2 \xrightarrow{} C_2H + H}$. This reaction is missing in UMIST, which might explain why \cite{walsh_molecular_2015} also do not highlight this reaction in their formation scheme of C$_2$H$_2$. Due to a different reactivity between two close-lying electronic states of C$_2$, NIST considers an activation barrier of 4000 K, inconsistent with 1420 K in \cite{harada_new_2010}, the smaller one being the right barrier under inner disk conditions (M. van Hemert, private communication)\footnote{See Appendix \ref{focus_C2_H2} for more details about this inconsistency.}. More work has been done on the endothermicities \citep{tinacci_gretobape_2023}, which enables discrimination between strongly endothermic reactions such as $\rm{C_3 + H_2 \xrightarrow[]{} C_3H + H}$ (see Appendix \ref{appendix:endo_tinacci}). Still, most of these reactions need to be studied to estimate the possible activation energies using quantum calculations or experimental measurements, which might ultimately change the abundances obtained with this chemical network. 

        \begin{figure}[ht]
        \centering
        \includegraphics[scale=0.12, trim={0cm 3cm 0 2cm}, clip]{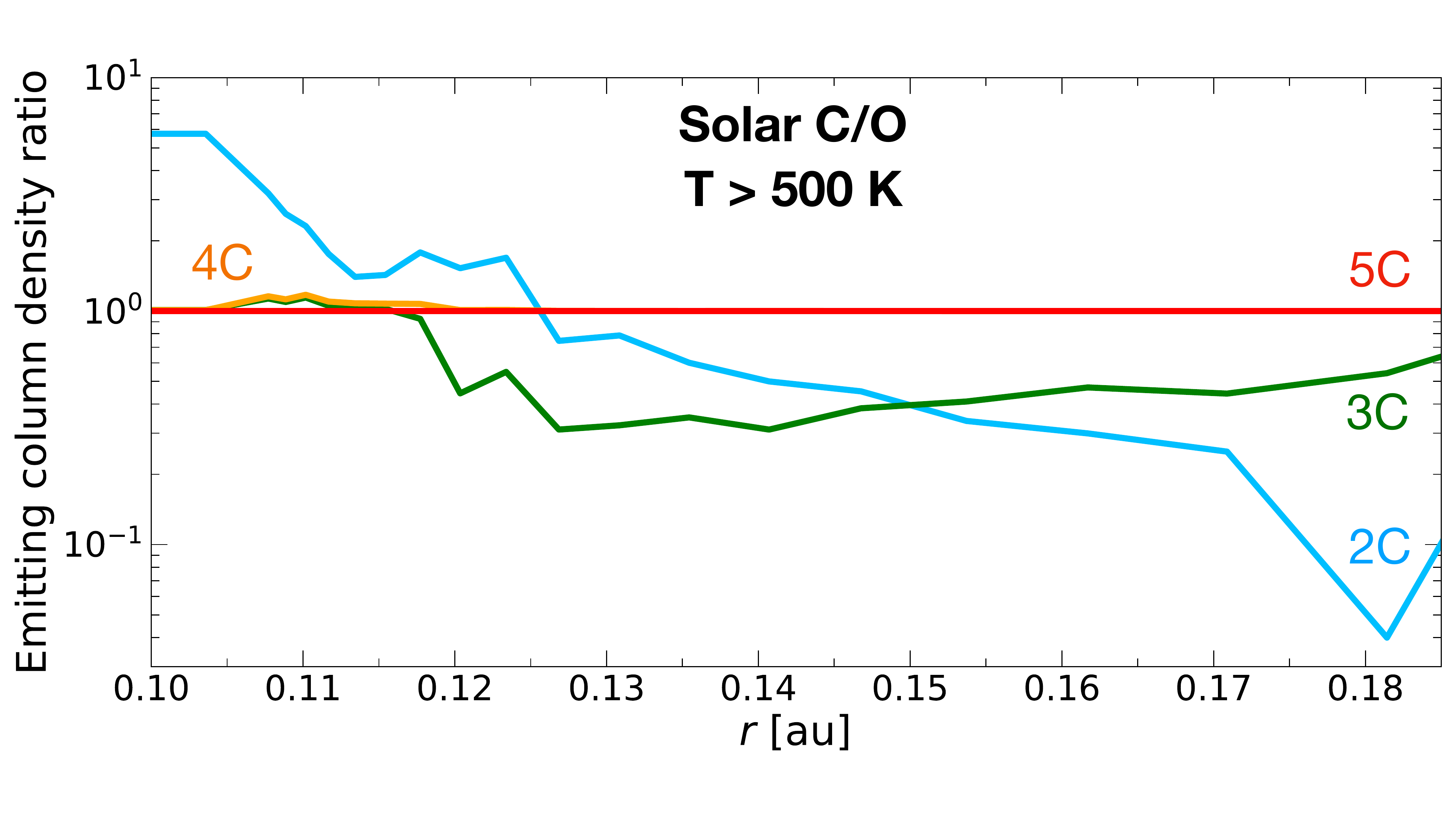}
        \caption{Emitting column density ratio (above the surface $\tau_{\rm{}dust} = 1$ at 14 $\mu$m and $T_{\rm{}gas} > 500$ K) of C$_2$H$_2$ obtained for chemical networks including various maximum number of carbon in hydrocarbons. {The reference is the fiducial network including 5 atoms of carbon (5C, in red).} }
        \label{fig: length network}
        \end{figure}
        In addition, we examine the sensitivity of acetylene abundance to the length of carbon chains included in the chemical network. We construct three other networks, “2C,” “3C,” and “4C,” with a maximum number of carbon atoms in hydrocarbons of, respectively, 2, 3, and 4  (the fiducial network includes hydrocarbons with a maximum of five carbon atoms). Figure \ref{fig: length network} depicts the variation in the emitting column density of C$_2$H$_2$ in these {three networks compared to our fiducial network.} {This sensitivity analysis shows that the network with a maximum of four atoms of carbon is sufficient to predict robust acetylene abundance. By extension, this would suggest that modeling the emission of an hydrocarbon with $N_C$ atoms of carbon requires the chemical network to extend at least up to $N_C+2$ atoms of carbon. The variation in the C$_2$H$_2$ column density as a function of the length of the network is due to the successive opening of complex destruction and formation routes, which can enhance or reduce C$_2$H$_2$ abundance. For example, the 2C network underpredicts the amount of C$_2$H$_2$ outside of 0.13~au due to the reaction
        \begin{equation}
        \label{H2OC}
            \mathrm{H_2O + C \xrightarrow[]{} HCO + H}
        ,\end{equation}
        which efficiently recycles C back to CO due to the lack of an efficient route to integrate free carbon into carbon chains.}
        In the 3C network, {the reaction (\ref{H2OC}) competes with the two main reactions that lock free carbon in carbon chains} \citep{Chastaing_1999PCCP....1.2247C,kanwar_hydrocarbon_2024}:
        \begin{equation}
            \mathrm{C_2H_2 + C \xrightarrow[]{} C_3 + H_2}\\
            \mathrm{C_2H_2 + C \xrightarrow[]{} C_3H + H}
        ,\end{equation}
        {but these reactions also destroy C$_2$H$_2$. Therefore, the resulting abundance of C$_2$H$_2$ is not systematically higher than for the 2C network as most of the carbon is converted into C$_3$ (around 80\%).} The column density of C$_2$H$_2$ converges with the 4C network. The increase in C$_2$H$_2$ comes notably from a new formation pathway \citep{LOISON_2017MNRAS.470.4075L}: 
        \begin{equation}
            \mathrm{C_4H_3^+ + e^- \xrightarrow[]{} C_2H_2 + C_2H}
        .\end{equation}
        This is an example of a loop that sustains C$_2$H$_2$ even if we increase the number of carbon in species of the network. It also reveals the central place of acetylene in carbon chemistry, making it particularly abundant in warm layers of inner disks.
    \subsection{Hydrocarbons beyond acetylene} 
    
    \begin{figure}[ht]
    \centering
    \includegraphics[scale=0.13, trim={0cm 0 0 0cm}, clip]{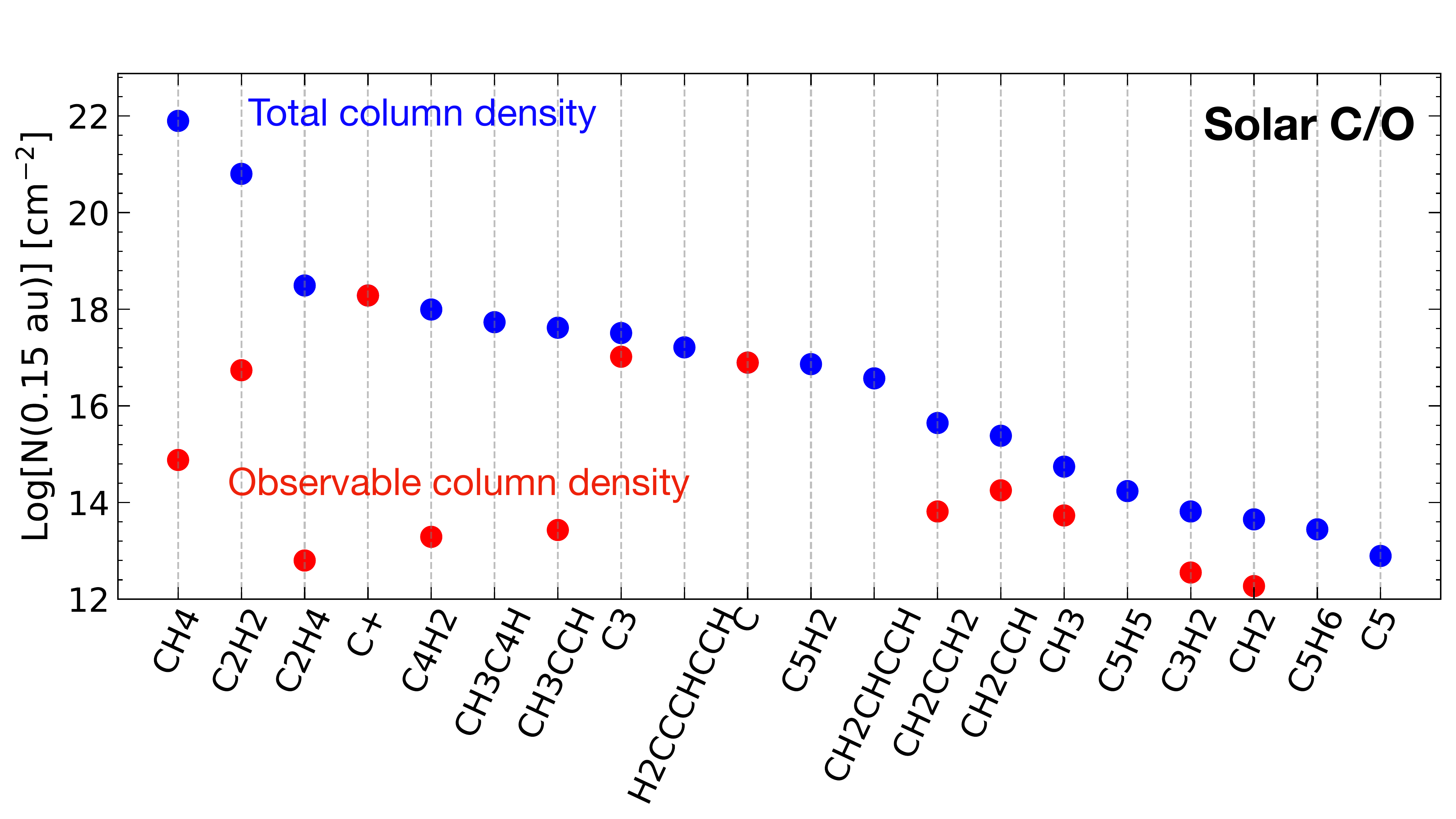}
    \caption{Vertical column densities at $r=0.15$ au for hydrocarbons. The total vertical column density is shown in blue. The red dots represent the emitting column density (above the surface $\tau_{\rm{}dust} = 1$ at 14 $\mu$m and $T_{\rm{}gas} > 500$ K). While CH$_4$ is the major hydrocarbon, C$_2$H$_2$ is dominant in the emitting layers. }
    \label{fig: N_stable_CO=0.47}
    \end{figure}
    
    C$_2$H$_2$ is detected in nearly all disks around T Tauri, according to \textit{Spitzer} \citep{pontoppidan_spitzer_2010} and JWST observations \citep{Arulanantham_2025AJ....170...67A,Grant_2025A&A...702A.126G}. However, the other hydrocarbons have not been detected so far in T Tauri disks, except C$_4$H$_2$ recently \citep{colmenares_jwstmiri_2024}. Here, we propose a broader view of hydrocarbons to try to understand why species other than C$_2$H$_2$ are hardly detectable.
    With a solar C/O, Fig.  \ref{fig: N_stable_CO=0.47} shows that most of the free carbon is contained in methane and acetylene (total column density in blue), which clearly stand out compared to the other species. Interestingly, the emitting column density of methane is orders of magnitude smaller than C$_2$H$_2$ (red dots): most of the methane is hidden in deep layers of the disk, whereas acetylene is very abundant in upper and warm layers of the disk (see the abundance map Fig. \ref{fig: Fid_model}). This may explain why methane is hardly detectable (non detections listed in \citealt{Temmink_2025A&A...699A.134T} as example) while C$_2$H$_2$ is bright in disks around T Tauri. 
    
    Regarding the other hydrocarbons, C$_3$ is surprisingly abundant in emissive layers, although never detected so far in disks (difficult to detect in the MIR because of its main feature overlapping with CO at 4.7 $\mu$m). Next, C$_4$H$_2$ has a relatively high emitting column density and has been detected for the first time around a surprisingly carbon-rich T Tauri disk in \cite{colmenares_jwstmiri_2024}. However, the two isomers of C$_3$H$_4$ (CH$_2$CCH$_2$ and CH$_3$CCH) have roughly the same emitting column densities as C$_4$H$_2$ but have never been observed by JWST around T Tauri stars, possibly due to a unfavorable spectroscopy compared to C$_4$H$_2$.  Interestingly, these are the same hydrocarbons that have been recently detected for the first time around VLMSs \citep{Tabone_2023NatAs...7..805T,arabhavi_abundant_2024} and brown dwarfs \citep{Arabhavi_2025A&A...699A.194A,MoralesCalderon_2025arXiv250805155M}. Their greater stability compared to other hydrocarbons prevents them from reacting with H$_2$, which may explain their abundance in the emitting layers. {Given that the cross sections for most of these species are not known, the exact emitting column densities should be interpreted with caution. They strongly depend on the shape of the UV cross section, especially for $\lambda > 180$ nm where UV photons are not absorbed by water. Adopting the water cross section for these species increases their emitting column density by an order of magnitude as their photodissociation is suppressed by water UV shielding, but it does not favor long carbon chains.}
    
    \subsection{Dust settling prescriptions}
    \begin{figure}[ht]
    \centering
    \includegraphics[scale=0.13, trim={0 0 0 0cm}, clip]{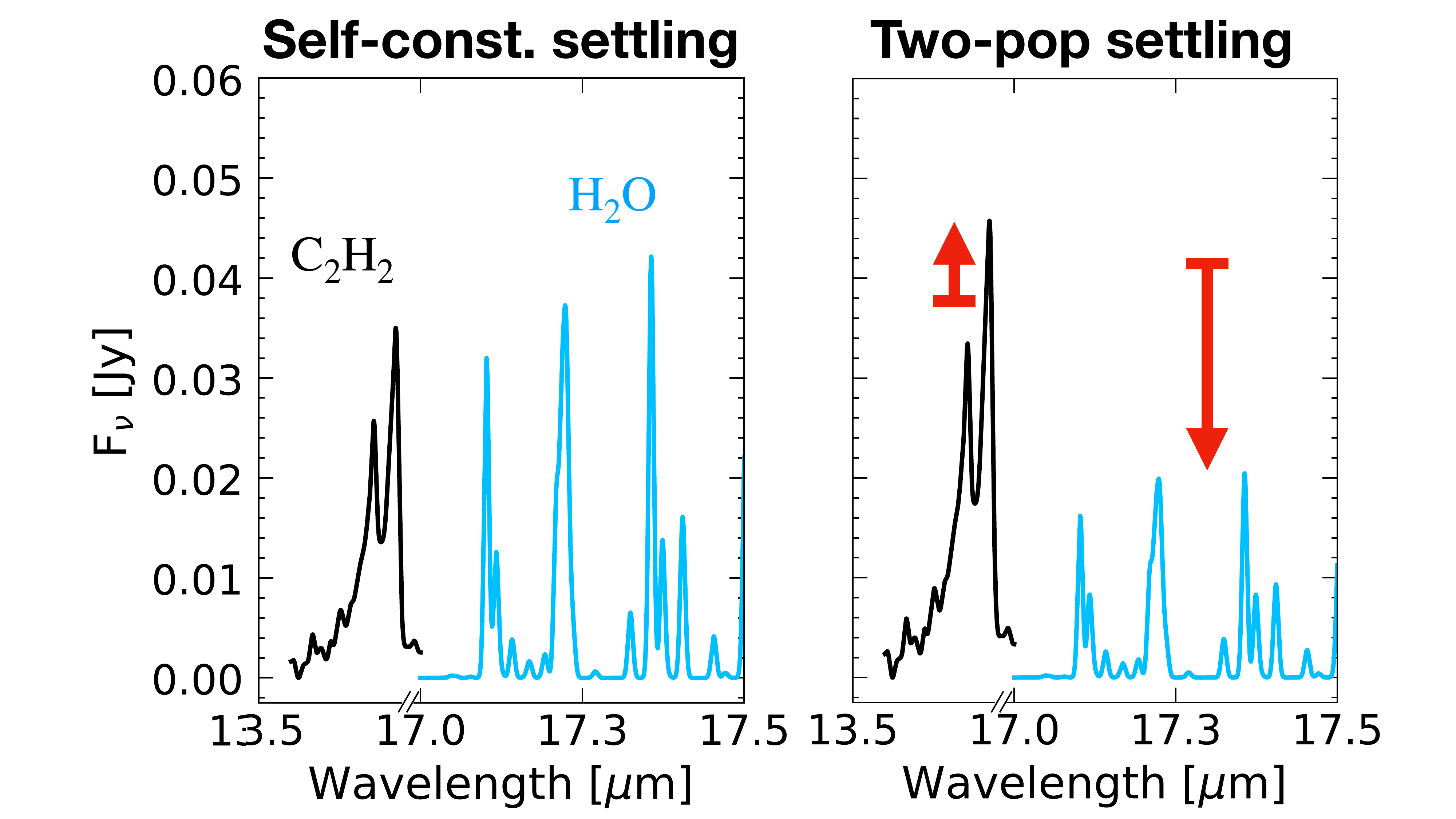}
    \caption{DALI synthetic spectra for two dust settling prescriptions. Left: Fiducial spectrum with self-consistent dust settling (Riols settling; \citealt{RiolsLesur2018}). Right: Spectrum obtained with the two-pop settling prescription, based on two dust populations (small and large). The two-pop settling reduces water emission by a factor of 2.} 
    \label{fig: diagnostic_plot_obs_settling}
    \end{figure}

    Our results show that molecular emissions are sensitive to dust properties. Therefore, $a_{min}$, $a_{max}$ or $q$ defining the dust distribution prove to be important parameters to consider for the emission of acetylene and water. Another modeling consideration that could change the dust distribution is the prescription of the dust settling. So far, we used a self-consistent dust settling, following "Riols settling" hereafter. We explore here if the standard DALI prescription (called "two-pop settling" hereafter, as reference to the two populations of grains considered) would change the results. In this prescription, dust is modeled by 2 components: small grains (5 nm - 1 $\mu$m) and large grains (5 nm - 1 mm; following \citealt{alessio2006ApJ}, consistent with \citealt{2011ApJ...732...42A}) with the same composition (mixture of 60\% silicate and 40\% graphite, \citealt{Weingartner&Draine2001ApJ...548..296W}). The scale height for the small population is set to $h$ (the same as the gas), while the scale height of large grains is reduced to $\chi h$, with $\chi = 0.2$. As for Riols settling, we consider a gas-to-dust mass ratio of $gd=10^3$ with a fraction of large grains over small grains, $f_{large/small} = 0.90$. It corresponds to $gd=10^4$ in the surface layers, consistent with the gas-to-dust ratio where H$_2$O and C$_2$H$_2$ emit (see Fig. \ref{fig: Fid_model}b).
    
    Figure \ref{fig: diagnostic_plot_obs_settling} shows the two spectra obtained with these two prescriptions, focusing on the $Q$ branch of acetylene and water lines around 17 $\mu$m. The two-pop settling underpredicts water emission by a factor of 2 compared to Riols settling, whereas the acetylene remains almost the same, with a slight increase. This different behavior arises from the two distinct line-emitting regions of these molecules. As mentioned in Sect. \ref{sec: Fid_results}, the gas-to-dust ratio is lower where C$_2$H$_2$ emits ($gd\sim9\times10^3$) than where water emits ($gd\sim5\times10^4$). We also show that molecular emission is increased when the grains are depleted (see Sect. \ref{Sec: modelgrid results}). Consequently, the gas-to-dust ratio of $10^4$ in the upper layers of the two-pop settling model would reduce the emission of water, since the gas-to-dust ratio is locally lower (there is more dust) than in the Riols settling prescription. In contrast, where C$_2$H$_2$ emits, the difference in the gas-to-dust ratio is small; hence, this increase is negligible. Extending this result to other species, since the gradient of the gas-to-dust ratio is very strong in the IR emitting layers, we expect that the two-pop settling would reduce the emission of species with higher line formation regions, such as OH, CO, and H$_2$O, but should increase or have a negligible impact on species emitting from deeper layers. Future studies are needed to confirm this.
        
    \subsection{Comparison with observations}
    \begin{figure*}[ht]
    \centering
    \resizebox{\hsize}{!}{
    \includegraphics[scale=0.15, trim={0 0 0 0 cm}, clip]{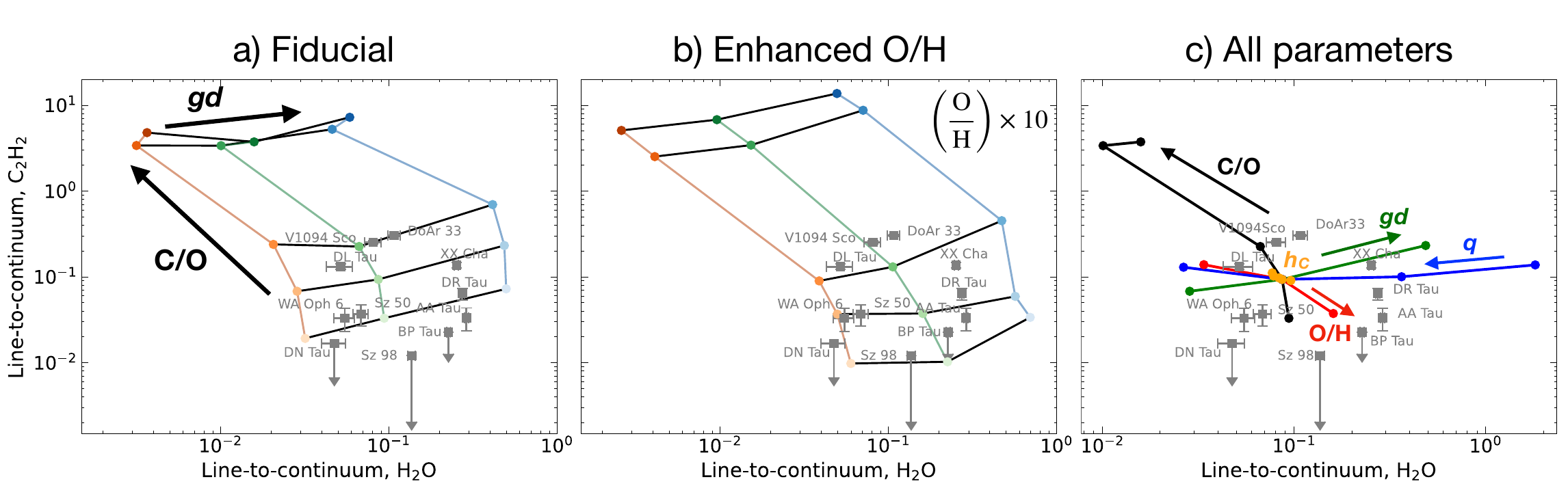}}
    \caption{Same as Fig. \ref{fig: diagnostic_plot} but showing the line-to-continuum ratios for acetylene and water. Panel (c): Influence of the parameters covered in this work: black for C/O, red for O/H, green for $gd$, blue for $q$, and orange for $h_C$. MINDS observations are shown with gray squares, with fluxes exported from \cite{Grant_2025A&A...702A.126G}. We include only T Tauri full disks, without cavities. This plot suggests that a C/O > 1 is excluded in T Tauri disks, and most of the disks are better reproduced by a high O/H or low C/O.}
    \label{fig: diagnostic_plot_obs}
    \end{figure*}
    
    The comparison of the model grid with observations from the MIRI Mid-INfrared Disk Survey (MINDS) GTO program (PID: 1282, PI: T. Henning; see \citealt{henning_minds_2024}) {and DoAr 33 \citep{colmenares_jwstmiri_2024}} is shown in Fig. \ref{fig: diagnostic_plot_obs}. For this comparison, we select only the disks with similar characteristics as the DALI models (full T Tauri disks orbiting a star of luminosity close to solar). We export the line flux from \citet{Grant_2025A&A...702A.126G}. The ten selected disks span over an appreciable range of luminosities (0.41 - 1.9 $L_{\odot}$) so we plot the line-to-continuum ratio instead of the absolute fluxes to free ourselves from this dependency. The line-to-continuum ratios are calculated with the same method as \cite{Grant_2025A&A...702A.126G}, by dividing the peak flux of C$_2$H$_2$ ($Q$ branch) and H$_2$O (between 17.3514 $\mu$m and 17.36 $\mu$m) by the continuum flux at the same wavelength. We refer to \citet{Grant_2025A&A...702A.126G} for further information on these disks.
    
    The fiducial grid is presented in Fig. \ref{fig: diagnostic_plot_obs}a. The absolute fluxes of C$_2$H$_2$ and H$_2$O predicted by DALI are in line with the observations. Our fiducial model (with solar C/O and O/H, $gd=10^3$) reproduces disks with relatively strong C$_2$H$_2$, in between Sz 50 and DL Tau. {The enhanced O/H grid (Fig. \ref{fig: diagnostic_plot_obs}b) covers most of the JWST observations, indicating that the oxygen enrichment seems to be a common feature in T Tauri disks. Figure \ref{fig: diagnostic_plot_obs}c also suggests T Tauri disks showing the strongest water emission (with regard to to the continuum) might experience grain growth, possibly combined with a stronger settling leading to a depleted atmosphere, as for XX Cha or DR Tau. Interestingly, the region with C/O > 1 is always empty regardless of which parameter we take. {Even disks showing the strongest C$_2$H$_2$ emission, such as V1094Sco or DoAr33 \citep{colmenares_jwstmiri_2024}, are not in this region.} This might be evidence that T Tauri disks with a C/O > 1 are rare, consistent with evolutionary models by {\cite{mah_close-ice_2023,Sellek_2025A&A...701A.239S}}. This result is also consistent with the ProDiMo results from \cite{Arabhavi_2026A&A...708A..82A}, in which the region with C/O > 1 is orders of magnitude away from the observations.

    Our models also reproduce the observed line flux ratio C$_2$H$_2$/H$_2$O (Fig. \ref{fig: c2h2_H2O_ratio_obs}). This figure clearly shows that the spread seen in observations \citep{Grant_2025A&A...702A.126G} can be explained by either the elemental abundances (C/O and O/H) or the power law index of the dust size distribution ($q$), while  $gd$ and $h_C$ have a negligible impact. However, as mentioned above, C/O > 1 clearly overshoots this spread, reinforcing that T Tauri disks are likely to have a C/O below 1 in their inner regions.
    \begin{figure}[ht]
    \centering
    \resizebox{\hsize}{!}{
    \includegraphics[]{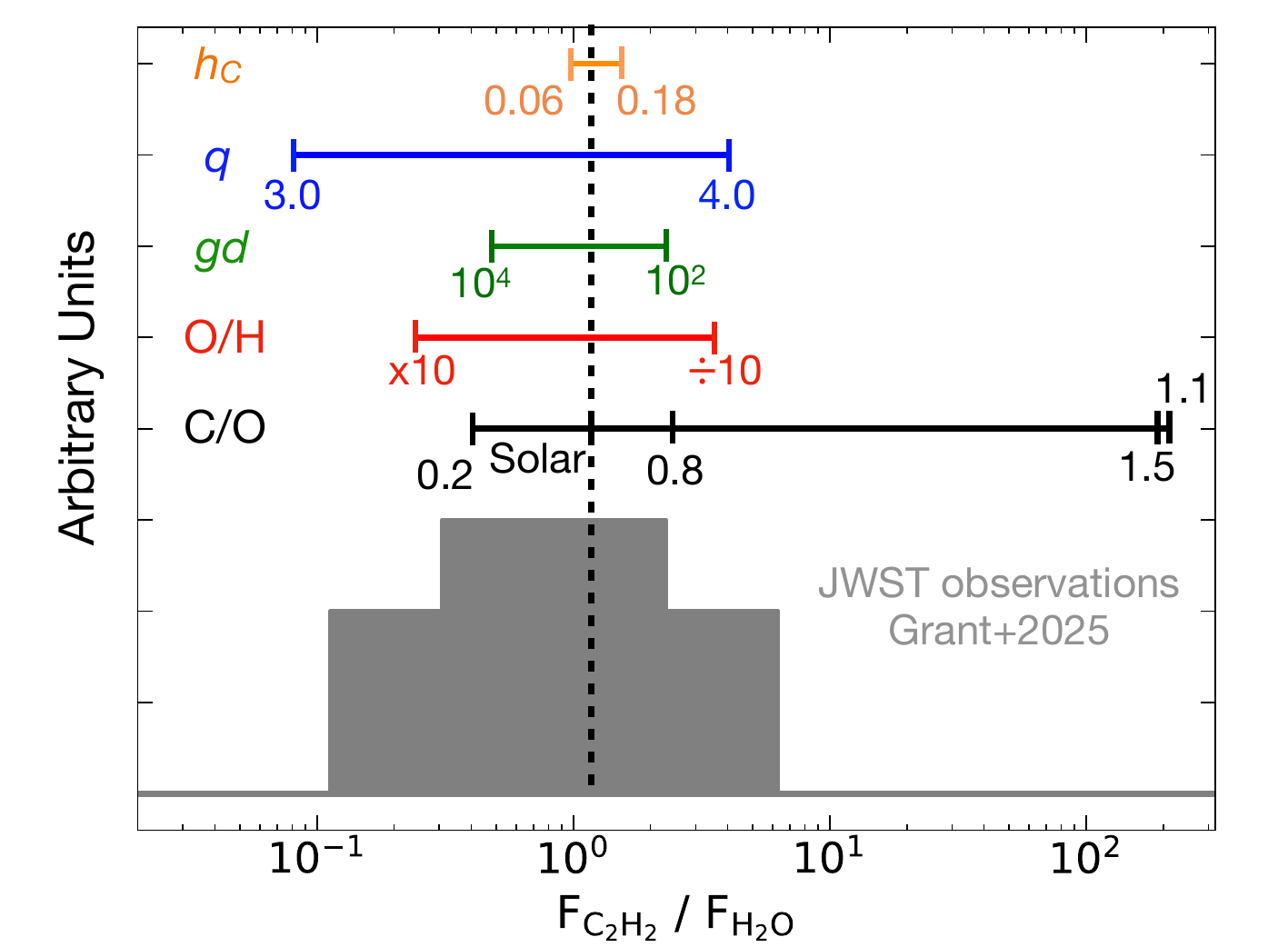}}
    \caption{Distribution of the line flux ratio C$_2$H$_2$/H$_2$O observed by JWST \citep{Grant_2025A&A...702A.126G} compared to the parameters explored in this work. Three parameters stand out to explain this spread: C/O, O/H or $q$. A C/O > 1 seems excluded for inner disks of T Tauri stars. }
    \label{fig: c2h2_H2O_ratio_obs}
    \end{figure}

\subsection{Retrieved excitation conditions}
\label{Sec: excited_cond}
\begin{figure}[ht]
    \centering
    \includegraphics[scale=0.17, trim={0 0 0 0 cm}, clip]{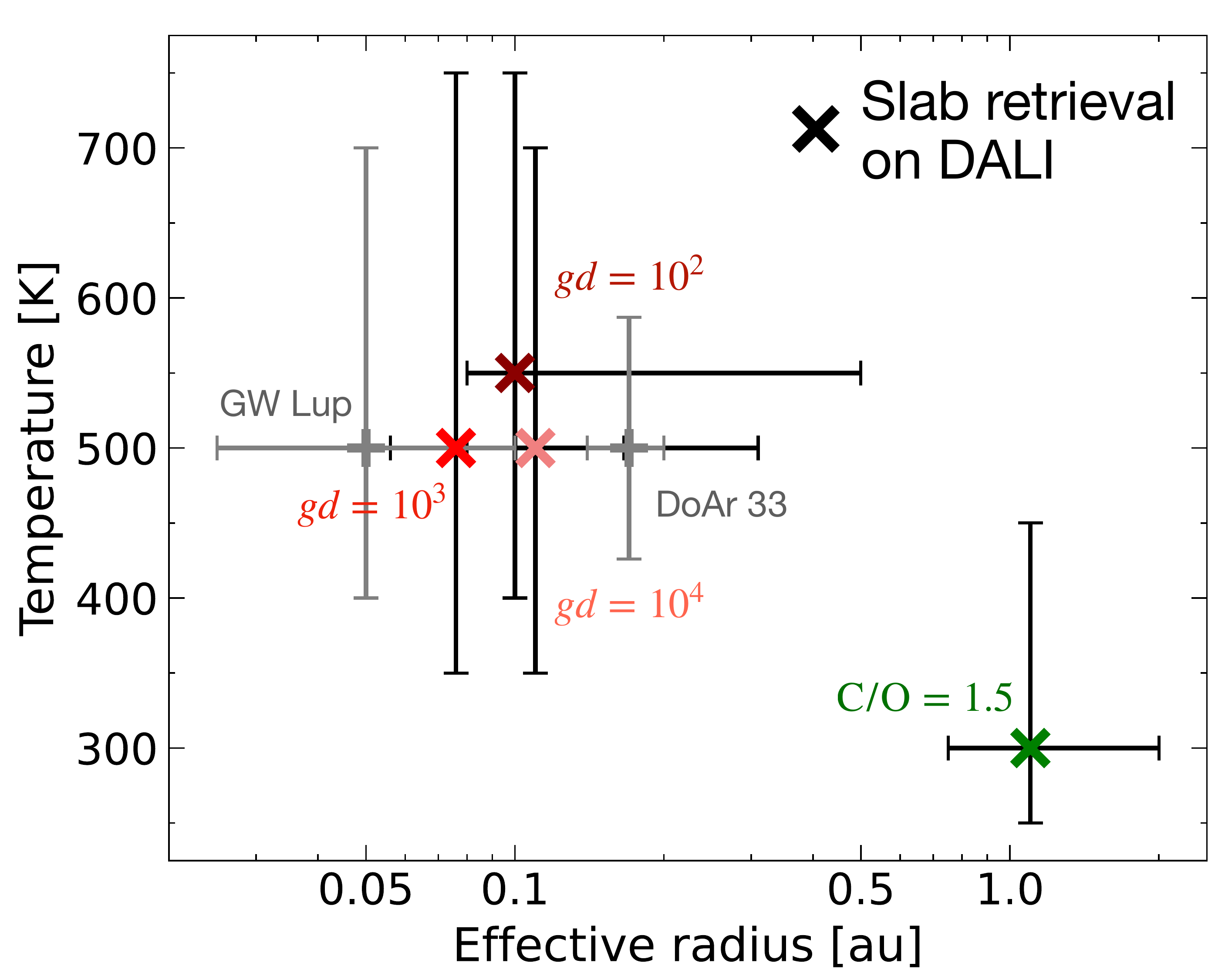}
    \caption{Temperature and effective radius (corresponding to an emitting area $\pi R^2$) of C$_2$H$_2$ emission retrieved from slab models on DALI predicted spectra. Red crosses indicate DALI models with solar C/O and $gd=10^2$, $10^3$, $10^4$, while the green cross corresponds to C/O = 1.5 and $gd=10^3$. JWST observations of 2 disks are shown in gray \citep{Grant_2023ApJ...947L...6G, colmenares_jwstmiri_2024}.} 
    \label{fig: dali vs slab}
\end{figure}
    
    In this section we investigate the consistency between the results of DALI and the 0D slab retrievals used to interpret JWST observations \citep[see code and fitting procedure in][]{Tabone_2023NatAs...7..805T}. Figure \ref{fig: dali vs slab} shows that the excitation conditions retrieved for C$_2$H$_2$ on DALI for a solar C/O using slab models (red crosses, $T_{\rm{}ex}\sim500$ K and effective radius corresponding to an emitting area $\pi R^2$ of $R \sim 0.1$ au) are consistent with the temperature and effective radius retrieved from JWST observations of GW Lup and DoAr 33 \citep{Grant_2023ApJ...947L...6G, colmenares_jwstmiri_2024}. {Our thermochemical model is able to reproduce} both the line flux and the excitation conditions of the C$_2$H$_2$ observed in disks. Figure \ref{fig: dali vs slab} also suggests that the effective radii and temperatures do not change significantly with the gas-to-dust ratio. Interestingly, despite the large uncertainties, the conditions retrieved from the DALI model with C/O = 1.5 are distinctly colder ($T_{\rm{}ex}\sim$ 300 K) and more extended ($R\sim$ 1.1 au), confirming that the chemical structure changes drastically when C/O > 1. This is also reflected in the abundance map of acetylene for a C/O > 1, which is dramatically different from C/O < 1 (shown in Appendix \ref{appendix:model_co1.5}). {Our models predict that DoAr33 is one of the richest T Tauri disk in carbon C/O $\sim$ 0.9 according to Fig. \ref{fig: diagnostic_plot_obs} with a large gas-to-dust ratio. However, the detailed analysis of DoAr 33 by \cite{colmenares_jwstmiri_2024} determined a C/O ratio between 2-4. This difference probably arises from the chemical network: their network is based on UMIST only
    which underproduces acetylene (see Fig. \ref{fig: Umistvskida}) compared to our network. To compensate, they likely need to increase the C/O ratio. In addition, our study relies on molecular fluxes, whereas they compared the predicted column densities to those retrieved from single-zone slab models. Radiative transfer effects can play a role, especially in the enhanced C/O models where C$_2$H$_2$ and dust are optically thick in the inner disk. Detailed radiative transfer is therefore needed to obtain a robust estimate of elemental abundances.}
    
\subsection{Evidence of radial drift ?}\label{Sec: radial drift}
\noindent The comparison of our model grid with JWST observations suggests that most of the inner disks of T Tauri have a solar or subsolar C/O ratio, and possibly combined with an increase in the elemental oxygen O/H. These two conditions are not contradictory and can be naturally fulfilled by the radial dust drift. Transport models show that oxygen-dominated ices (CO, CO$_2$, and H$_2$O) carried by grains coming from the outer disk would sublimate in the inner disk and enrich the gas phase in oxygen {\citep{Krijt_2016ApJ...833..285K,booth_chemical_2017,Kalyaan_2021ApJ...921...84K,Houge_2025aMNRAS.537..691H}}. This increase in the gas-phase O/H would therefore lower the C/O ratio of the inner disk {\citep{oberg_effects_2011, Oberg_2021ApJS..257....1O, mah_close-ice_2023, sellek_co2_2025, Williams_2025MNRAS.544.3562W}}. As a result, water emission would increase dramatically at the expense of C$_2$H$_2$ and other organic species, the latter being destroyed in greater quantity. These results were further supported observationally by \cite{banzatti_hints_2020, banzatti_jwst_2023, banzatti_water_2025}. However, ALMA reveals that most disks show substructures with rings and gaps \citep{Andrews_2018ApJ...869L..41A}. In the case of deep gaps in the disk, this drift would stop, preventing the grains from enriching the gas in oxygen. According to {hydrodynamic simulations} {\citep{Lubow_dAngelo_2006ApJ...641..526L,Bergez-Casalou_2020A&A...643A.133B}}, the gas from the outer disk would still cross the gap. This would elevate the C/O ratio while reducing O/H of the inner disk, since the gas from the outer disk is known to be oxygen-poor and have a high C/O ratio \citep{bergin_hydrocarbon_2016, miotello_lupus_2017, Sturm_2022A&A...660A.126S}. For this situation, according to our models, MIR spectra would show a prominent $Q$ branch of acetylene with a reduced water emission. Nevertheless, several studies did not necessarily find a relation between the substructures seen with ALMA and the molecular features of the inner disk \citep{gasman_minds_2025-1, Temmink_2025A&A...699A.134T}, revealing that this interpretation might be too simplistic. Indeed, micron-size grains can still cross gaps and reach the inner disk {\citep{Rice_2006MNRAS.373.1619R, Zhu_2012ApJ...755....6Z, Weber_2018ApJ...854..153W, Stammler_2023A&A...670L...5S}}. This would suggest that it might be the leakiness of gaps that governs the balance between the drift of dust and gas \citep{krijt_cosmic_2025}, setting the elemental abundances in the inner disk \citep{tabone_2026}.

Following this scenario, the spread of the line flux ratio C$_2$H$_2$/H$_2$O might be a consequence of a change in elemental abundances due to {the diversity of leakiness of gaps during the radial drift}. Consequently, constraining the elemental abundances of the inner disk is fundamental to understanding the relation between inner and outer disks.
\section{Conclusion}
\label{sec: conclusion}

\noindent The JWST is revolutionizing the characterization of the gas content in the inner disk of T Tauri stars. Our work aims to model the MIR emission of water and acetylene, two molecules ubiquitously detected by JWST, to quantify the underlying information they contain about this region. To better model inner disks, we improve the thermochemical model DALI by extending the chemical network, refining the UV self-shielding of molecules and including the line overlap in the ray-tracing. With a realistic disk geometry and a solar C/O ratio, we are able to reproduce the observed C$_2$H$_2$ emission in T Tauri disks reasonably well. We explore parameters related to the elemental abundances, the dust properties, and the geometry to conclude that:\\
\begin{itemize}
    \item The abundance of hydrocarbons {for C/O < 1} is set by a balance between formation seeded by CO X-ray induced dissociation and destruction by atomic O. Therefore, a change in these abundances, in particular O/H, strongly influences this balance. C$_2$H$_2$ is the most abundant hydrocarbon in emitting layers because it is a small and stable molecule, especially against reactions with H$_2$, and also faster to form compared to CH$_4$. It is also a key molecule in carbon chemistry as it is the reaction intermediate to form long carbon chains.
    \item Our models predict observable C$_2$H$_2$ {for C/O < 1} thanks to water UV shielding and our new chemical network, which includes many reactions between hydrocarbons and H$_2$, both increasing the emission of acetylene.
    \item The emission of C$_2$H$_2$ and H$_2$O are good tracers of the elemental composition of disks. They both vary strongly with the C/O and the O/H ratios. In particular, enhanced elemental abundances reduce C$_2$H$_2$ emission due to the excess of atomic oxygen, which destroys carbon chains. 
   \item Dust properties also have a significant impact on molecular features. A shallower dust size distribution (lower power law index, $q$) increases water emission but decreases acetylene emission, which can thus be a key parameter in the C$_2$H$_2$/H$_2$O line flux ratio. Depleting grains in the atmosphere increases both C$_2$H$_2$ and H$_2$O emissions, and does not favor C$_2$H$_2$ as speculated in recent observational works (in the limit of the parameter space explored).  
    \item The spread of the line flux ratio C$_2$H$_2$/H$_2$O observed in T Tauri disks naturally arises from the elemental abundances (C/O and O/H), or from the dust size distribution. This highlights the importance of considering both gas and dust emission when interpreting JWST data. It also underscores the benefit of combining JWST data with inner disk dust observations to better constrain dust properties and thereby the elemental composition of the gas. {Still, C/O > 1 is excluded to explain JWST observation, even for the disks showing a prominent C$_2$H$_2$ feature.} Distinguishing between a low C/O or enhanced O/H with only C$_2$H$_2$/H$_2$O seems difficult, but other species such as CO$_2$ could help break the degeneracy according to \cite{Arabhavi_2026A&A...708A..82A}.
\end{itemize}
The results of the model grid suggest that the gas in the inner disk of T Tauri has enhanced elemental abundances, with a C/O < 1, consistent with recent work {\citep{mah_close-ice_2023,Sellek_2025A&A...701A.239S,tabone_2026}}. JWST is now revealing the chemical composition of the atmospheres of close-in gas giant exoplanets, for which the C/O ratio and the metallicities appear to be in agreement with these results. A more detailed comparison and population analysis should be carried out to confirm this tentative link between disk and exoplanet composition. We finally stress that the estimates of the elemental composition of inner disks hinge on our knowledge of rate coefficients of gas-phase reactions at high temperature, which remain poorly studied for specific yet key types of reactions, such as the hydrogenation of hydrocarbons by H$_2$.   
\begin{acknowledgements}
   The authors thank the referee for a constructive report that  improved the quality of the paper. P.E and B.T thank E. Roueff, M. van Hemert, and G. Pineau des Forêts for helping us clarify the C$_2$+H$_2$ reaction rates and V. Wakelam for her support on the use of KIDA networks, and S. Facchini for his useful help with the dust settling module in DALI. P.E. and B.T also thank M. J. Colmenares for providing us the spectrum of DoAr33 and for interesting discussions regarding chemical networks. 
\end{acknowledgements}

\bibliographystyle{aa}
\bibliography{Chemistry_bibtex}

@article{krijt_cosmic_2025,
	title = {Cosmic {Cascades}: {How} {Disk} {Substructure} {Regulates} the {Flow} of {Water} to {Inner} {Planetary} {Systems}},
	volume = {990},
	issn = {2041-8205, 2041-8213},
	shorttitle = {Cosmic {Cascades}},
	url = {https://iopscience.iop.org/article/10.3847/2041-8213/adfbe3},
	doi = {10.3847/2041-8213/adfbe3},
	language = {en},
	number = {2},
	urldate = {2026-05-12},
	journal = {\apjl},
	author = {Krijt, Sebastiaan and Banzatti, Andrea and Zhang, Ke and Pinilla, Paola and Kaeufer, Till and Bergin, Edwin A. and Salyk, Colette and Pontoppidan, Klaus and Blake, Geoffrey A. and Long, Feng and Huang, Jane and Colmenares, María José and Williams, Joe and Houge, Adrien and Narang, Mayank and Vioque, Miguel and Lambrechts, Michiel and Cleeves, L. Ilsedore and Öberg, Karin and {The JDISCS Collaboration}},
	month = sep,
	year = {2025},
	pages = {L72},
}

@ARTICLE{tabone_2026,
       author = {{Tabone}, Beno{\^\i}t and {Temmink}, Milou and {Waters}, Laurens B.~F.~M. and {van Dishoeck}, Ewine F. and {Sellek}, Andrew and {Est{\`e}ve}, Pac{\^o}me and {Kurtovic}, Nicolas T. and {Kamp}, Inga and {Henning}, Thomas and {Gasman}, Danny and {Grant}, Sierra L. and {Varga}, J{\'o}zsef and {Guerras}, Alice and {Semenov}, Dmitry and {Arabhavi}, Aditya M. and {Garatti}, Alessio Caratti o and {Dutrey}, Anne and {Chapillon}, Edwige and {Guilloteau}, St{\'e}phane and {G{\"u}del}, Manuel and {Jang}, Hyerin and {Kaeufer}, Till and {Kanwar}, Jayatee and {Olofsson}, G{\"o}ran and {Perotti}, Giulia and {Pi{\'e}tu}, Vincent and {Ray}, Thomas P. and {Vlasblom}, Marissa},
        title = "{MINDS: Intertwined evolution of dust and gas in large planet-forming disks. A diversity driven by halted pebble drift?}",
      journal = {arXiv e-prints},
         year = 2026,
        month = apr,
          eid = {arXiv:2604.21803},
        pages = {arXiv:2604.21803},
          doi = {10.48550/arXiv.2604.21803},
archivePrefix = {arXiv},
       eprint = {2604.21803},
 primaryClass = {astro-ph.EP},
       adsurl = {https://ui.adsabs.harvard.edu/abs/2026arXiv260421803T}
}

@article{millar_umist_2023,
	title = {The {UMIST} database for astrochemistry 2022},
	copyright = {https://www.edpsciences.org/en/authors/copyright-and-licensing},
	issn = {0004-6361, 1432-0746},
	url = {https://www.aanda.org/10.1051/0004-6361/202346908},
	doi = {10.1051/0004-6361/202346908},
	language = {en},
	urldate = {2024-09-12},
	journal = {\aap},
	author = {Millar, T. J. and Walsh, C. and Van De Sande, M. and Markwick, A. J.},
	month = nov,
	year = {2023},
}

@article{wakelam_2024_2024,
	title = {The 2024 {KIDA} network for interstellar chemistry},
	volume = {689},
	copyright = {https://creativecommons.org/licenses/by/4.0},
	issn = {0004-6361, 1432-0746},
	url = {https://www.aanda.org/10.1051/0004-6361/202450606},
	doi = {10.1051/0004-6361/202450606},
	language = {en},
	urldate = {2024-09-12},
	journal = {\aap},
	author = {Wakelam, V. and Gratier, P. and Loison, J.-C. and Hickson, K. M. and Penguen, J. and Mechineau, A.},
	month = sep,
	year = {2024},
	pages = {A63},
}

@article{tinacci_gretobape_2023,
	title = {The {GRETOBAPE} {Gas}-phase {Reaction} {Network}: {The} {Importance} of {Being} {Exothermic}},
	volume = {266},
	issn = {0067-0049, 1538-4365},
	shorttitle = {The {GRETOBAPE} {Gas}-phase {Reaction} {Network}},
	url = {https://iopscience.iop.org/article/10.3847/1538-4365/accae9},
	doi = {10.3847/1538-4365/accae9},
	language = {en},
	number = {2},
	urldate = {2024-09-16},
	journal = {\apjs},
	author = {Tinacci, Lorenzo and Ferrada-Chamorro, Simón and Ceccarelli, Cecilia and Pantaleone, Stefano and Ascenzi, Daniela and Maranzana, Andrea and Balucani, Nadia and Ugliengo, Piero},
	month = jun,
	year = {2023},
	pages = {38},
}

@article{bosman_water_2022,
	title = {Water {UV}-shielding in the {Terrestrial} {Planet}-forming {Zone}: {Implications} for {Carbon} {Dioxide} {Emission}},
	volume = {933},
	issn = {2041-8205, 2041-8213},
	shorttitle = {Water {UV}-shielding in the {Terrestrial} {Planet}-forming {Zone}},
	url = {https://iopscience.iop.org/article/10.3847/2041-8213/ac7d9f},
	doi = {10.3847/2041-8213/ac7d9f},
	language = {en},
	number = {2},
	urldate = {2024-09-18},
	journal = {\apjl},
	author = {Bosman, Arthur D. and Bergin, Edwin A. and Calahan, Jenny K. and Duval, Sara E.},
	month = jul,
	year = {2022},
	pages = {L40},
}

@article{bosman_water_2022-1,
	title = {Water {UV}-shielding in the {Terrestrial} {Planet}-forming {Zone}: {Implications} from {Water} {Emission}},
	volume = {930},
	issn = {2041-8205, 2041-8213},
	shorttitle = {Water {UV}-shielding in the {Terrestrial} {Planet}-forming {Zone}},
	url = {https://iopscience.iop.org/article/10.3847/2041-8213/ac66ce},
	doi = {10.3847/2041-8213/ac66ce},
	language = {en},
	number = {2},
	urldate = {2024-09-18},
	journal = {\apjl},
	author = {Bosman, Arthur D. and Bergin, Edwin A. and Calahan, Jenny and Duval, Sara E.},
	month = may,
	year = {2022},
	pages = {L26},
}

@article{duval_water_2022,
	title = {Water {Shielding} in the {Terrestrial} {Planet}-forming {Zone}: {Implication} for {Inner} {Disk} {Organics}},
	volume = {934},
	issn = {2041-8205, 2041-8213},
	shorttitle = {Water {Shielding} in the {Terrestrial} {Planet}-forming {Zone}},
	url = {https://iopscience.iop.org/article/10.3847/2041-8213/ac822b},
	doi = {10.3847/2041-8213/ac822b},
	language = {en},
	number = {2},
	urldate = {2024-09-18},
	journal = {\apjl},
	author = {Duval, Sara E. and Bosman, Arthur D. and Bergin, Edwin A.},
	month = aug,
	year = {2022},
	pages = {L25},
}

@article{kanwar_hydrocarbon_2024,
	title = {Hydrocarbon chemistry in the inner regions of planet-forming disks},
	volume = {681},
	copyright = {https://creativecommons.org/licenses/by/4.0},
	issn = {0004-6361, 1432-0746},
	url = {https://www.aanda.org/10.1051/0004-6361/202346262},
	doi = {10.1051/0004-6361/202346262},
	language = {en},
	urldate = {2024-10-22},
	journal = {\aap},
	author = {Kanwar, J. and Kamp, I. and Woitke, P. and Rab, Ch. and Thi, W. F. and Min, M.},
	month = jan,
	year = {2024},
	pages = {A22},
}

@article{mah_close-ice_2023,
	title = {Close-in ice lines and the super-stellar {C}/{O} ratio in discs around very low-mass stars},
	volume = {677},
	copyright = {https://creativecommons.org/licenses/by/4.0},
	issn = {0004-6361, 1432-0746},
	url = {https://www.aanda.org/10.1051/0004-6361/202347169},
	doi = {10.1051/0004-6361/202347169},
	language = {en},
	urldate = {2024-10-23},
	journal = {\aap},
	author = {Mah, Jingyi and Bitsch, Bertram and Pascucci, Ilaria and Henning, Thomas},
	month = sep,
	year = {2023},
	pages = {L7},
}

@article{van_hemert_photodissociation_2008,
	title = {Photodissociation of small carbonaceous molecules of astrophysical interest},
	volume = {343},
	copyright = {https://www.elsevier.com/tdm/userlicense/1.0/},
	issn = {03010104},
	url = {https://linkinghub.elsevier.com/retrieve/pii/S0301010407003667},
	doi = {10.1016/j.chemphys.2007.08.011},
	language = {en},
	number = {2-3},
	urldate = {2024-10-30},
	journal = {Chemical Physics},
	author = {van Hemert, M.C. and van Dishoeck, E.F.},
	month = jan,
	year = {2008},
	pages = {292--302},
}

@article{bruderer_ro-vibrational_2015,
	title = {Ro-vibrational excitation of an organic molecule ({HCN}) in protoplanetary disks},
	volume = {575},
	issn = {0004-6361, 1432-0746},
	url = {http://www.aanda.org/10.1051/0004-6361/201425009},
	doi = {10.1051/0004-6361/201425009},
	language = {en},
	urldate = {2024-10-31},
	journal = {\aap},
	author = {Bruderer, Simon and Harsono, Daniel and van Dishoeck, Ewine F.},
	month = mar,
	year = {2015},
	pages = {A94},
}

@article{kamp_consistent_2017,
	title = {Consistent dust and gas models for protoplanetary disks: {II}. {Chemical} networks and rates},
	volume = {607},
	issn = {0004-6361, 1432-0746},
	shorttitle = {Consistent dust and gas models for protoplanetary disks},
	url = {http://www.aanda.org/10.1051/0004-6361/201730388},
	doi = {10.1051/0004-6361/201730388},
	language = {en},
	urldate = {2024-11-04},
	journal = {\aap},
	author = {Kamp, I. and Thi, W.-F. and Woitke, P. and Rab, C. and Bouma, S. and Ménard, F.},
	month = nov,
	year = {2017},
	pages = {A41},
}

@ARTICLE{oberg_astrochemistry_2021,
       author = {{{\"O}berg}, Karin I. and {Bergin}, Edwin A.},
        title = "{Astrochemistry and compositions of planetary systems}",
      journal = {\physrep},
         year = 2021,
        month = jan,
       volume = {893},
        pages = {1-48},
          doi = {10.1016/j.physrep.2020.09.004},
archivePrefix = {arXiv},
       eprint = {2010.03529},
 primaryClass = {astro-ph.EP},
       adsurl = {https://ui.adsabs.harvard.edu/abs/2021PhR...893....1O}
}

@article{bosman_co_2018,
	title = {{CO} destruction in protoplanetary disk midplanes: {Inside} versus outside the {CO} snow surface},
	volume = {618},
	copyright = {https://www.edpsciences.org/en/authors/copyright-and-licensing},
	issn = {0004-6361, 1432-0746},
	shorttitle = {{CO} destruction in protoplanetary disk midplanes},
	url = {https://www.aanda.org/10.1051/0004-6361/201833497},
	doi = {10.1051/0004-6361/201833497},
	language = {en},
	urldate = {2024-12-09},
	journal = {\aap},
	author = {Bosman, Arthur D. and Walsh, Catherine and van Dishoeck, Ewine F.},
	month = oct,
	year = {2018},
	pages = {A182},
}

@article{mcelroy_umist_2013,
	title = {The {UMIST} database for astrochemistry 2012},
	volume = {550},
	issn = {0004-6361, 1432-0746},
	url = {http://www.aanda.org/10.1051/0004-6361/201220465},
	doi = {10.1051/0004-6361/201220465},
	language = {en},
	urldate = {2024-12-09},
	journal = {\aap},
	author = {McElroy, D. and Walsh, C. and Markwick, A. J. and Cordiner, M. A. and Smith, K. and Millar, T. J.},
	month = feb,
	year = {2013},
	pages = {A36},
}

@article{colmenares_jwstmiri_2024,
	title = {{JWST}/{MIRI} {Detection} of a {Carbon}-rich {Chemistry} in the {Disk} of a {Solar} {Nebula} {Analog}},
	volume = {977},
	issn = {0004-637X, 1538-4357},
	url = {https://iopscience.iop.org/article/10.3847/1538-4357/ad8b4f},
	doi = {10.3847/1538-4357/ad8b4f},
	language = {en},
	number = {2},
	urldate = {2024-12-18},
	journal = {\apj},
	author = {Colmenares, María José and Bergin, Edwin A. and Salyk, Colette and Pontoppidan, Klaus M. and Arulanantham, Nicole and Calahan, Jenny and Banzatti, Andrea and Andrews, Sean and Blake, Geoffrey A. and Ciesla, Fred and Green, Joel and Long 龙, Feng 凤 and Lambrechts, Michiel and Najita, Joan and Pascucci, Ilaria and Pinilla, Paola and Krijt, Sebastiaan and Trapman, Leon and {the JDISCS Collaboration}},
	month = dec,
	year = {2024},
	pages = {173},
}

@article{hrodmarsson_photodissociation_2023,
	title = {Photodissociation and photoionization of molecules of astronomical interest: {Updates} to the {Leiden} photodissociation and photoionization cross section database},
	volume = {675},
	copyright = {https://creativecommons.org/licenses/by/4.0},
	issn = {0004-6361, 1432-0746},
	shorttitle = {Photodissociation and photoionization of molecules of astronomical interest},
	url = {https://www.aanda.org/10.1051/0004-6361/202346645},
	doi = {10.1051/0004-6361/202346645},
	language = {en},
	urldate = {2024-12-20},
	journal = {\aap},
	author = {Hrodmarsson, H. R. and van Dishoeck, E. F.},
	month = jul,
	year = {2023},
	pages = {A25},
}

@ARTICLE{Vlasbom_CXtau_2025A&A...693A.278V,
       author = {{Vlasblom}, Marissa and {Temmink}, Milou and {Grant}, Sierra L. and {Kurtovic}, Nicolas and {Sellek}, Andrew D. and {van Dishoeck}, Ewine F. and {G{\"u}del}, Manuel and {Henning}, Thomas and {Lagage}, Pierre-Olivier and {Barrado}, David and {Caratti o Garatti}, Alessio and {Glauser}, Adrian M. and {Kamp}, Inga and {Lahuis}, Fred and {Olofsson}, G{\"o}ran and {Arabhavi}, Aditya M. and {Christiaens}, Valentin and {Gasman}, Danny and {Jang}, Hyerin and {Morales-Calder{\'o}n}, Maria and {Perotti}, Giulia and {Schwarz}, Kamber and {Tabone}, Beno{\^\i}t},
        title = "{MINDS. JWST-MIRI reveals a peculiar CO$_{2}$-rich chemistry in the drift-dominated disk CX Tau}",
      journal = {\aap},
         year = 2025,
        month = jan,
       volume = {693},
          eid = {A278},
        pages = {A278},
          doi = {10.1051/0004-6361/202450863},
archivePrefix = {arXiv},
       eprint = {2412.12715},
 primaryClass = {astro-ph.EP},
       adsurl = {https://ui.adsabs.harvard.edu/abs/2025A&A...693A.278V}
}

@article{anderson_observing_2021,
	title = {Observing {Carbon} and {Oxygen} {Carriers} in {Protoplanetary} {Disks} at {Mid}-infrared {Wavelengths}},
	volume = {909},
	issn = {0004-637X, 1538-4357},
	url = {https://iopscience.iop.org/article/10.3847/1538-4357/abd9c1},
	doi = {10.3847/1538-4357/abd9c1},
	language = {en},
	number = {1},
	urldate = {2025-02-05},
	journal = {\apj},
	author = {Anderson, Dana E. and Blake, Geoffrey A. and Cleeves, L. Ilsedore and Bergin, Edwin A. and Zhang, Ke and Schwarz, Kamber R. and Salyk, Colette and Bosman, Arthur D.},
	month = mar,
	year = {2021},
	pages = {55},
}

@article{pitts_temperature_1982,
	title = {Temperature dependence of the {C2}({X1Σg}+) reaction with {H2} and {CH4} and {C2}({X1Σg}+ and a {3Πu} equilibrated states) with {O2}},
	volume = {68},
	copyright = {https://www.elsevier.com/tdm/userlicense/1.0/},
	issn = {03010104},
	url = {https://linkinghub.elsevier.com/retrieve/pii/030101048287050X},
	doi = {10.1016/0301-0104(82)87050-X},
	language = {en},
	number = {3},
	urldate = {2025-02-20},
	journal = {Chemical Physics},
	author = {Pitts, Willam M. and Pasternack, L. and McDonald, J.R.},
	month = jul,
	year = {1982},
	pages = {417--422},
}

@article{kruse_kinetics_1997,
	title = {Kinetics of {C2} {Reactions} during {High}-{Temperature} {Pyrolysis} of {Acetylene}},
	volume = {101},
	issn = {1089-5639},
	url = {https://doi.org/10.1021/jp963373o},
	doi = {10.1021/jp963373o},
	number = {11},
	urldate = {2025-02-20},
	journal = {The Journal of Physical Chemistry A},
	author = {Kruse, T. and Roth, P.},
	month = mar,
	year = {1997},
	pages = {2138--2146},
}

@article{pasternack_temperature_1981,
	title = {Temperature dependence of reactions and intersystem crossing of {C2a3Πu} with hydrogen and small hydrocarbons from 300–600 {K}},
	volume = {57},
	copyright = {https://www.elsevier.com/tdm/userlicense/1.0/},
	issn = {03010104},
	url = {https://linkinghub.elsevier.com/retrieve/pii/0301010481800171},
	doi = {10.1016/0301-0104(81)80017-1},
	language = {en},
	number = {1-2},
	urldate = {2025-02-20},
	journal = {Chemical Physics},
	author = {Pasternack, L. and Pitts, W.M. and McDonald, J.R.},
	month = may,
	year = {1981},
	pages = {19--28},
}

@article{heays_photodissociation_2017,
	title = {Photodissociation and photoionisation of atoms and molecules of astrophysical interest},
	volume = {602},
	issn = {0004-6361, 1432-0746},
	url = {http://www.aanda.org/10.1051/0004-6361/201628742},
	doi = {10.1051/0004-6361/201628742},
	language = {en},
	urldate = {2025-02-25},
	journal = {\aap},
	author = {Heays, A. N. and Bosman, A. D. and van Dishoeck, E. F.},
	month = jun,
	year = {2017},
	pages = {A105},
}

@article{hebrard_how_2009,
	title = {How {Measurements} of {Rate} {Coefficients} at {Low} {Temperature} {Increase} the {Predictivity} of {Photochemical} {Models} of {Titan}’s {Atmosphere}},
	volume = {113},
	issn = {1089-5639, 1520-5215},
	url = {https://pubs.acs.org/doi/10.1021/jp905524e},
	doi = {10.1021/jp905524e},
	language = {en},
	number = {42},
	urldate = {2025-03-05},
	journal = {The Journal of Physical Chemistry A},
	author = {Hébrard, E. and Dobrijevic, M. and Pernot, P. and Carrasco, N. and Bergeat, A. and Hickson, K. M. and Canosa, A. and Le Picard, S. D. and Sims, I. R.},
	month = oct,
	year = {2009},
	pages = {11227--11237},
}

@article{banzatti_water_2025,
	title = {Water in {Protoplanetary} {Disks} with {JWST}-{MIRI}: {Spectral} {Excitation} {Atlas} and {Radial} {Distribution} from {Temperature} {Diagnostic} {Diagrams} and {Doppler} {Mapping}},
	volume = {169},
	issn = {0004-6256, 1538-3881},
	shorttitle = {Water in {Protoplanetary} {Disks} with {JWST}-{MIRI}},
	url = {https://iopscience.iop.org/article/10.3847/1538-3881/ada962},
	doi = {10.3847/1538-3881/ada962},
	language = {en},
	number = {3},
	urldate = {2025-03-17},
	journal = {\aj},
	author = {Banzatti, Andrea and Salyk, Colette and Pontoppidan, Klaus M. and Carr, John S. and Zhang, Ke and Arulanantham, Nicole and Krijt, Sebastiaan and Öberg, Karin I. and Cleeves, L. Ilsedore and Najita, Joan R. and Pascucci, Ilaria and Blake, Geoffrey A. and Romero-Mirza, Carlos E. and Bergin, Edwin A. and Cieza, Lucas A. and Pinilla, Paola and Long, Feng and Mallaney, Patrick and Xie, Chengyan and Waggoner, Abygail R. and Kaeufer, Till and {the JDISCS collaboration}},
	month = mar,
	year = {2025},
	pages = {165},
}

@article{tabone_oh_2024,
	title = {{OH} mid-infrared emission as a diagnostic of {H}$_{\textrm{2}}$ {O} {UV} photodissociation: {III}. {Application} to planet-forming disks},
	volume = {691},
	copyright = {https://creativecommons.org/licenses/by/4.0},
	issn = {0004-6361, 1432-0746},
	shorttitle = {{OH} mid-infrared emission as a diagnostic of {H}$_{\textrm{2}}$ {O} {UV} photodissociation},
	url = {https://www.aanda.org/10.1051/0004-6361/202348487},
	doi = {10.1051/0004-6361/202348487},
	language = {en},
	urldate = {2025-04-14},
	journal = {\aap},
	author = {Tabone, Benoît and van Dishoeck, Ewine F. and Black, John H.},
	month = nov,
	year = {2024},
	pages = {A11},
}

@article{woitke_2d_2024,
	title = {{2D} disc modelling of the {JWST} line spectrum of {EX} {Lupi}},
	volume = {683},
	copyright = {https://creativecommons.org/licenses/by/4.0},
	issn = {0004-6361, 1432-0746},
	url = {https://www.aanda.org/10.1051/0004-6361/202347730},
	doi = {10.1051/0004-6361/202347730},
	language = {en},
	urldate = {2025-04-22},
	journal = {\aap},
	author = {Woitke, P. and Thi, W.-F. and Arabhavi, A. M. and Kamp, I. and Kóspál, Á. and Ábrahám, P.},
	month = mar,
	year = {2024},
	pages = {A219},
}

@article{harada_new_2010,
	title = {A {NEW} {NETWORK} {FOR} {HIGHER}-{TEMPERATURE} {GAS}-{PHASE} {CHEMISTRY}. {I}. {A} {PRELIMINARY} {STUDY} {OF} {ACCRETION} {DISKS} {IN} {ACTIVE} {GALACTIC} {NUCLEI}},
	volume = {721},
	issn = {0004-637X, 1538-4357},
	url = {https://iopscience.iop.org/article/10.1088/0004-637X/721/2/1570},
	doi = {10.1088/0004-637X/721/2/1570},
	language = {en},
	number = {2},
	urldate = {2025-05-07},
	journal = {\apj},
	author = {Harada, Nanase and Herbst, Eric and Wakelam, Valentine},
	month = oct,
	year = {2010},
	pages = {1570--1578},
}

@article{gasman_minds_2025-1,
	title = {{MINDS}: {The} influence of outer dust disc structure on the volatile delivery to the inner disc},
	volume = {694},
	copyright = {https://creativecommons.org/licenses/by/4.0},
	issn = {0004-6361, 1432-0746},
	shorttitle = {{MINDS}},
	url = {https://www.aanda.org/10.1051/0004-6361/202452152},
	doi = {10.1051/0004-6361/202452152},
	language = {en},
	urldate = {2025-05-19},
	journal = {\aap},
	author = {Gasman, Danny and Temmink, Milou and van Dishoeck, Ewine F. and Kurtovic, Nicolas T. and Grant, Sierra L. and Sellek, Andrew and Tabone, Benoît and Henning, Thomas and Kamp, Inga and Güdel, Manuel and Barrado, David and Caratti O Garatti, Alessio and Glauser, Adrian M. and Waters, Laurens B. F. M. and Arabhavi, Aditya M. and Jang, Hyerin and Kanwar, Jayatee and Lienert, Julia L. and Perotti, Giulia and Schwarz, Kamber and Vlasblom, Marissa},
	month = feb,
	year = {2025},
	pages = {A147},
}

@article{sellek_co2_2025,
	title = {{CO}$_{\textrm{2}}$ -rich protoplanetary discs as a probe of dust radial drift and trapping},
	volume = {694},
	copyright = {https://creativecommons.org/licenses/by/4.0},
	issn = {0004-6361, 1432-0746},
	url = {https://www.aanda.org/10.1051/0004-6361/202451137},
	doi = {10.1051/0004-6361/202451137},
	language = {en},
	urldate = {2025-05-19},
	journal = {\aap},
	author = {Sellek, Andrew D. and Vlasblom, Marissa and van Dishoeck, Ewine F.},
	month = feb,
	year = {2025},
	pages = {A79},
}

@ARTICLE{Arulanantham_2025AJ....170...67A,
       author = {{Arulanantham}, Nicole and {Salyk}, Colette and {Pontoppidan}, Klaus and {Banzatti}, Andrea and {Zhang}, Ke and {{\"O}berg}, Karin and {Long}, Feng and {Carr}, John and {Najita}, Joan and {Pascucci}, Ilaria and {Colmenares}, Mar{\'\i}a Jos{\'e} and {Xie}, Chengyan and {Huang}, Jane and {Green}, Joel and {Andrews}, Sean M. and {Blake}, Geoffrey A. and {Bergin}, Edwin A. and {Pinilla}, Paola and {Vioque}, Miguel and {Dahl}, Emma and {Raul}, Eshan and {Krijt}, Sebastiaan and {The Jdiscs Collaboration}},
        title = "{The JDISC Survey: Linking the Physics and Chemistry of Inner and Outer Protoplanetary Disk Zones}",
      journal = {\aj},
         year = 2025,
        month = aug,
       volume = {170},
       number = {2},
          eid = {67},
        pages = {67},
          doi = {10.3847/1538-3881/addd01},
archivePrefix = {arXiv},
       eprint = {2505.07562},
 primaryClass = {astro-ph.SR},
       adsurl = {https://ui.adsabs.harvard.edu/abs/2025AJ....170...67A}
}

@article{ligterink_mind_2024,
	title = {Mind the trap: {Non}-negligible effect of volatile trapping in ice on {C}/{O} ratios in protoplanetary disks and exoplanetary atmospheres},
	volume = {687},
	copyright = {https://creativecommons.org/licenses/by/4.0},
	issn = {0004-6361, 1432-0746},
	shorttitle = {Mind the trap},
	url = {https://www.aanda.org/10.1051/0004-6361/202450405},
	doi = {10.1051/0004-6361/202450405},
	language = {en},
	urldate = {2025-05-22},
	journal = {\aap},
	author = {Ligterink, N. F. W. and Kipfer, K. A. and Gavino, S.},
	month = jul,
	year = {2024},
	pages = {A224},
}

@ARTICLE{Borderies_2025A&A...694A..89B,
       author = {{Borderies}, Antonin and {Commer{\c{c}}on}, Beno{\^\i}t and {Bourdon}, Bernard},
        title = "{Dust evolution by chemisputtering during protostellar formation}",
      journal = {\aap},
         year = 2025,
        month = feb,
       volume = {694},
          eid = {A89},
        pages = {A89},
          doi = {10.1051/0004-6361/202452228},
archivePrefix = {arXiv},
       eprint = {2501.17937},
 primaryClass = {astro-ph.GA},
       adsurl = {https://ui.adsabs.harvard.edu/abs/2025A&A...694A..89B}
}

@article{bergin_hydrocarbon_2016,
	title = {{HYDROCARBON} {EMISSION} {RINGS} {IN} {PROTOPLANETARY} {DISKS} {INDUCED} {BY} {DUST} {EVOLUTION}},
	volume = {831},
	issn = {0004-637X, 1538-4357},
	url = {https://iopscience.iop.org/article/10.3847/0004-637X/831/1/101},
	doi = {10.3847/0004-637X/831/1/101},
	language = {en},
	number = {1},
	urldate = {2025-06-03},
	journal = {\apj},
	author = {Bergin, Edwin A. and Du, Fujun and Cleeves, L. Ilsedore and Blake, G. A. and Schwarz, K. and Visser, R. and Zhang, K.},
	month = nov,
	year = {2016},
	pages = {101},
}

@article{miotello_lupus_2017,
	title = {Lupus disks with faint {CO} isotopologues: low gas/dust or high carbon depletion?},
	volume = {599},
	issn = {0004-6361, 1432-0746},
	shorttitle = {Lupus disks with faint {CO} isotopologues},
	url = {http://www.aanda.org/10.1051/0004-6361/201629556},
	doi = {10.1051/0004-6361/201629556},
	language = {en},
	urldate = {2025-06-12},
	journal = {\aap},
	author = {Miotello, A. and van Dishoeck, E. F. and Williams, J. P. and Ansdell, M. and Guidi, G. and Hogerheijde, M. and Manara, C. F. and Tazzari, M. and Testi, L. and Van Der Marel, N. and Van Terwisga, S.},
	month = mar,
	year = {2017},
	pages = {A113},
}

@article{bosman_co2_2017,
	title = {{CO}$_{\textrm{2}}$ infrared emission as a diagnostic of planet-forming regions of disks},
	volume = {601},
	issn = {0004-6361, 1432-0746},
	url = {http://www.aanda.org/10.1051/0004-6361/201629946},
	doi = {10.1051/0004-6361/201629946},
	language = {en},
	urldate = {2025-06-12},
	journal = {\aap},
	author = {Bosman, Arthur D. and Bruderer, Simon and van Dishoeck, Ewine F.},
	month = may,
	year = {2017},
	pages = {A36},
}

@ARTICLE{Arabhavi_2025A&A...699A.194A,
       author = {{Arabhavi}, A.~M. and {Kamp}, I. and {Henning}, Th. and {van Dishoeck}, E.~F. and {Jang}, H. and {Waters}, L.~B.~F.~M. and {Christiaens}, V. and {Gasman}, D. and {Pascucci}, I. and {Perotti}, G. and {Grant}, S.~L. and {G{\"u}del}, M. and {Lagage}, P.-O. and {Barrado}, D. and {Caratti o Garatti}, A. and {Lahuis}, F. and {Kaeufer}, T. and {Kanwar}, J. and {Morales-Calder{\'o}n}, M. and {Schwarz}, K. and {Sellek}, A.~D. and {Tabone}, B. and {Temmink}, M. and {Vlasblom}, M. and {Patapis}, P.},
        title = "{MINDS: The very low-mass star and brown dwarf sample: Detections and trends in the inner disk gas}",
      journal = {\aap},
         year = 2025,
        month = jul,
       volume = {699},
          eid = {A194},
        pages = {A194},
          doi = {10.1051/0004-6361/202554109},
archivePrefix = {arXiv},
       eprint = {2506.02748},
 primaryClass = {astro-ph.EP},
       adsurl = {https://ui.adsabs.harvard.edu/abs/2025A&A...699A.194A}
}

@article{woitke_consistent_2016,
	title = {Consistent dust and gas models for protoplanetary disks: {I}. {Disk} shape, dust settling, opacities, and {PAHs}},
	volume = {586},
	issn = {0004-6361, 1432-0746},
	shorttitle = {Consistent dust and gas models for protoplanetary disks},
	url = {http://www.aanda.org/10.1051/0004-6361/201526538},
	doi = {10.1051/0004-6361/201526538},
	language = {en},
	urldate = {2025-06-12},
	journal = {\aap},
	author = {Woitke, P. and Min, M. and Pinte, C. and Thi, W.-F. and Kamp, I. and Rab, C. and Anthonioz, F. and Antonellini, S. and Baldovin-Saavedra, C. and Carmona, A. and Dominik, C. and Dionatos, O. and Greaves, J. and Güdel, M. and Ilee, J. D. and Liebhart, A. and Ménard, F. and Rigon, L. and Waters, L. B. F. M. and Aresu, G. and Meijerink, R. and Spaans, M.},
	month = feb,
	year = {2016},
	pages = {A103},
}

@article{schneider_how_2021,
	title = {How drifting and evaporating pebbles shape giant planets: {I}. {Heavy} element content and atmospheric {C}/{O}},
	volume = {654},
	copyright = {https://creativecommons.org/licenses/by/4.0},
	issn = {0004-6361, 1432-0746},
	shorttitle = {How drifting and evaporating pebbles shape giant planets},
	url = {https://www.aanda.org/10.1051/0004-6361/202039640},
	doi = {10.1051/0004-6361/202039640},
	language = {en},
	urldate = {2025-06-12},
	journal = {\aap},
	author = {Schneider, Aaron David and Bitsch, Bertram},
	month = oct,
	year = {2021},
	pages = {A71},
}

@article{visser_nitrogen_2018,
	title = {Nitrogen isotope fractionation in protoplanetary disks},
	volume = {615},
	copyright = {https://www.edpsciences.org/en/authors/copyright-and-licensing},
	issn = {0004-6361, 1432-0746},
	url = {https://www.aanda.org/10.1051/0004-6361/201731898},
	doi = {10.1051/0004-6361/201731898},
	language = {en},
	urldate = {2025-06-12},
	journal = {\aap},
	author = {Visser, Ruud and Bruderer, Simon and Cazzoletti, Paolo and Facchini, Stefano and Heays, Alan N. and van Dishoeck, Ewine F.},
	month = jul,
	year = {2018},
	pages = {A75},
}

@article{facchini_different_2017,
	title = {Different dust and gas radial extents in protoplanetary disks: consistent models of grain growth and {CO} emission},
	volume = {605},
	issn = {0004-6361},
	shorttitle = {Different dust and gas radial extents in protoplanetary disks},
	url = {https://ui.adsabs.harvard.edu/abs/2017A&A...605A..16F},
	doi = {10.1051/0004-6361/201630329},
	urldate = {2025-07-09},
	journal = {\aap},
	author = {Facchini, S. and Birnstiel, T. and Bruderer, S. and van Dishoeck, E. F.},
	month = sep,
	year = {2017},
	note = {Publisher: EDP
ADS Bibcode: 2017A\&A...605A..16F},
	pages = {A16},
}

@article{birnstiel_dust_2015,
	title = {Dust {Evolution} {Can} {Produce} {Scattered} {Light} {Gaps} in {Protoplanetary} {Disks}},
	volume = {813},
	issn = {0004-637X},
	url = {https://ui.adsabs.harvard.edu/abs/2015ApJ...813L..14B},
	doi = {10.1088/2041-8205/813/1/L14},
	urldate = {2025-07-09},
	journal = {\apj},
	author = {Birnstiel, Tilman and Andrews, Sean M. and Pinilla, Paola and Kama, Mihkel},
	month = nov,
	year = {2015},
	pages = {L14},
}

@article{birnstiel_dust_2011,
	title = {Dust size distributions in coagulation/fragmentation equilibrium: numerical solutions and analytical fits},
	volume = {525},
	issn = {0004-6361},
	shorttitle = {Dust size distributions in coagulation/fragmentation equilibrium},
	url = {https://ui.adsabs.harvard.edu/abs/2011A&A...525A..11B},
	doi = {10.1051/0004-6361/201015228},
	urldate = {2025-07-09},
	journal = {\aap},
	author = {Birnstiel, T. and Ormel, C. W. and Dullemond, C. P.},
	month = jan,
	year = {2011},
	pages = {A11},
}

@article{antonellini_mid-infrared_2023,
	title = {Mid-infrared blends and continuum signatures of dust drift and accretion in protoplanetary disks},
	volume = {672},
	copyright = {https://creativecommons.org/licenses/by/4.0},
	issn = {0004-6361, 1432-0746},
	url = {https://www.aanda.org/10.1051/0004-6361/202244773},
	doi = {10.1051/0004-6361/202244773},
	language = {en},
	urldate = {2025-07-29},
	journal = {\aap},
	author = {Antonellini, S. and Kamp, I. and Waters, L. B. F. M.},
	month = apr,
	year = {2023},
	pages = {A92},
}

@ARTICLE{2003ApJ...587..278W,
       author = {{Wolfire}, Mark G. and {McKee}, Christopher F. and {Hollenbach}, David and {Tielens}, A.~G.~G.~M.},
        title = "{Neutral Atomic Phases of the Interstellar Medium in the Galaxy}",
      journal = {\apj},
         year = 2003,
        month = apr,
       volume = {587},
       number = {1},
        pages = {278-311},
          doi = {10.1086/368016},
archivePrefix = {arXiv},
       eprint = {astro-ph/0207098},
 primaryClass = {astro-ph},
       adsurl = {https://ui.adsabs.harvard.edu/abs/2003ApJ...587..278W}
}

@ARTICLE{alessio2006ApJ,
       author = {{D'Alessio}, Paola and {Calvet}, Nuria and {Hartmann}, Lee and {Franco-Hern{\'a}ndez}, Ramiro and {Serv{\'\i}n}, Hermelinda},
        title = "{Effects of Dust Growth and Settling in T Tauri Disks}",
      journal = {\apj},
         year = 2006,
        month = feb,
       volume = {638},
       number = {1},
        pages = {314-335},
          doi = {10.1086/498861},
archivePrefix = {arXiv},
       eprint = {astro-ph/0511564},
 primaryClass = {astro-ph},
       adsurl = {https://ui.adsabs.harvard.edu/abs/2006ApJ...638..314D}
}

@ARTICLE{2011ApJ...732...42A,
       author = {{Andrews}, Sean M. and {Wilner}, David J. and {Espaillat}, Catherine and {Hughes}, A.~M. and {Dullemond}, C.~P. and {McClure}, M.~K. and {Qi}, Chunhua and {Brown}, J.~M.},
        title = "{Resolved Images of Large Cavities in Protoplanetary Transition Disks}",
      journal = {\apj},
         year = 2011,
        month = may,
       volume = {732},
       number = {1},
          eid = {42},
        pages = {42},
          doi = {10.1088/0004-637X/732/1/42},
archivePrefix = {arXiv},
       eprint = {1103.0284},
 primaryClass = {astro-ph.GA},
       adsurl = {https://ui.adsabs.harvard.edu/abs/2011ApJ...732...42A}
}

@ARTICLE{RiolsLesur2018,
       author = {{Riols}, A. and {Lesur}, G.},
        title = "{Dust settling and rings in the outer regions of protoplanetary discs subject to ambipolar diffusion}",
      journal = {\aap},
         year = 2018,
        month = sep,
       volume = {617},
          eid = {A117},
        pages = {A117},
          doi = {10.1051/0004-6361/201833212},
archivePrefix = {arXiv},
       eprint = {1805.00458},
 primaryClass = {astro-ph.EP},
       adsurl = {https://ui.adsabs.harvard.edu/abs/2018A&A...617A.117R}
}

@ARTICLE{Tabone_2023NatAs...7..805T,
       author = {{Tabone}, B. and {Bettoni}, G. and {van Dishoeck}, E.~F. and {Arabhavi}, A.~M. and {Grant}, S. and {Gasman}, D. and {Henning}, Th. and {Kamp}, I. and {G{\"u}del}, M. and {Lagage}, P.~O. and {Ray}, T. and {Vandenbussche}, B. and {Abergel}, A. and {Absil}, O. and {Argyriou}, I. and {Barrado}, D. and {Boccaletti}, A. and {Bouwman}, J. and {Caratti o Garatti}, A. and {Geers}, V. and {Glauser}, A.~M. and {Justannont}, K. and {Lahuis}, F. and {Mueller}, M. and {Nehm{\'e}}, C. and {Olofsson}, G. and {Pantin}, E. and {Scheithauer}, S. and {Waelkens}, C. and {Waters}, L.~B.~F.~M. and {Black}, J.~H. and {Christiaens}, V. and {Guadarrama}, R. and {Morales-Calder{\'o}n}, M. and {Jang}, H. and {Kanwar}, J. and {Pawellek}, N. and {Perotti}, G. and {Perrin}, A. and {Rodgers-Lee}, D. and {Samland}, M. and {Schreiber}, J. and {Schwarz}, K. and {Colina}, L. and {{\"O}stlin}, G. and {Wright}, G.},
        title = "{A rich hydrocarbon chemistry and high C to O ratio in the inner disk around a very low-mass star}",
      journal = {Nature Astronomy},
         year = 2023,
        month = jul,
       volume = {7},
        pages = {805-814},
          doi = {10.1038/s41550-023-01965-3},
archivePrefix = {arXiv},
       eprint = {2304.05954},
 primaryClass = {astro-ph.EP},
       adsurl = {https://ui.adsabs.harvard.edu/abs/2023NatAs...7..805T}
}

@ARTICLE{Miotello_2019A&A...631A..69M,
       author = {{Miotello}, A. and {Facchini}, S. and {van Dishoeck}, E.~F. and {Cazzoletti}, P. and {Testi}, L. and {Williams}, J.~P. and {Ansdell}, M. and {van Terwisga}, S. and {van der Marel}, N.},
        title = "{Bright C$_{2}$H emission in protoplanetary discs in Lupus: high volatile C/O > 1 ratios}",
      journal = {\aap},
         year = 2019,
        month = nov,
       volume = {631},
          eid = {A69},
        pages = {A69},
          doi = {10.1051/0004-6361/201935441},
archivePrefix = {arXiv},
       eprint = {1909.04477},
 primaryClass = {astro-ph.SR},
       adsurl = {https://ui.adsabs.harvard.edu/abs/2019A&A...631A..69M}
}

@ARTICLE{Miotello_2014A&A...572A..96M,
       author = {{Miotello}, A. and {Bruderer}, S. and {van Dishoeck}, E.~F.},
        title = "{Protoplanetary disk masses from CO isotopologue line emission}",
      journal = {\aap},
         year = 2014,
        month = dec,
       volume = {572},
          eid = {A96},
        pages = {A96},
          doi = {10.1051/0004-6361/201424712},
archivePrefix = {arXiv},
       eprint = {1410.2093},
 primaryClass = {astro-ph.SR},
       adsurl = {https://ui.adsabs.harvard.edu/abs/2014A&A...572A..96M}
}

@ARTICLE{Temmink_2024A&A...686A.117T,
       author = {{Temmink}, Milou and {van Dishoeck}, Ewine F. and {Grant}, Sierra L. and {Tabone}, Beno{\^\i}t and {Gasman}, Danny and {Christiaens}, Valentin and {Samland}, Matthias and {Argyriou}, Ioannis and {Perotti}, Giulia and {G{\"u}del}, Manuel and {Henning}, Thomas and {Lagage}, Pierre-Olivier and {Abergel}, Alain and {Absil}, Olivier and {Barrado}, David and {Caratti o Garatti}, Alessio and {Glauser}, Adrian M. and {Kamp}, Inga and {Lahuis}, Fred and {Olofsson}, G{\"o}ran and {Ray}, Tom P. and {Scheithauer}, Silvia and {Vandenbussche}, Bart and {Waters}, L.~B.~F.~M. and {Arabhavi}, Aditya M. and {Jang}, Hyerin and {Kanwar}, Jayatee and {Morales-Calder{\'o}n}, Maria and {Rodgers-Lee}, Donna and {Schreiber}, J{\"u}rgen and {Schwarz}, Kamber and {Colina}, Luis},
        title = "{MINDS: The DR Tau disk. I. Combining JWST-MIRI data with high-resolution CO spectra to characterise the hot gas}",
      journal = {\aap},
         year = 2024,
        month = jun,
       volume = {686},
          eid = {A117},
        pages = {A117},
          doi = {10.1051/0004-6361/202348911},
archivePrefix = {arXiv},
       eprint = {2403.13591},
 primaryClass = {astro-ph.EP},
       adsurl = {https://ui.adsabs.harvard.edu/abs/2024A&A...686A.117T}
}

@ARTICLE{Schwarz_2024ApJ...962....8S,
       author = {{Schwarz}, Kamber R. and {Henning}, Thomas and {Christiaens}, Valentin and {Gasman}, Danny and {Samland}, Matthias and {Perotti}, Giulia and {Jang}, Hyerin and {Grant}, Sierra L. and {Tabone}, Beno{\^\i}t and {Morales-Calder{\'o}n}, Maria and {Kamp}, Inga and {van Dishoeck}, Ewine F. and {G{\"u}del}, Manuel and {Lagage}, Pierre-Olivier and {Barrado}, David and {Caratti o Garatti}, Alessio and {Glauser}, Adrian M. and {Ray}, Tom P. and {Vandenbussche}, Bart and {Waters}, L.~B.~F.~M. and {Arabhavi}, Aditya M. and {Kanwar}, Jayatee and {Olofsson}, G{\"o}ran and {Rodgers-Lee}, Donna and {Schreiber}, J{\"u}rgen and {Temmink}, Milou},
        title = "{MINDS. JWST/MIRI Reveals a Dynamic Gas-rich Inner Disk inside the Cavity of SY Cha}",
      journal = {\apj},
         year = 2024,
        month = feb,
       volume = {962},
       number = {1},
          eid = {8},
        pages = {8},
          doi = {10.3847/1538-4357/ad1393},
archivePrefix = {arXiv},
       eprint = {2312.07135},
 primaryClass = {astro-ph.EP},
       adsurl = {https://ui.adsabs.harvard.edu/abs/2024ApJ...962....8S}
}

@ARTICLE{Gamsan_2023A&A...679A.117G,
       author = {{Gasman}, Danny and {van Dishoeck}, Ewine F. and {Grant}, Sierra L. and {Temmink}, Milou and {Tabone}, Beno{\^\i}t and {Henning}, Thomas and {Kamp}, Inga and {G{\"u}del}, Manuel and {Lagage}, Pierre-Olivier and {Perotti}, Giulia and {Christiaens}, Valentin and {Samland}, Matthias and {Arabhavi}, Aditya M. and {Argyriou}, Ioannis and {Abergel}, Alain and {Absil}, Olivier and {Barrado}, David and {Boccaletti}, Anthony and {Bouwman}, Jeroen and {Caratti o Garatti}, Alessio and {Geers}, Vincent and {Glauser}, Adrian M. and {Guadarrama}, Rodrigo and {Jang}, Hyerin and {Kanwar}, Jayatee and {Lahuis}, Fred and {Morales-Calder{\'o}n}, Maria and {Mueller}, Michael and {Nehm{\'e}}, Cyrine and {Olofsson}, G{\"o}ran and {Pantin}, {\'E}ric and {Pawellek}, Nicole and {Ray}, Tom P. and {Rodgers-Lee}, Donna and {Scheithauer}, Silvia and {Schreiber}, J{\"u}rgen and {Schwarz}, Kamber and {Vandenbussche}, Bart and {Vlasblom}, Marissa and {Waters}, Rens L.~B.~F.~M. and {Wright}, Gillian and {Colina}, Luis and {Greve}, Thomas R. and {{\"O}stlin}, G{\"o}ran},
        title = "{MINDS. Abundant water and varying C/O across the disk of Sz 98 as seen by JWST/MIRI}",
      journal = {\aap},
         year = 2023,
        month = nov,
       volume = {679},
          eid = {A117},
        pages = {A117},
          doi = {10.1051/0004-6361/202347005},
archivePrefix = {arXiv},
       eprint = {2307.09301},
 primaryClass = {astro-ph.EP},
       adsurl = {https://ui.adsabs.harvard.edu/abs/2023A&A...679A.117G}
}

@ARTICLE{Grant_2023ApJ...947L...6G,
       author = {{Grant}, Sierra L. and {van Dishoeck}, Ewine F. and {Tabone}, Beno{\^\i}t and {Gasman}, Danny and {Henning}, Thomas and {Kamp}, Inga and {G{\"u}del}, Manuel and {Lagage}, Pierre-Olivier and {Bettoni}, Giulio and {Perotti}, Giulia and {Christiaens}, Valentin and {Samland}, Matthias and {Arabhavi}, Aditya M. and {Argyriou}, Ioannis and {Abergel}, Alain and {Absil}, Olivier and {Barrado}, David and {Boccaletti}, Anthony and {Bouwman}, Jeroen and {Caratti o Garatti}, Alessio and {Geers}, Vincent and {Glauser}, Adrian M. and {Guadarrama}, Rodrigo and {Jang}, Hyerin and {Kanwar}, Jayatee and {Lahuis}, Fred and {Morales-Calder{\'o}n}, Maria and {Mueller}, Michael and {Nehm{\'e}}, Cyrine and {Olofsson}, G{\"o}ran and {Pantin}, Eric and {Pawellek}, Nicole and {Ray}, Tom P. and {Rodgers-Lee}, Donna and {Scheithauer}, Silvia and {Schreiber}, J{\"u}rgen and {Schwarz}, Kamber and {Temmink}, Milou and {Vandenbussche}, Bart and {Vlasblom}, Marissa and {Waters}, L.~B.~F.~M. and {Wright}, Gillian and {Colina}, Luis and {Greve}, Thomas R. and {Justannont}, Kay and {{\"O}stlin}, G{\"o}ran},
        title = "{MINDS. The Detection of $^{13}$CO$_{2}$ with JWST-MIRI Indicates Abundant CO$_{2}$ in a Protoplanetary Disk}",
      journal = {\apjl},
         year = 2023,
        month = apr,
       volume = {947},
       number = {1},
          eid = {L6},
        pages = {L6},
          doi = {10.3847/2041-8213/acc44b},
archivePrefix = {arXiv},
       eprint = {2212.08047},
 primaryClass = {astro-ph.SR},
       adsurl = {https://ui.adsabs.harvard.edu/abs/2023ApJ...947L...6G}
}

@ARTICLE{vanDishoeck_2013ChRv..113.9043V,
       author = {{van Dishoeck}, Ewine F. and {Herbst}, Eric and {Neufeld}, David A.},
        title = "{Interstellar Water Chemistry: From Laboratory to Observations}",
      journal = {Chemical Reviews},
         year = 2013,
        month = dec,
       volume = {113},
       number = {12},
        pages = {9043-9085},
          doi = {10.1021/cr4003177},
archivePrefix = {arXiv},
       eprint = {1312.4684},
 primaryClass = {astro-ph.GA},
       adsurl = {https://ui.adsabs.harvard.edu/abs/2013ChRv..113.9043V}
}

@article{woitke_modelling_2018,
	title = {Modelling mid-infrared molecular emission lines from {T} {Tauri} stars},
	volume = {618},
	copyright = {https://www.edpsciences.org/en/authors/copyright-and-licensing},
	issn = {0004-6361, 1432-0746},
	url = {https://www.aanda.org/10.1051/0004-6361/201731460},
	doi = {10.1051/0004-6361/201731460},
	language = {en},
	urldate = {2024-05-01},
	journal = {\aap},
	author = {Woitke, P. and Min, M. and Thi, W.-F. and Roberts, C. and Carmona, A. and Kamp, I. and Ménard, F. and Pinte, C.},
	month = oct,
	year = {2018},
	pages = {A57},
}

@article{van_dishoeck_diverse_2023,
	title = {The diverse chemistry of protoplanetary disks as revealed by {JWST}},
	volume = {245},
	issn = {1359-6640, 1364-5498},
	url = {http://arxiv.org/abs/2307.11817},
	doi = {10.1039/d3fd00010a},
	language = {en},
	urldate = {2024-05-01},
	journal = {Faraday Discussions},
	author = {van Dishoeck, Ewine F. and Grant, S. and Tabone, B. and van Gelder, M. and Francis, L. and Tychoniec, L. and Bettoni, G. and Arabhavi, A. M. and Gasman, D. and Nazari, P. and Vlasblom, M. and Kavanagh, P. and Christiaens, V. and Klaassen, P. and Beuther, H. and Henning, Th and Kamp, I.},
	year = {2023},
	pages = {52--79},
}

@article{kamp_chemical_2023,
	title = {The {Chemical} {Inventory} of the {Inner} {Regions} of {Planet}-forming {Disks} – {The} {JWST}/{MINDS} {Program}},
	volume = {245},
	issn = {1359-6640, 1364-5498},
	url = {http://arxiv.org/abs/2307.16729},
	doi = {10.1039/D3FD00013C},
	language = {en},
	urldate = {2024-05-01},
	journal = {Faraday Discussions},
	author = {Kamp, Inga and Henning, Thomas and Arabhavi, Aditya M. and Bettoni, Giulio and Christiaens, Valentin and Gasman, Danny and Grant, Sierra L. and Morales-Calderón, Maria and Tabone, Benoît and Abergel, Alain and Absil, Olivier and Argyriou, Ioannis and Barrado, David and Boccaletti, Anthony and Bouwman, Jeroen and Garatti, Alessio Caratti o and van Dishoeck, Ewine F. and Geers, Vincent and Glauser, Adrian M. and Güdel, Manuel and Guadarrama, Rodrigo and Jang, Hyerin and Kanwar, Jayatee and Lagage, Pierre-Olivier and Lahuis, Fred and Mueller, Michael and Nehmé, Cyrine and Olofsson, Göran and Pantin, Eric and Pawellek, Nicole and Perotti, Giulia and Ray, Tom P. and Rodgers-Lee, Donna and Samland, Matthias and Scheithauer, Silvia and Schreiber, Jürgen and Schwarz, Kamber and Temmink, Milou and Vandenbussche, Bart and Vlasblom, Marissa and Waelkens, Christoffel and Waters, L. B. F. M. and Wright, Gillian},
	year = {2023},
	pages = {112--137},
}

@article{bruderer_warm_2012,
	title = {The warm gas atmosphere of the {HD} 100546 disk seen by \textit{{Herschel}}: {Evidence} of a gas-rich, carbon-poor atmosphere?},
	volume = {541},
	issn = {0004-6361, 1432-0746},
	shorttitle = {The warm gas atmosphere of the {HD} 100546 disk seen by \textit{{Herschel}}},
	url = {http://www.aanda.org/10.1051/0004-6361/201118218},
	doi = {10.1051/0004-6361/201118218},
	language = {en},
	urldate = {2024-05-01},
	journal = {\aap},
	author = {Bruderer, S. and van Dishoeck, E. F. and Doty, S. D. and Herczeg, G. J.},
	month = may,
	year = {2012},
	pages = {A91},
}

@article{banzatti_hints_2020,
	title = {Hints for {Icy} {Pebble} {Migration} {Feeding} an {Oxygen}-rich {Chemistry} in the {Inner} {Planet}-forming {Region} of {Disks}},
	volume = {903},
	issn = {0004-637X, 1538-4357},
	url = {https://iopscience.iop.org/article/10.3847/1538-4357/abbc1a},
	doi = {10.3847/1538-4357/abbc1a},
	language = {en},
	number = {2},
	urldate = {2024-05-01},
	journal = {\apj},
	author = {Banzatti, Andrea and Pascucci, Ilaria and Bosman, Arthur D. and Pinilla, Paola and Salyk, Colette and Herczeg, Gregory J. and Pontoppidan, Klaus M. and Vazquez, Ivan and Watkins, Andrew and Krijt, Sebastiaan and Hendler, Nathan and Long, Feng},
	month = nov,
	year = {2020},
	pages = {124},
}

@article{pontoppidan_spitzer_2010,
	title = {A \textit{{SPITZER}} {SURVEY} {OF} {MID}-{INFRARED} {MOLECULAR} {EMISSION} {FROM} {PROTOPLANETARY} {DISKS}. {I}. {DETECTION} {RATES}},
	volume = {720},
	issn = {0004-637X, 1538-4357},
	url = {https://iopscience.iop.org/article/10.1088/0004-637X/720/1/887},
	doi = {10.1088/0004-637X/720/1/887},
	language = {en},
	number = {1},
	urldate = {2024-05-18},
	journal = {\apj},
	author = {Pontoppidan, Klaus M. and Salyk, Colette and Blake, Geoffrey A. and Meijerink, Rowin and Carr, John S. and Najita, Joan},
	month = sep,
	year = {2010},
	pages = {887--903},
}

@article{salyk_h_2008,
	title = {H ₂ {O} and {OH} {Gas} in the {Terrestrial} {Planet}-forming {Zones} of {Protoplanetary} {Disks}},
	volume = {676},
	issn = {0004-637X, 1538-4357},
	url = {https://iopscience.iop.org/article/10.1086/586894},
	doi = {10.1086/586894},
	language = {en},
	number = {1},
	urldate = {2024-05-20},
	journal = {\apj},
	author = {Salyk, Colette and Pontoppidan, Klaus M. and Blake, Geoffrey A. and Lahuis, Fred and van Dishoeck, Ewine F. and Evans Ii, Neal J.},
	month = mar,
	year = {2008},
	pages = {L49--L52},
}

@article{carr_organic_2008,
	title = {Organic {Molecules} and {Water} in the {Planet} {Formation} {Region} of {Young} {Circumstellar} {Disks}},
	volume = {319},
	issn = {0036-8075, 1095-9203},
	url = {https://www.science.org/doi/10.1126/science.1153807},
	doi = {10.1126/science.1153807},
	language = {en},
	number = {5869},
	urldate = {2024-05-20},
	journal = {Science},
	author = {Carr, John S. and Najita, Joan R.},
	month = mar,
	year = {2008},
	pages = {1504--1506},
}

@article{madhusudhan_exoplanetary_2016,
	title = {Exoplanetary {Atmospheres} - {Chemistry}, {Formation} {Conditions}, and {Habitability}},
	volume = {205},
	issn = {0038-6308, 1572-9672},
	url = {http://arxiv.org/abs/1604.06092},
	doi = {10.1007/s11214-016-0254-3},
	language = {en},
	number = {1-4},
	urldate = {2024-05-23},
	journal = {Space Science Reviews},
	author = {Madhusudhan, Nikku and Agúndez, Marcelino and Moses, Julianne I. and Hu, Yongyun},
	month = dec,
	year = {2016},
	pages = {285--348},
}

@article{carr_organic_2011,
	title = {{ORGANIC} {MOLECULES} {AND} {WATER} {IN} {THE} {INNER} {DISKS} {OF} {T} {TAURI} {STARS}},
	volume = {733},
	issn = {0004-637X, 1538-4357},
	url = {https://iopscience.iop.org/article/10.1088/0004-637X/733/2/102},
	doi = {10.1088/0004-637X/733/2/102},
	language = {en},
	number = {2},
	urldate = {2024-05-23},
	journal = {\apj},
	author = {Carr, John S. and Najita, Joan R.},
	month = jun,
	year = {2011},
	pages = {102},
}

@article{walsh_molecular_2015,
	title = {The molecular composition of the planet-forming regions of protoplanetary disks across the luminosity regime},
	volume = {582},
	issn = {0004-6361, 1432-0746},
	url = {http://www.aanda.org/10.1051/0004-6361/201526751},
	doi = {10.1051/0004-6361/201526751},
	language = {en},
	urldate = {2024-05-27},
	journal = {\aap},
	author = {Walsh, Catherine and Nomura, Hideko and van Dishoeck, Ewine},
	month = oct,
	year = {2015},
	pages = {A88},
}

@article{woitke_radiation_2009,
	title = {Radiation thermo-chemical models of protoplanetary disks: {I}. {Hydrostatic} disk structure and inner rim},
	volume = {501},
	issn = {0004-6361, 1432-0746},
	shorttitle = {Radiation thermo-chemical models of protoplanetary disks},
	url = {http://www.aanda.org/10.1051/0004-6361/200911821},
	doi = {10.1051/0004-6361/200911821},
	language = {en},
	number = {1},
	urldate = {2024-06-02},
	journal = {\aap},
	author = {Woitke, P. and Kamp, I. and Thi, W.-F.},
	month = jul,
	year = {2009},
	pages = {383--406},
}

@article{henning_minds_2024,
	title = {{MINDS}: {The} {JWST} {MIRI} {Mid}-{INfrared} {Disk} {Survey}},
	volume = {136},
	issn = {0004-6280, 1538-3873},
	shorttitle = {{MINDS}},
	url = {https://iopscience.iop.org/article/10.1088/1538-3873/ad3455},
	doi = {10.1088/1538-3873/ad3455},
	language = {en},
	number = {5},
	urldate = {2024-06-02},
	journal = {Publications of the Astronomical Society of the Pacific},
	author = {Henning, Thomas and Kamp, Inga and Samland, Matthias and Arabhavi, Aditya M. and Kanwar, Jayatee and van Dishoeck, Ewine F. and Güdel, Manuel and Lagage, Pierre-Olivier and Waelkens, Christoffel and Abergel, Alain and Absil, Olivier and Barrado, David and Boccaletti, Anthony and Bouwman, Jeroen and Caratti O Garatti, Alessio and Geers, Vincent and Glauser, Adrian M. and Lahuis, Fred and Mueller, Michael and Nehmé, Cyrine and Olofsson, Göran and Pantin, Eric and Ray, Tom P. and Scheithauer, Silvia and Vandenbussche, Bart and Waters, L. B. F. M. and Wright, Gillian and Argyriou, Ioannis and Christiaens, Valentin and Franceschi, Riccardo and Gasman, Danny and Grant, Sierra L. and Guadarrama, Rodrigo and Jang, Hyerin and Morales-Calderón, Maria and Pawellek, Nicole and Perotti, Giulia and Rodgers-Lee, Donna and Schreiber, Jürgen and Schwarz, Kamber and Tabone, Benoît and Temmink, Milou and Vlasblom, Marissa and Colina, Luis and Greve, Thomas R. and Östlin, Göran},
	month = may,
	year = {2024},
	pages = {054302},
}

@article{arabhavi_abundant_2024,
	title = {Abundant hydrocarbons in the disk around a very-low-mass star},
	volume = {384},
	issn = {0036-8075, 1095-9203},
	url = {https://www.science.org/doi/10.1126/science.adi8147},
	doi = {10.1126/science.adi8147},
	language = {en},
	number = {6700},
	urldate = {2024-06-25},
	journal = {Science},
	author = {Arabhavi, A. M. and Kamp, I. and Henning, Th. and van Dishoeck, E. F. and Christiaens, V. and Gasman, D. and Perrin, A. and Güdel, M. and Tabone, B. and Kanwar, J. and Waters, L. B. F. M. and Pascucci, I. and Samland, M. and Perotti, G. and Bettoni, G. and Grant, S. L. and Lagage, P. O. and Ray, T. P. and Vandenbussche, B. and Absil, O. and Argyriou, I. and Barrado, D. and Boccaletti, A. and Bouwman, J. and Caratti O Garatti, A. and Glauser, A. M. and Lahuis, F. and Mueller, M. and Olofsson, G. and Pantin, E. and Scheithauer, S. and Morales-Calderón, M. and Franceschi, R. and Jang, H. and Pawellek, N. and Rodgers-Lee, D. and Schreiber, J. and Schwarz, K. and Temmink, M. and Vlasblom, M. and Wright, G. and Colina, L. and Östlin, G.},
	month = jun,
	year = {2024},
	pages = {1086--1090},
}

@article{cridland_connecting_2019b,
	title = {Connecting planet formation and astrochemistry: {Refractory} carbon depletion leading to super-stellar {C}/{O} in giant planetary atmospheres},
	volume = {627},
	copyright = {https://www.edpsciences.org/en/authors/copyright-and-licensing},
	issn = {0004-6361, 1432-0746},
	shorttitle = {Connecting planet formation and astrochemistry},
	url = {https://www.aanda.org/10.1051/0004-6361/201834378},
	doi = {10.1051/0004-6361/201834378},
	language = {en},
	urldate = {2024-06-27},
	journal = {\aap},
	author = {Cridland, Alexander J. and Eistrup, Christian and van Dishoeck, Ewine F.},
	month = jul,
	year = {2019},
	pages = {A127},
}

@ARTICLE{oberg_effects_2011,
       author = {{{\"O}berg}, Karin I. and {Murray-Clay}, Ruth and {Bergin}, Edwin A.},
        title = "{The Effects of Snowlines on C/O in Planetary Atmospheres}",
      journal = {\apjl},
         year = 2011,
        month = dec,
       volume = {743},
       number = {1},
          eid = {L16},
        pages = {L16},
          doi = {10.1088/2041-8205/743/1/L16},
archivePrefix = {arXiv},
       eprint = {1110.5567},
 primaryClass = {astro-ph.GA},
       adsurl = {https://ui.adsabs.harvard.edu/abs/2011ApJ...743L..16O}
}

@article{booth_chemical_2017,
	title = {Chemical enrichment of giant planets and discs due to pebble drift},
	volume = {469},
	issn = {0035-8711, 1365-2966},
	url = {https://academic.oup.com/mnras/article-lookup/doi/10.1093/mnras/stx1103},
	doi = {10.1093/mnras/stx1103},
	language = {en},
	number = {4},
	urldate = {2024-06-27},
	journal = {\mnras},
	author = {Booth, Richard A. and Clarke, Cathie J. and Madhusudhan, Nikku and Ilee, John D.},
	month = aug,
	year = {2017},
	pages = {3994--4011},
}

@article{banzatti_jwst_2023,
	title = {{JWST} {Reveals} {Excess} {Cool} {Water} near the {Snow} {Line} in {Compact} {Disks}, {Consistent} with {Pebble} {Drift}},
	volume = {957},
	issn = {2041-8205, 2041-8213},
	url = {https://iopscience.iop.org/article/10.3847/2041-8213/acf5ec},
	doi = {10.3847/2041-8213/acf5ec},
	language = {en},
	number = {2},
	urldate = {2024-06-27},
	journal = {\apjl},
	author = {Banzatti, Andrea and Pontoppidan, Klaus M. and Carr, John S. and Jellison, Evan and Pascucci, Ilaria and Najita, Joan R. and Muñoz-Romero, Carlos E. and Öberg, Karin I. and Kalyaan, Anusha and Pinilla, Paola and Krijt, Sebastiaan and Long, Feng and Lambrechts, Michiel and Rosotti, Giovanni and Herczeg, Gregory J. and Salyk, Colette and Zhang, Ke and Bergin, Edwin A. and Ballering, Nicholas P. and Meyer, Michael R. and Bruderer, Simon and {The JDISCS Collaboration}},
	month = nov,
	year = {2023},
	pages = {L22},
}

@article{bruderer_survival_2013,
	title = {Survival of molecular gas in cavities of transition disks: {I}. {CO}},
	volume = {559},
	issn = {0004-6361, 1432-0746},
	shorttitle = {Survival of molecular gas in cavities of transition disks},
	url = {http://www.aanda.org/10.1051/0004-6361/201321171},
	doi = {10.1051/0004-6361/201321171},
	language = {en},
	urldate = {2024-06-27},
	journal = {\aap},
	author = {Bruderer, Simon},
	month = nov,
	year = {2013},
	pages = {A46},
}

@article{lynden-bell_evolution_1974,
	title = {The evolution of viscous discs and the origin of the nebular variables.},
	volume = {168},
	issn = {0035-8711},
	url = {https://ui.adsabs.harvard.edu/abs/1974MNRAS.168..603L},
	doi = {10.1093/mnras/168.3.603},
	urldate = {2024-06-27},
	journal = {\mnras},
	author = {Lynden-Bell, D. and Pringle, J. E.},
	month = sep,
	year = {1974},
	pages = {603--637},
}

@ARTICLE{Bast_2013A&A...551A.118B,
       author = {{Bast}, J.~E. and {Lahuis}, F. and {van Dishoeck}, E.~F. and {Tielens}, A.~G.~G.~M.},
        title = "{Exploring organic chemistry in planet-forming zones}",
      journal = {\aap},
         year = 2013,
        month = mar,
       volume = {551},
          eid = {A118},
        pages = {A118},
          doi = {10.1051/0004-6361/201219908},
archivePrefix = {arXiv},
       eprint = {1212.3297},
 primaryClass = {astro-ph.SR},
       adsurl = {https://ui.adsabs.harvard.edu/abs/2013A&A...551A.118B}
}

@ARTICLE{Grant_2025A&A...702A.126G,
       author = {{Grant}, S.~L. and {Temmink}, M. and {van Dishoeck}, E.~F. and {Gasman}, D. and {Arabhavi}, A.~M. and {Tabone}, B. and {Henning}, T. and {Kamp}, I. and {Caratti o Garatti}, A. and {Christiaens}, V. and {Esteve}, P. and {G{\"u}del}, M. and {Jang}, H. and {Kaeufer}, T. and {Kurtovic}, N.~T. and {Morales-Calder{\'o}n}, M. and {Perotti}, G. and {Schwarz}, K. and {Sellek}, A.~D. and {Stapper}, L.~M. and {Vlasblom}, M. and {Waters}, L.~B.~F.~M.},
        title = "{MINDS: A transition from H$_{2}$O to C$_{2}$H$_{2}$ dominated disk spectra with decreasing stellar luminosity}",
      journal = {\aap},
         year = 2025,
        month = oct,
       volume = {702},
          eid = {A126},
        pages = {A126},
          doi = {10.1051/0004-6361/202555862},
archivePrefix = {arXiv},
       eprint = {2508.04692},
 primaryClass = {astro-ph.EP},
       adsurl = {https://ui.adsabs.harvard.edu/abs/2025A&A...702A.126G}
}

@ARTICLE{Kanwar_2025A&A...698A.294K,
       author = {{Kanwar}, Jayatee and {Woitke}, Peter and {Kamp}, Inga and {Rimmer}, Paul and {Helling}, Christiane},
        title = "{Can thermodynamic equilibrium be established in planet-forming disks?}",
      journal = {\aap},
         year = 2025,
        month = jun,
       volume = {698},
          eid = {A294},
        pages = {A294},
          doi = {10.1051/0004-6361/202452249},
archivePrefix = {arXiv},
       eprint = {2505.13705},
 primaryClass = {astro-ph.EP},
       adsurl = {https://ui.adsabs.harvard.edu/abs/2025A&A...698A.294K}
}

@ARTICLE{MoralesCalderon_2025arXiv250805155M,
       author = {{Morales-Calder{\'o}n}, Mar{\'\i}a and {Jang}, Hyerin and {Arabhavi}, Aditya M. and {Christiaens}, Valentin and {Barrado}, David and {Kamp}, Inga and {van Dishoeck}, Ewine F. and {Henning}, Thomas and {Waters}, L.~B.~F.~M. and {Temmink}, Milou and {G{\"u}del}, Manuel and {Lagage}, Pierre-Olivier and {Caratti o Garatti}, Alessio and {Glauser}, Adrian M. and {Ray}, Tom P. and {Franceschi}, Riccardo and {Gasman}, Danny and {Grant}, Sierra L. and {Kaeufer}, Till and {Kanwar}, Jayatee and {Perotti}, Giulia and {Samland}, Matthias and {Schwarz}, Kamber and {Vlasblom}, Marissa and {Colina}, Luis and {{\"O}stlin}, G{\"o}ran},
        title = "{MINDS: Cha H{\ensuremath{\alpha}} 1, a brown dwarf with a hydrocarbon-rich disk}",
      journal = {\aap},
         year = 2025,
        month = nov,
       volume = {703},
          eid = {A18},
        pages = {A18},
          doi = {10.1051/0004-6361/202555621},
archivePrefix = {arXiv},
       eprint = {2508.05155},
 primaryClass = {astro-ph.SR},
       adsurl = {https://ui.adsabs.harvard.edu/abs/2025A&A...703A..18M}
}

@ARTICLE{Lenzuni_1995ApJ...447..848L,
       author = {{Lenzuni}, Paolo and {Gail}, Hans-Peter and {Henning}, Thomas},
        title = "{Dust Evaporation in Protostellar Cores}",
      journal = {\apj},
         year = 1995,
        month = jul,
       volume = {447},
        pages = {848},
          doi = {10.1086/175922},
       adsurl = {https://ui.adsabs.harvard.edu/abs/1995ApJ...447..848L}
}

@ARTICLE{Oberg_2021ApJS..257....1O,
       author = {{{\"O}berg}, Karin I. and {Guzm{\'a}n}, Viviana V. and {Walsh}, Catherine and {Aikawa}, Yuri and {Bergin}, Edwin A. and {Law}, Charles J. and {Loomis}, Ryan A. and {Alarc{\'o}n}, Felipe and {Andrews}, Sean M. and {Bae}, Jaehan and {Bergner}, Jennifer B. and {Boehler}, Yann and {Booth}, Alice S. and {Bosman}, Arthur D. and {Calahan}, Jenny K. and {Cataldi}, Gianni and {Cleeves}, L. Ilsedore and {Czekala}, Ian and {Furuya}, Kenji and {Huang}, Jane and {Ilee}, John D. and {Kurtovic}, Nicolas T. and {Le Gal}, Romane and {Liu}, Yao and {Long}, Feng and {M{\'e}nard}, Fran{\c{c}}ois and {Nomura}, Hideko and {P{\'e}rez}, Laura M. and {Qi}, Chunhua and {Schwarz}, Kamber R. and {Sierra}, Anibal and {Teague}, Richard and {Tsukagoshi}, Takashi and {Yamato}, Yoshihide and {van't Hoff}, Merel L.~R. and {Waggoner}, Abygail R. and {Wilner}, David J. and {Zhang}, Ke},
        title = "{Molecules with ALMA at Planet-forming Scales (MAPS). I. Program Overview and Highlights}",
      journal = {\apjs},
         year = 2021,
        month = nov,
       volume = {257},
       number = {1},
          eid = {1},
        pages = {1},
          doi = {10.3847/1538-4365/ac1432},
archivePrefix = {arXiv},
       eprint = {2109.06268},
 primaryClass = {astro-ph.EP},
       adsurl = {https://ui.adsabs.harvard.edu/abs/2021ApJS..257....1O}
}

@ARTICLE{ALma_HLTau_2015ApJ...808L...3A,
       author = {{ALMA Partnership} and {Brogan}, C.~L. and {P{\'e}rez}, L.~M. and {Hunter}, T.~R. and {Dent}, W.~R.~F. and {Hales}, A.~S. and {Hills}, R.~E. and {Corder}, S. and {Fomalont}, E.~B. and {Vlahakis}, C. and {Asaki}, Y. and {Barkats}, D. and {Hirota}, A. and {Hodge}, J.~A. and {Impellizzeri}, C.~M.~V. and {Kneissl}, R. and {Liuzzo}, E. and {Lucas}, R. and {Marcelino}, N. and {Matsushita}, S. and {Nakanishi}, K. and {Phillips}, N. and {Richards}, A.~M.~S. and {Toledo}, I. and {Aladro}, R. and {Broguiere}, D. and {Cortes}, J.~R. and {Cortes}, P.~C. and {Espada}, D. and {Galarza}, F. and {Garcia-Appadoo}, D. and {Guzman-Ramirez}, L. and {Humphreys}, E.~M. and {Jung}, T. and {Kameno}, S. and {Laing}, R.~A. and {Leon}, S. and {Marconi}, G. and {Mignano}, A. and {Nikolic}, B. and {Nyman}, L. -A. and {Radiszcz}, M. and {Remijan}, A. and {Rod{\'o}n}, J.~A. and {Sawada}, T. and {Takahashi}, S. and {Tilanus}, R.~P.~J. and {Vila Vilaro}, B. and {Watson}, L.~C. and {Wiklind}, T. and {Akiyama}, E. and {Chapillon}, E. and {de Gregorio-Monsalvo}, I. and {Di Francesco}, J. and {Gueth}, F. and {Kawamura}, A. and {Lee}, C. -F. and {Nguyen Luong}, Q. and {Mangum}, J. and {Pietu}, V. and {Sanhueza}, P. and {Saigo}, K. and {Takakuwa}, S. and {Ubach}, C. and {van Kempen}, T. and {Wootten}, A. and {Castro-Carrizo}, A. and {Francke}, H. and {Gallardo}, J. and {Garcia}, J. and {Gonzalez}, S. and {Hill}, T. and {Kaminski}, T. and {Kurono}, Y. and {Liu}, H. -Y. and {Lopez}, C. and {Morales}, F. and {Plarre}, K. and {Schieven}, G. and {Testi}, L. and {Videla}, L. and {Villard}, E. and {Andreani}, P. and {Hibbard}, J.~E. and {Tatematsu}, K.},
        title = "{The 2014 ALMA Long Baseline Campaign: First Results from High Angular Resolution Observations toward the HL Tau Region}",
      journal = {\apjl},
         year = 2015,
        month = jul,
       volume = {808},
       number = {1},
          eid = {L3},
        pages = {L3},
          doi = {10.1088/2041-8205/808/1/L3},
archivePrefix = {arXiv},
       eprint = {1503.02649},
 primaryClass = {astro-ph.SR},
       adsurl = {https://ui.adsabs.harvard.edu/abs/2015ApJ...808L...3A}
}

@ARTICLE{Temmink_2025A&A...699A.134T,
       author = {{Temmink}, Milou and {Sellek}, Andrew D. and {Gasman}, Danny and {van Dishoeck}, Ewine F. and {Vlasblom}, Marissa and {Pranger}, Ang{\`e}l and {G{\"u}del}, Manuel and {Henning}, Thomas and {Lagage}, Pierre-Olivier and {Caratti o Garatti}, Alessio and {Kamp}, Inga and {Olofsson}, G{\"o}ran and {Arabhavi}, Aditya M. and {Grant}, Sierra L. and {Kaeufer}, Till and {Kurtovic}, Nicolas T. and {Perotti}, Giulia and {Samland}, Matthias and {Schwarz}, Kamber and {Tabone}, Beno{\^\i}t},
        title = "{MINDS: Water reservoirs of compact planet-forming dust discs: A diversity of H$_{2}$O distributions}",
      journal = {\aap},
         year = 2025,
        month = jul,
       volume = {699},
          eid = {A134},
        pages = {A134},
          doi = {10.1051/0004-6361/202554213},
archivePrefix = {arXiv},
       eprint = {2505.15237},
 primaryClass = {astro-ph.EP},
       adsurl = {https://ui.adsabs.harvard.edu/abs/2025A&A...699A.134T}
}

@ARTICLE{Bosman_2021ApJS..257....7B,
       author = {{Bosman}, Arthur D. and {Alarc{\'o}n}, Felipe and {Bergin}, Edwin A. and {Zhang}, Ke and {van't Hoff}, Merel L.~R. and {{\"O}berg}, Karin I. and {Guzm{\'a}n}, Viviana V. and {Walsh}, Catherine and {Aikawa}, Yuri and {Andrews}, Sean M. and {Bergner}, Jennifer B. and {Booth}, Alice S. and {Cataldi}, Gianni and {Cleeves}, L. Ilsedore and {Czekala}, Ian and {Furuya}, Kenji and {Huang}, Jane and {Ilee}, John D. and {Law}, Charles J. and {Le Gal}, Romane and {Liu}, Yao and {Long}, Feng and {Loomis}, Ryan A. and {M{\'e}nard}, Fran{\c{c}}ois and {Nomura}, Hideko and {Qi}, Chunhua and {Schwarz}, Kamber R. and {Teague}, Richard and {Tsukagoshi}, Takashi and {Yamato}, Yoshihide and {Wilner}, David J.},
        title = "{Molecules with ALMA at Planet-forming Scales (MAPS). VII. Substellar O/H and C/H and Superstellar C/O in Planet-feeding Gas}",
      journal = {\apjs},
         year = 2021,
        month = nov,
       volume = {257},
       number = {1},
          eid = {7},
        pages = {7},
          doi = {10.3847/1538-4365/ac1435},
archivePrefix = {arXiv},
       eprint = {2109.06221},
 primaryClass = {astro-ph.EP},
       adsurl = {https://ui.adsabs.harvard.edu/abs/2021ApJS..257....7B}
}

@ARTICLE{Sturm_2022A&A...660A.126S,
       author = {{Sturm}, J.~A. and {McClure}, M.~K. and {Harsono}, D. and {Facchini}, S. and {Long}, F. and {Kama}, M. and {Bergin}, E.~A. and {van Dishoeck}, E.~F.},
        title = "{Tracing pebble drift and trapping using radial carbon depletion profiles in protoplanetary disks}",
      journal = {\aap},
         year = 2022,
        month = apr,
       volume = {660},
          eid = {A126},
        pages = {A126},
          doi = {10.1051/0004-6361/202141860},
archivePrefix = {arXiv},
       eprint = {2201.04089},
 primaryClass = {astro-ph.EP},
       adsurl = {https://ui.adsabs.harvard.edu/abs/2022A&A...660A.126S}
}

@ARTICLE{Sellek_2025A&A...701A.239S,
       author = {{Sellek}, Andrew D. and {van Dishoeck}, Ewine F.},
        title = "{Chemical transformation of CO in evolving protoplanetary discs across stellar masses: A route to C-rich inner regions}",
      journal = {\aap},
         year = 2025,
        month = sep,
       volume = {701},
          eid = {A239},
        pages = {A239},
          doi = {10.1051/0004-6361/202555195},
       adsurl = {https://ui.adsabs.harvard.edu/abs/2025A&A...701A.239S}
}

@ARTICLE{Maijereink_2009ApJ...704.1471M,
       author = {{Meijerink}, R. and {Pontoppidan}, K.~M. and {Blake}, G.~A. and {Poelman}, D.~R. and {Dullemond}, C.~P.},
        title = "{Radiative Transfer Models of Mid-Infrared H$_{2}$O Lines in the Planet-Forming Region of Circumstellar Disks}",
      journal = {\apj},
         year = 2009,
        month = oct,
       volume = {704},
       number = {2},
        pages = {1471-1481},
          doi = {10.1088/0004-637X/704/2/1471},
archivePrefix = {arXiv},
       eprint = {0909.0975},
 primaryClass = {astro-ph.EP},
       adsurl = {https://ui.adsabs.harvard.edu/abs/2009ApJ...704.1471M}
}

@ARTICLE{Glassgold_2009ApJ...701..142G,
       author = {{Glassgold}, A.~E. and {Meijerink}, R. and {Najita}, J.~R.},
        title = "{Formation of Water in the Warm Atmospheres of Protoplanetary Disks}",
      journal = {\apj},
         year = 2009,
        month = aug,
       volume = {701},
       number = {1},
        pages = {142-153},
          doi = {10.1088/0004-637X/701/1/142},
archivePrefix = {arXiv},
       eprint = {0905.4523},
 primaryClass = {astro-ph.GA},
       adsurl = {https://ui.adsabs.harvard.edu/abs/2009ApJ...701..142G}
}

@ARTICLE{Antonellini_2015A&A...582A.105A,
       author = {{Antonellini}, S. and {Kamp}, I. and {Riviere-Marichalar}, P. and {Meijerink}, R. and {Woitke}, P. and {Thi}, W.-F. and {Spaans}, M. and {Aresu}, G. and {Lee}, E.},
        title = "{Understanding the water emission in the mid- and far-IR from protoplanetary disks around T Tauri stars}",
      journal = {\aap},
         year = 2015,
        month = oct,
       volume = {582},
          eid = {A105},
        pages = {A105},
          doi = {10.1051/0004-6361/201525724},
archivePrefix = {arXiv},
       eprint = {1510.01482},
 primaryClass = {astro-ph.SR},
       adsurl = {https://ui.adsabs.harvard.edu/abs/2015A&A...582A.105A}
}

@ARTICLE{Najita_2013ApJ...766..134N,
       author = {{Najita}, Joan R. and {Carr}, John S. and {Pontoppidan}, Klaus M. and {Salyk}, Colette and {van Dishoeck}, Ewine F. and {Blake}, Geoffrey A.},
        title = "{The HCN-Water Ratio in the Planet Formation Region of Disks}",
      journal = {\apj},
         year = 2013,
        month = apr,
       volume = {766},
       number = {2},
          eid = {134},
        pages = {134},
          doi = {10.1088/0004-637X/766/2/134},
archivePrefix = {arXiv},
       eprint = {1303.2692},
 primaryClass = {astro-ph.SR},
       adsurl = {https://ui.adsabs.harvard.edu/abs/2013ApJ...766..134N}
}

@ARTICLE{Kalyaan_2021ApJ...921...84K,
       author = {{Kalyaan}, Anusha and {Pinilla}, Paola and {Krijt}, Sebastiaan and {Mulders}, Gijs D. and {Banzatti}, Andrea},
        title = "{Linking Outer Disk Pebble Dynamics and Gaps to Inner Disk Water Enrichment}",
      journal = {\apj},
         year = 2021,
        month = nov,
       volume = {921},
       number = {1},
          eid = {84},
        pages = {84},
          doi = {10.3847/1538-4357/ac1e96},
archivePrefix = {arXiv},
       eprint = {2109.02687},
 primaryClass = {astro-ph.EP},
       adsurl = {https://ui.adsabs.harvard.edu/abs/2021ApJ...921...84K}
}

@ARTICLE{Greenwood_dustevo_2019A&A...626A...6G,
       author = {{Greenwood}, A.~J. and {Kamp}, I. and {Waters}, L.~B.~F.~M. and {Woitke}, P. and {Thi}, W.-F.},
        title = "{Effects of dust evolution on protoplanetary disks in the mid-infrared}",
      journal = {\aap},
         year = 2019,
        month = jun,
       volume = {626},
          eid = {A6},
        pages = {A6},
          doi = {10.1051/0004-6361/201834365},
archivePrefix = {arXiv},
       eprint = {1903.12649},
 primaryClass = {astro-ph.EP},
       adsurl = {https://ui.adsabs.harvard.edu/abs/2019A&A...626A...6G}
}

@ARTICLE{Najita_2011ApJ...743..147N,
       author = {{Najita}, Joan R. and {{\'A}d{\'a}mkovics}, M{\'a}t{\'e} and {Glassgold}, Alfred E.},
        title = "{Formation of Organic Molecules and Water in Warm Disk Atmospheres}",
      journal = {\apj},
         year = 2011,
        month = dec,
       volume = {743},
       number = {2},
          eid = {147},
        pages = {147},
          doi = {10.1088/0004-637X/743/2/147},
archivePrefix = {arXiv},
       eprint = {1109.6673},
 primaryClass = {astro-ph.SR},
       adsurl = {https://ui.adsabs.harvard.edu/abs/2011ApJ...743..147N}
}

@ARTICLE{Greenwood_2019A&A...631A..81G,
       author = {{Greenwood}, A.~J. and {Kamp}, I. and {Waters}, L.~B.~F.~M. and {Woitke}, P. and {Thi}, W.-F.},
        title = "{The infrared line-emitting regions of T Tauri protoplanetary disks}",
      journal = {\aap},
         year = 2019,
        month = nov,
       volume = {631},
          eid = {A81},
        pages = {A81},
          doi = {10.1051/0004-6361/201834175},
archivePrefix = {arXiv},
       eprint = {1910.05400},
 primaryClass = {astro-ph.EP},
       adsurl = {https://ui.adsabs.harvard.edu/abs/2019A&A...631A..81G}
}

@ARTICLE{Chastaing_1999PCCP....1.2247C,
       author = {{Chastaing}, Delphine and {James}, Philip L. and {Sims}, Ian R. and {Smith}, Ian W.~M.},
        title = "{Neutral neutral reactions at the temperatures of interstellar clouds: Rate coefficients for reactions of atomic carbon, C(3P), with O2, C2H2, C2H4 and C3H6 down to 15 K}",
      journal = {Physical Chemistry Chemical Physics (Incorporating Faraday Transactions)},
         year = 1999,
        month = jan,
       volume = {1},
       number = {9},
        pages = {2247-2256},
          doi = {10.1039/A900449A},
       adsurl = {https://ui.adsabs.harvard.edu/abs/1999PCCP....1.2247C}
}

@ARTICLE{LOISON_2017MNRAS.470.4075L,
       author = {{Loison}, Jean-Christophe and {Ag{\'u}ndez}, Marcelino and {Wakelam}, Valentine and {Roueff}, Evelyne and {Gratier}, Pierre and {Marcelino}, N{\'u}ria and {Reyes}, Dianailys Nu{\~n}ez and {Cernicharo}, Jos{\'e} and {Gerin}, Maryvonne},
        title = "{The interstellar chemistry of C$_{3}$H and C$_{3}$H$_{2}$ isomers}",
      journal = {\mnras},
         year = 2017,
        month = oct,
       volume = {470},
       number = {4},
        pages = {4075-4088},
          doi = {10.1093/mnras/stx1265},
archivePrefix = {arXiv},
       eprint = {1707.07926},
 primaryClass = {astro-ph.GA},
       adsurl = {https://ui.adsabs.harvard.edu/abs/2017MNRAS.470.4075L}
}

@ARTICLE{Alata_2014A&A...569A.119A,
       author = {{Alata}, I. and {Cruz-Diaz}, G.~A. and {Mu{\~n}oz Caro}, G.~M. and {Dartois}, E.},
        title = "{Vacuum ultraviolet photolysis of hydrogenated amorphous carbons . I. Interstellar H$_{2}$ and CH$_{4}$ formation rates}",
      journal = {\aap},
         year = 2014,
        month = sep,
       volume = {569},
          eid = {A119},
        pages = {A119},
          doi = {10.1051/0004-6361/201323118},
       adsurl = {https://ui.adsabs.harvard.edu/abs/2014A&A...569A.119A}
}

@ARTICLE{vandishoeck_black_1982ApJ...258..533V,
       author = {{van Dishoeck}, E.~F. and {Black}, J.~H.},
        title = "{The excitation of interstellar C2.}",
      journal = {\apj},
         year = 1982,
        month = jul,
       volume = {258},
        pages = {533-547},
          doi = {10.1086/160104},
       adsurl = {https://ui.adsabs.harvard.edu/abs/1982ApJ...258..533V}
}

@ARTICLE{Andrews_2018ApJ...869L..41A,
       author = {{Andrews}, Sean M. and {Huang}, Jane and {P{\'e}rez}, Laura M. and {Isella}, Andrea and {Dullemond}, Cornelis P. and {Kurtovic}, Nicol{\'a}s T. and {Guzm{\'a}n}, Viviana V. and {Carpenter}, John M. and {Wilner}, David J. and {Zhang}, Shangjia and {Zhu}, Zhaohuan and {Birnstiel}, Tilman and {Bai}, Xue-Ning and {Benisty}, Myriam and {Hughes}, A. Meredith and {{\"O}berg}, Karin I. and {Ricci}, Luca},
        title = "{The Disk Substructures at High Angular Resolution Project (DSHARP). I. Motivation, Sample, Calibration, and Overview}",
      journal = {\apjl},
         year = 2018,
        month = dec,
       volume = {869},
       number = {2},
          eid = {L41},
        pages = {L41},
          doi = {10.3847/2041-8213/aaf741},
archivePrefix = {arXiv},
       eprint = {1812.04040},
 primaryClass = {astro-ph.SR},
       adsurl = {https://ui.adsabs.harvard.edu/abs/2018ApJ...869L..41A}
}

@ARTICLE{Booth_2024AJ....167..164B,
       author = {{Booth}, Alice S. and {Leemker}, Margot and {van Dishoeck}, Ewine F. and {Evans}, Lucy and {Ilee}, John D. and {Kama}, Mihkel and {Keyte}, Luke and {Law}, Charles J. and {van der Marel}, Nienke and {Nomura}, Hideko and {Notsu}, Shota and {{\"O}berg}, Karin and {Temmink}, Milou and {Walsh}, Catherine},
        title = "{An ALMA Molecular Inventory of Warm Herbig Ae Disks. I. Molecular Rings, Asymmetries, and Complexity in the HD 100546 Disk}",
      journal = {\aj},
         year = 2024,
        month = apr,
       volume = {167},
       number = {4},
          eid = {164},
        pages = {164},
          doi = {10.3847/1538-3881/ad2700},
archivePrefix = {arXiv},
       eprint = {2402.04001},
 primaryClass = {astro-ph.EP},
       adsurl = {https://ui.adsabs.harvard.edu/abs/2024AJ....167..164B}
}

@ARTICLE{Andrews_2020ARA&A..58..483A,
       author = {{Andrews}, Sean M.},
        title = "{Observations of Protoplanetary Disk Structures}",
      journal = {\araa},
         year = 2020,
        month = aug,
       volume = {58},
        pages = {483-528},
          doi = {10.1146/annurev-astro-031220-010302},
archivePrefix = {arXiv},
       eprint = {2001.05007},
 primaryClass = {astro-ph.EP},
       adsurl = {https://ui.adsabs.harvard.edu/abs/2020ARA&A..58..483A}
}

@ARTICLE{Krijt_2020ApJ...899..134K,
       author = {{Krijt}, Sebastiaan and {Bosman}, Arthur D. and {Zhang}, Ke and {Schwarz}, Kamber R. and {Ciesla}, Fred J. and {Bergin}, Edwin A.},
        title = "{CO Depletion in Protoplanetary Disks: A Unified Picture Combining Physical Sequestration and Chemical Processing}",
      journal = {\apj},
         year = 2020,
        month = aug,
       volume = {899},
       number = {2},
          eid = {134},
        pages = {134},
          doi = {10.3847/1538-4357/aba75d},
archivePrefix = {arXiv},
       eprint = {2007.09517},
 primaryClass = {astro-ph.SR},
       adsurl = {https://ui.adsabs.harvard.edu/abs/2020ApJ...899..134K}
}

@ARTICLE{Agundez_2008A&A...483..831A,
       author = {{Ag{\'u}ndez}, M. and {Cernicharo}, J. and {Goicoechea}, J.~R.},
        title = "{Formation of simple organic molecules in inner T Tauri disks}",
      journal = {\aap},
         year = 2008,
        month = jun,
       volume = {483},
       number = {3},
        pages = {831-837},
          doi = {10.1051/0004-6361:20077927},
archivePrefix = {arXiv},
       eprint = {0803.0938},
 primaryClass = {astro-ph},
       adsurl = {https://ui.adsabs.harvard.edu/abs/2008A&A...483..831A}
}

@ARTICLE{Notsu_2021A&A...650A.180N,
       author = {{Notsu}, Shota and {van Dishoeck}, Ewine F. and {Walsh}, Catherine and {Bosman}, Arthur D. and {Nomura}, Hideko},
        title = "{X-ray-induced chemistry of water and related molecules in low-mass protostellar envelopes}",
      journal = {\aap},
         year = 2021,
        month = jun,
       volume = {650},
          eid = {A180},
        pages = {A180},
          doi = {10.1051/0004-6361/202140667},
archivePrefix = {arXiv},
       eprint = {2104.06878},
 primaryClass = {astro-ph.GA},
       adsurl = {https://ui.adsabs.harvard.edu/abs/2021A&A...650A.180N}
}

@ARTICLE{Weber_2018ApJ...854..153W,
       author = {{Weber}, Philipp and {Ben{\'\i}tez-Llambay}, Pablo and {Gressel}, Oliver and {Krapp}, Leonardo and {Pessah}, Martin E.},
        title = "{Characterizing the Variable Dust Permeability of Planet-induced Gaps}",
      journal = {\apj},
         year = 2018,
        month = feb,
       volume = {854},
       number = {2},
          eid = {153},
        pages = {153},
          doi = {10.3847/1538-4357/aaab63},
archivePrefix = {arXiv},
       eprint = {1801.07971},
 primaryClass = {astro-ph.EP},
       adsurl = {https://ui.adsabs.harvard.edu/abs/2018ApJ...854..153W}
}

@ARTICLE{Salyk_2011ApJ...731..130S,
       author = {{Salyk}, C. and {Pontoppidan}, K.~M. and {Blake}, G.~A. and {Najita}, J.~R. and {Carr}, J.~S.},
        title = "{A Spitzer Survey of Mid-infrared Molecular Emission from Protoplanetary Disks. II. Correlations and Local Thermal Equilibrium Models}",
      journal = {\apj},
         year = 2011,
        month = apr,
       volume = {731},
       number = {2},
          eid = {130},
        pages = {130},
          doi = {10.1088/0004-637X/731/2/130},
archivePrefix = {arXiv},
       eprint = {1104.0948},
 primaryClass = {astro-ph.GA},
       adsurl = {https://ui.adsabs.harvard.edu/abs/2011ApJ...731..130S}
}

@ARTICLE{Horne_1986MNRAS.218..761H,
       author = {{Horne}, K. and {Marsh}, T.~R.},
        title = "{Emission line formation in accretion discs}",
      journal = {\mnras},
         year = 1986,
        month = feb,
       volume = {218},
        pages = {761-773},
          doi = {10.1093/mnras/218.4.761},
       adsurl = {https://ui.adsabs.harvard.edu/abs/1986MNRAS.218..761H}
}

@ARTICLE{Bethell_2009Sci...326.1675B,
       author = {{Bethell}, Thomas and {Bergin}, Edwin},
        title = "{Formation and Survival of Water Vapor in the Terrestrial Planet-Forming Region}",
      journal = {Science},
         year = 2009,
        month = dec,
       volume = {326},
       number = {5960},
        pages = {1675},
          doi = {10.1126/science.1176879},
       adsurl = {https://ui.adsabs.harvard.edu/abs/2009Sci...326.1675B}
}

@ARTICLE{Adamkovics_2014ApJ...786..135A,
       author = {{{\'A}d{\'a}mkovics}, M{\'a}t{\'e} and {Glassgold}, Alfred E. and {Najita}, Joan R.},
        title = "{Shielding by Water and OH in FUV and X-Ray Irradiated Protoplanetary Disks}",
      journal = {\apj},
         year = 2014,
        month = may,
       volume = {786},
       number = {2},
          eid = {135},
        pages = {135},
          doi = {10.1088/0004-637X/786/2/135},
archivePrefix = {arXiv},
       eprint = {1403.8131},
 primaryClass = {astro-ph.EP},
       adsurl = {https://ui.adsabs.harvard.edu/abs/2014ApJ...786..135A}
}

@ARTICLE{Faure_Josselin_2008A&A...492..257F,
       author = {{Faure}, A. and {Josselin}, E.},
        title = "{Collisional excitation of water in warm astrophysical media. I. Rate coefficients for rovibrationally excited states}",
      journal = {\aap},
         year = 2008,
        month = dec,
       volume = {492},
       number = {1},
        pages = {257-264},
          doi = {10.1051/0004-6361:200810717},
       adsurl = {https://ui.adsabs.harvard.edu/abs/2008A&A...492..257F}
}

@ARTICLE{Jonkheid_2006A&A...453..163J,
       author = {{Jonkheid}, B. and {Kamp}, I. and {Augereau}, J.-C. and {van Dishoeck}, E.~F.},
        title = "{Modeling the gas-phase chemistry of the transitional disk around HD 141569A}",
      journal = {\aap},
         year = 2006,
        month = jul,
       volume = {453},
       number = {1},
        pages = {163-171},
          doi = {10.1051/0004-6361:20054769},
archivePrefix = {arXiv},
       eprint = {astro-ph/0603515},
 primaryClass = {astro-ph},
       adsurl = {https://ui.adsabs.harvard.edu/abs/2006A&A...453..163J}
}

@incollection{ehrig_1999,
	address = {Berlin, Heidelberg},
	title = {Advanced {Extrapolation} {Methods} for {Large} {Scale} {Differential} {Algebraic} {Problems}},
	volume = {8},
	isbn = {978-3-540-65730-9 978-3-642-60155-2},
	url = {http://link.springer.com/10.1007/978-3-642-60155-2_20},
	urldate = {2025-12-20},
	booktitle = {High {Performance} {Scientific} and {Engineering} {Computing}},
	publisher = {Springer Berlin Heidelberg},
	author = {Ehrig, R. and Nowak, U. and Oeverdieck, L. and Deuflhard, P.},
	editor = {Griebel, M. and Keyes, D. E. and Nieminen, R. M. and Roose, D. and Schlick, T. and Bungartz, Hans-Joachim and Durst, Franz and Zenger, Christoph},
	year = {1999},
	doi = {10.1007/978-3-642-60155-2_20},
	note = {Series Title: Lecture Notes in Computational Science and Engineering},
	pages = {233--241},
}

@ARTICLE{Mayor_1995Natur.378..355M,
       author = {{Mayor}, Michel and {Queloz}, Didier},
        title = "{A Jupiter-mass companion to a solar-type star}",
      journal = {\nat},
         year = 1995,
        month = nov,
       volume = {378},
       number = {6555},
        pages = {355-359},
          doi = {10.1038/378355a0},
       adsurl = {https://ui.adsabs.harvard.edu/abs/1995Natur.378..355M}
}

@ARTICLE{Bitsch_2015bA&A...582A.112B,
       author = {{Bitsch}, Bertram and {Lambrechts}, Michiel and {Johansen}, Anders},
        title = "{The growth of planets by pebble accretion in evolving protoplanetary discs}",
      journal = {\aap},
         year = 2015,
        month = oct,
       volume = {582},
          eid = {A112},
        pages = {A112},
          doi = {10.1051/0004-6361/201526463},
archivePrefix = {arXiv},
       eprint = {1507.05209},
 primaryClass = {astro-ph.EP},
       adsurl = {https://ui.adsabs.harvard.edu/abs/2015A&A...582A.112B}
}

@ARTICLE{Bitsch_2022A&A...665A.138B,
       author = {{Bitsch}, Bertram and {Schneider}, Aaron David and {Kreidberg}, Laura},
        title = "{How drifting and evaporating pebbles shape giant planets. III. The formation of WASP-77A b and {\ensuremath{\tau}} Bo{\"o}tis b}",
      journal = {\aap},
         year = 2022,
        month = sep,
       volume = {665},
          eid = {A138},
        pages = {A138},
          doi = {10.1051/0004-6361/202243345},
archivePrefix = {arXiv},
       eprint = {2207.06077},
 primaryClass = {astro-ph.EP},
       adsurl = {https://ui.adsabs.harvard.edu/abs/2022A&A...665A.138B}
}

@ARTICLE{Kalyaan_2023ApJ...954...66K,
       author = {{Kalyaan}, Anusha and {Pinilla}, Paola and {Krijt}, Sebastiaan and {Banzatti}, Andrea and {Rosotti}, Giovanni and {Mulders}, Gijs D. and {Lambrechts}, Michiel and {Long}, Feng and {Herczeg}, Gregory J.},
        title = "{The Effect of Dust Evolution and Traps on Inner Disk Water Enrichment}",
      journal = {\apj},
         year = 2023,
        month = sep,
       volume = {954},
       number = {1},
          eid = {66},
        pages = {66},
          doi = {10.3847/1538-4357/ace535},
archivePrefix = {arXiv},
       eprint = {2307.01789},
 primaryClass = {astro-ph.EP},
       adsurl = {https://ui.adsabs.harvard.edu/abs/2023ApJ...954...66K}
}

@ARTICLE{Houge_2025aMNRAS.537..691H,
       author = {{Houge}, Adrien and {Krijt}, Sebastiaan and {Banzatti}, Andrea and {Blake}, Geoffrey A and {Pinilla}, Paola and {Pontoppidan}, Klaus M and {Trapman}, Leon and {Williams}, Joe and {Zhang}, Ke},
        title = "{Smuggling unnoticed: towards a 2D view of water and dust delivery to the inner regions of protoplanetary discs}",
      journal = {\mnras},
         year = 2025,
        month = feb,
       volume = {537},
       number = {2},
        pages = {691-704},
          doi = {10.1093/mnras/staf057},
archivePrefix = {arXiv},
       eprint = {2501.05881},
 primaryClass = {astro-ph.EP},
       adsurl = {https://ui.adsabs.harvard.edu/abs/2025MNRAS.537..691H},
      sortkey = {2023-02}
}

@ARTICLE{Williams_2025MNRAS.544.3562W,
       author = {{Williams}, Joe and {Krijt}, Sebastiaan and {Bitsch}, Bertram and {Houge}, Adrien and {Bergner}, Jennifer},
        title = "{Locked in ice: how pebble drift and volatile entrapment can significantly impact carbon and oxygen ratios in evolving protoplanetary discs}",
      journal = {\mnras},
         year = 2025,
        month = dec,
       volume = {544},
       number = {4},
        pages = {3562-3578},
          doi = {10.1093/mnras/staf1839},
archivePrefix = {arXiv},
       eprint = {2510.18587},
 primaryClass = {astro-ph.EP},
       adsurl = {https://ui.adsabs.harvard.edu/abs/2025MNRAS.544.3562W}
}

@ARTICLE{Houge_2025bA&A...699A.227H,
       author = {{Houge}, Adrien and {Johansen}, Anders and {Bergin}, Edwin and {Ciesla}, Fred J. and {Bitsch}, Bertram and {Lambrechts}, Michiel and {Henning}, Thomas and {Perotti}, Giulia},
        title = "{Burned to ashes: How the thermal decomposition of refractory organics in the inner protoplanetary disc impacts the gas-phase C/O ratio}",
      journal = {\aap},
         year = 2025,
        month = jul,
       volume = {699},
          eid = {A227},
        pages = {A227},
          doi = {10.1051/0004-6361/202555164},
archivePrefix = {arXiv},
       eprint = {2505.20427},
 primaryClass = {astro-ph.EP},
       adsurl = {https://ui.adsabs.harvard.edu/abs/2025A&A...699A.227H},
      sortkey = {2023-07}
}

@ARTICLE{Pinilla_2012A&A...538A.114P,
       author = {{Pinilla}, P. and {Birnstiel}, T. and {Ricci}, L. and {Dullemond}, C.~P. and {Uribe}, A.~L. and {Testi}, L. and {Natta}, A.},
        title = "{Trapping dust particles in the outer regions of protoplanetary disks}",
      journal = {\aap},
         year = 2012,
        month = feb,
       volume = {538},
          eid = {A114},
        pages = {A114},
          doi = {10.1051/0004-6361/201118204},
archivePrefix = {arXiv},
       eprint = {1112.2349},
 primaryClass = {astro-ph.EP},
       adsurl = {https://ui.adsabs.harvard.edu/abs/2012A&A...538A.114P}
}

@ARTICLE{Zhu_2012ApJ...755....6Z,
       author = {{Zhu}, Zhaohuan and {Nelson}, Richard P. and {Dong}, Ruobing and {Espaillat}, Catherine and {Hartmann}, Lee},
        title = "{Dust Filtration by Planet-induced Gap Edges: Implications for Transitional Disks}",
      journal = {\apj},
         year = 2012,
        month = aug,
       volume = {755},
       number = {1},
          eid = {6},
        pages = {6},
          doi = {10.1088/0004-637X/755/1/6},
archivePrefix = {arXiv},
       eprint = {1205.5042},
 primaryClass = {astro-ph.SR},
       adsurl = {https://ui.adsabs.harvard.edu/abs/2012ApJ...755....6Z}
}

@ARTICLE{Drazkowska_2019ApJ...885...91D,
       author = {{Dr{\k{a}}{\.z}kowska}, Joanna and {Li}, Shengtai and {Birnstiel}, Til and {Stammler}, Sebastian M. and {Li}, Hui},
        title = "{Including Dust Coagulation in Hydrodynamic Models of Protoplanetary Disks: Dust Evolution in the Vicinity of a Jupiter-mass Planet}",
      journal = {\apj},
         year = 2019,
        month = nov,
       volume = {885},
       number = {1},
          eid = {91},
        pages = {91},
          doi = {10.3847/1538-4357/ab46b7},
archivePrefix = {arXiv},
       eprint = {1909.10526},
 primaryClass = {astro-ph.EP},
       adsurl = {https://ui.adsabs.harvard.edu/abs/2019ApJ...885...91D}
}

@ARTICLE{Stammler_2023A&A...670L...5S,
       author = {{Stammler}, Sebastian Markus and {Lichtenberg}, Tim and {Dr{\k{a}}{\.z}kowska}, Joanna and {Birnstiel}, Tilman},
        title = "{Leaky dust traps: How fragmentation impacts dust filtering by planets}",
      journal = {\aap},
         year = 2023,
        month = feb,
       volume = {670},
          eid = {L5},
        pages = {L5},
          doi = {10.1051/0004-6361/202245512},
archivePrefix = {arXiv},
       eprint = {2301.05505},
 primaryClass = {astro-ph.EP},
       adsurl = {https://ui.adsabs.harvard.edu/abs/2023A&A...670L...5S}
}

@ARTICLE{Kanwar_2024A&A...689A.231K,
       author = {{Kanwar}, Jayatee and {Kamp}, Inga and {Jang}, Hyerin and {Waters}, Laurens B.~F.~M. and {van Dishoeck}, Ewine F. and {Christiaens}, Valentin and {Arabhavi}, Aditya M. and {Henning}, Thomas and {G{\"u}del}, Manuel and {Woitke}, Peter and {Absil}, Olivier and {Barrado}, David and {Caratti o Garatti}, Alessio and {Glauser}, Adrian M. and {Lahuis}, Fred and {Scheithauer}, Silvia and {Vandenbussche}, Bart and {Gasman}, Danny and {Grant}, Sierra L. and {Kurtovic}, Nicolas T. and {Perotti}, Giulia and {Tabone}, Beno{\^\i}t and {Temmink}, Milou},
        title = "{MINDS. Hydrocarbons detected by JWST/MIRI in the inner disk of Sz28 consistent with a high C/O gas-phase chemistry}",
      journal = {\aap},
         year = 2024,
        month = sep,
       volume = {689},
          eid = {A231},
        pages = {A231},
          doi = {10.1051/0004-6361/202450078},
archivePrefix = {arXiv},
       eprint = {2407.14362},
 primaryClass = {astro-ph.EP},
       adsurl = {https://ui.adsabs.harvard.edu/abs/2024A&A...689A.231K}
}

@ARTICLE{Kanwar_2026A&A...705A.222K,
       author = {{Kanwar}, Jayatee and {Kamp}, Inga and {Woitke}, Peter and {van Dishoeck}, Ewine F. and {Henning}, Thomas and {Liu}, Yao and {Kaeufer}, Till and {Tabone}, Beno{\^\i}t and {G{\"u}del}, Manuel and {Barrado}, David and {Arabhavi}, Aditya M. and {Franceschi}, Riccardo and {Vlasblom}, Marissa},
        title = "{MINDS: Strong oxygen depletion in the inner regions of a very low-mass star disk?}",
      journal = {\aap},
         year = 2026,
        month = jan,
       volume = {705},
          eid = {A222},
        pages = {A222},
          doi = {10.1051/0004-6361/202451844},
archivePrefix = {arXiv},
       eprint = {2508.11761},
 primaryClass = {astro-ph.EP},
       adsurl = {https://ui.adsabs.harvard.edu/abs/2026A&A...705A.222K}
}

@ARTICLE{Gutler_2010A&A...513A..56G,
       author = {{G{\"u}ttler}, C. and {Blum}, J. and {Zsom}, A. and {Ormel}, C.~W. and {Dullemond}, C.~P.},
        title = "{The outcome of protoplanetary dust growth: pebbles, boulders, or planetesimals?. I. Mapping the zoo of laboratory collision experiments}",
      journal = {\aap},
         year = 2010,
        month = apr,
       volume = {513},
          eid = {A56},
        pages = {A56},
          doi = {10.1051/0004-6361/200912852},
archivePrefix = {arXiv},
       eprint = {0910.4251},
 primaryClass = {astro-ph.EP},
       adsurl = {https://ui.adsabs.harvard.edu/abs/2010A&A...513A..56G}
}

@ARTICLE{Blum_2008ARA&A..46...21B,
       author = {{Blum}, J. and {Wurm}, G.},
        title = "{The growth mechanisms of macroscopic bodies in protoplanetary disks.}",
      journal = {\araa},
         year = 2008,
        month = sep,
       volume = {46},
        pages = {21-56},
          doi = {10.1146/annurev.astro.46.060407.145152},
       adsurl = {https://ui.adsabs.harvard.edu/abs/2008ARA&A..46...21B}
}

@ARTICLE{Birnsitel_2024ARA&A..62..157B,
       author = {{Birnstiel}, Tilman},
        title = "{Dust Growth and Evolution in Protoplanetary Disks}",
      journal = {\araa},
         year = 2024,
        month = sep,
       volume = {62},
       number = {1},
        pages = {157-202},
          doi = {10.1146/annurev-astro-071221-052705},
archivePrefix = {arXiv},
       eprint = {2312.13287},
 primaryClass = {astro-ph.EP},
       adsurl = {https://ui.adsabs.harvard.edu/abs/2024ARA&A..62..157B}
}

@ARTICLE{Birnstiel_2012A&A...539A.148B,
       author = {{Birnstiel}, T. and {Klahr}, H. and {Ercolano}, B.},
        title = "{A simple model for the evolution of the dust population in protoplanetary disks}",
      journal = {\aap},
         year = 2012,
        month = mar,
       volume = {539},
          eid = {A148},
        pages = {A148},
          doi = {10.1051/0004-6361/201118136},
archivePrefix = {arXiv},
       eprint = {1201.5781},
 primaryClass = {astro-ph.EP},
       adsurl = {https://ui.adsabs.harvard.edu/abs/2012A&A...539A.148B}
}

@ARTICLE{Bitsch_Mah_2023A&A...679A..11B,
       author = {{Bitsch}, Bertram and {Mah}, Jingyi},
        title = "{Enriching inner discs and giant planets with heavy elements}",
      journal = {\aap},
         year = 2023,
        month = nov,
       volume = {679},
          eid = {A11},
        pages = {A11},
          doi = {10.1051/0004-6361/202347419},
archivePrefix = {arXiv},
       eprint = {2309.00509},
 primaryClass = {astro-ph.EP},
       adsurl = {https://ui.adsabs.harvard.edu/abs/2023A&A...679A..11B}
}

@ARTICLE{Zsom_2010A&A...513A..57Z,
       author = {{Zsom}, A. and {Ormel}, C.~W. and {G{\"u}ttler}, C. and {Blum}, J. and {Dullemond}, C.~P.},
        title = "{The outcome of protoplanetary dust growth: pebbles, boulders, or planetesimals? II. Introducing the bouncing barrier}",
      journal = {\aap},
         year = 2010,
        month = apr,
       volume = {513},
          eid = {A57},
        pages = {A57},
          doi = {10.1051/0004-6361/200912976},
archivePrefix = {arXiv},
       eprint = {1001.0488},
 primaryClass = {astro-ph.EP},
       adsurl = {https://ui.adsabs.harvard.edu/abs/2010A&A...513A..57Z}
}

@ARTICLE{Weidenschilling_1977MNRAS.180...57W,
       author = {{Weidenschilling}, S.~J.},
        title = "{Aerodynamics of solid bodies in the solar nebula.}",
      journal = {\mnras},
         year = 1977,
        month = jul,
       volume = {180},
        pages = {57-70},
          doi = {10.1093/mnras/180.2.57},
       adsurl = {https://ui.adsabs.harvard.edu/abs/1977MNRAS.180...57W}
}

@ARTICLE{Brauer_2008A&A...480..859B,
       author = {{Brauer}, F. and {Dullemond}, C.~P. and {Henning}, Th.},
        title = "{Coagulation, fragmentation and radial motion of solid particles in protoplanetary disks}",
      journal = {\aap},
         year = 2008,
        month = mar,
       volume = {480},
       number = {3},
        pages = {859-877},
          doi = {10.1051/0004-6361:20077759},
archivePrefix = {arXiv},
       eprint = {0711.2192},
 primaryClass = {astro-ph},
       adsurl = {https://ui.adsabs.harvard.edu/abs/2008A&A...480..859B}
}

@ARTICLE{Tazaki_2022A&A...663A..57T,
       author = {{Tazaki}, R. and {Dominik}, C.},
        title = "{The size of monomers of dust aggregates in planet-forming disks. Insights from quantitative optical and near-infrared polarimetry}",
      journal = {\aap},
         year = 2022,
        month = jul,
       volume = {663},
          eid = {A57},
        pages = {A57},
          doi = {10.1051/0004-6361/202243485},
archivePrefix = {arXiv},
       eprint = {2204.08506},
 primaryClass = {astro-ph.EP},
       adsurl = {https://ui.adsabs.harvard.edu/abs/2022A&A...663A..57T}
}

@ARTICLE{Mathis_1977ApJ...217..425M,
       author = {{Mathis}, J.~S. and {Rumpl}, W. and {Nordsieck}, K.~H.},
        title = "{The size distribution of interstellar grains.}",
      journal = {\apj},
         year = 1977,
        month = oct,
       volume = {217},
        pages = {425-433},
          doi = {10.1086/155591},
       adsurl = {https://ui.adsabs.harvard.edu/abs/1977ApJ...217..425M}
}

@ARTICLE{Krijt_2016ApJ...833..285K,
       author = {{Krijt}, Sebastiaan and {Ciesla}, Fred J. and {Bergin}, Edwin A.},
        title = "{Tracing Water Vapor and Ice During Dust Growth}",
      journal = {\apj},
         year = 2016,
        month = dec,
       volume = {833},
       number = {2},
          eid = {285},
        pages = {285},
          doi = {10.3847/1538-4357/833/2/285},
archivePrefix = {arXiv},
       eprint = {1610.06463},
 primaryClass = {astro-ph.EP},
       adsurl = {https://ui.adsabs.harvard.edu/abs/2016ApJ...833..285K}
}

@ARTICLE{Bergez-Casalou_2020A&A...643A.133B,
       author = {{Bergez-Casalou}, C. and {Bitsch}, B. and {Pierens}, A. and {Crida}, A. and {Raymond}, S.~N.},
        title = "{Influence of planetary gas accretion on the shape and depth of gaps in protoplanetary discs}",
      journal = {\aap},
         year = 2020,
        month = nov,
       volume = {643},
          eid = {A133},
        pages = {A133},
          doi = {10.1051/0004-6361/202038304},
archivePrefix = {arXiv},
       eprint = {2010.00485},
 primaryClass = {astro-ph.EP},
       adsurl = {https://ui.adsabs.harvard.edu/abs/2020A&A...643A.133B}
}

@ARTICLE{Lubow_dAngelo_2006ApJ...641..526L,
       author = {{Lubow}, S.~H. and {D'Angelo}, G.},
        title = "{Gas Flow across Gaps in Protoplanetary Disks}",
      journal = {\apj},
         year = 2006,
        month = apr,
       volume = {641},
       number = {1},
        pages = {526-533},
          doi = {10.1086/500356},
archivePrefix = {arXiv},
       eprint = {astro-ph/0512292},
 primaryClass = {astro-ph},
       adsurl = {https://ui.adsabs.harvard.edu/abs/2006ApJ...641..526L}
}

@ARTICLE{Rice_2006MNRAS.373.1619R,
       author = {{Rice}, W.~K.~M. and {Armitage}, Philip J. and {Wood}, Kenneth and {Lodato}, G.},
        title = "{Dust filtration at gap edges: implications for the spectral energy distributions of discs with embedded planets}",
      journal = {\mnras},
         year = 2006,
        month = dec,
       volume = {373},
       number = {4},
        pages = {1619-1626},
          doi = {10.1111/j.1365-2966.2006.11113.x},
archivePrefix = {arXiv},
       eprint = {astro-ph/0609808},
 primaryClass = {astro-ph},
       adsurl = {https://ui.adsabs.harvard.edu/abs/2006MNRAS.373.1619R}
}

@ARTICLE{Vlasblom_2024A&A...682A..91V,
       author = {{Vlasblom}, Marissa and {van Dishoeck}, Ewine F. and {Tabone}, Beno{\^\i}t and {Bruderer}, Simon},
        title = "{Mid-infrared spectra of T Tauri disks: Modeling the effects of a small inner cavity on CO$_{2}$ and H$_{2}$O emission}",
      journal = {\aap},
         year = 2024,
        month = feb,
       volume = {682},
          eid = {A91},
        pages = {A91},
          doi = {10.1051/0004-6361/202348224},
archivePrefix = {arXiv},
       eprint = {2311.12445},
 primaryClass = {astro-ph.EP},
       adsurl = {https://ui.adsabs.harvard.edu/abs/2024A&A...682A..91V}
}

@article{husain_reactions_1971,
	title = {Reactions of atomic carbon {C}({23P}) by kinetic absorption spectroscopy in the vacuum ultra-violet},
	volume = {67},
	url = {http://dx.doi.org/10.1039/TF9716702025},
	doi = {10.1039/TF9716702025},
	number = {0},
	journal = {Trans. Faraday Soc.},
	publisher = {The Royal Society of Chemistry},
	author = {Husain, D. and Kirsch, L. J.},
	year = {1971},
	pages = {2025--2035},
}

@ARTICLE{Li_CH2O_2017PCCP...1923280L,
       author = {{Li}, Jun and {Xie}, Changjian and {Guo}, Hua},
        title = "{Kinetics and dynamics of the C(3P) + H2O reaction on a full-dimensional accurate triplet state potential energy surface}",
      journal = {Physical Chemistry Chemical Physics (Incorporating Faraday Transactions)},
         year = 2017,
        month = aug,
       volume = {19},
       number = {34},
        pages = {23280-23288},
          doi = {10.1039/C7CP04578F},
       adsurl = {https://ui.adsabs.harvard.edu/abs/2017PCCP...1923280L}
}

@ARTICLE{Hickson_2016arXiv160808877H,
       author = {{Hickson}, Kevin M. and {Loison}, Jean-Christophe and {Nunez-Reyes}, Dianailys and {Mereau}, Raphael},
        title = "{Quantum Tunneling Enhancement of the C + H2O and C + D2O Reactions at Low Temperature}",
      journal = {arXiv e-prints},
         year = 2016,
        month = aug,
          eid = {arXiv:1608.08877},
        pages = {arXiv:1608.08877},
          doi = {10.48550/arXiv.1608.08877},
archivePrefix = {arXiv},
       eprint = {1608.08877},
 primaryClass = {physics.chem-ph},
       adsurl = {https://ui.adsabs.harvard.edu/abs/2016arXiv160808877H}
}

@ARTICLE{Schindhelm_2012ApJ...756L..23S,
       author = {{Schindhelm}, Rebecca and {France}, Kevin and {Herczeg}, Gregory J. and {Bergin}, Edwin and {Yang}, Hao and {Brown}, Alexander and {Brown}, Joanna M. and {Linsky}, Jeffrey L. and {Valenti}, Jeff},
        title = "{Ly{\ensuremath{\alpha}} Dominance of the Classical T Tauri Far-ultraviolet Radiation Field}",
      journal = {\apjl},
         year = 2012,
        month = sep,
       volume = {756},
       number = {1},
          eid = {L23},
        pages = {L23},
          doi = {10.1088/2041-8205/756/1/L23},
archivePrefix = {arXiv},
       eprint = {1208.2271},
 primaryClass = {astro-ph.SR},
       adsurl = {https://ui.adsabs.harvard.edu/abs/2012ApJ...756L..23S}
}

@ARTICLE{France_2014ApJ...784..127F,
       author = {{France}, Kevin and {Schindhelm}, Rebecca and {Bergin}, Edwin A. and {Roueff}, Evelyne and {Abgrall}, Herv{\'e}},
        title = "{High-resolution Ultraviolet Radiation Fields of Classical T Tauri Stars}",
      journal = {\apj},
         year = 2014,
        month = apr,
       volume = {784},
       number = {2},
          eid = {127},
        pages = {127},
          doi = {10.1088/0004-637X/784/2/127},
archivePrefix = {arXiv},
       eprint = {1402.6341},
 primaryClass = {astro-ph.SR},
       adsurl = {https://ui.adsabs.harvard.edu/abs/2014ApJ...784..127F}
}

@ARTICLE{Woitke_2022A&A...668A.164W,
       author = {{Woitke}, P. and {Arabhavi}, A.~M. and {Kamp}, I. and {Thi}, W.-F.},
        title = "{Mixing and diffusion in protoplanetary disc chemistry}",
      journal = {\aap},
         year = 2022,
        month = dec,
       volume = {668},
          eid = {A164},
        pages = {A164},
          doi = {10.1051/0004-6361/202244554},
archivePrefix = {arXiv},
       eprint = {2209.12233},
 primaryClass = {astro-ph.EP},
       adsurl = {https://ui.adsabs.harvard.edu/abs/2022A&A...668A.164W}
}

@ARTICLE{Arabhavi_2026A&A...708A..82A,
       author = {{Arabhavi}, Aditya M. and {Kamp}, Inga and {van Dishoeck}, Ewine F. and {Woitke}, Peter and {Rab}, Christian and {Thi}, Wing-Fai and {Kaeufer}, Till and {Kanwar}, Jayatee and {Tabone}, Beno{\^\i}t and {Esteve}, Pac{\^o}me and {Vlasblom}, Marissa},
        title = "{Molecular diagnostics for the mid-infrared emission of planet-forming disks: Carbon and oxygen elemental abundances}",
      journal = {\aap},
         year = 2026,
        month = mar,
       volume = {708},
          eid = {A82},
        pages = {A82},
          doi = {10.1051/0004-6361/202556335},
       adsurl = {https://ui.adsabs.harvard.edu/abs/2026A&A...708A..82A}
}

@ARTICLE{Kaufer_2023A&A...672A..30K,
       author = {{Kaeufer}, T. and {Woitke}, P. and {Min}, M. and {Kamp}, I. and {Pinte}, C.},
        title = "{Analysing the SEDs of protoplanetary disks with machine learning}",
      journal = {\aap},
         year = 2023,
        month = apr,
       volume = {672},
          eid = {A30},
        pages = {A30},
          doi = {10.1051/0004-6361/202245461},
archivePrefix = {arXiv},
       eprint = {2302.04629},
 primaryClass = {astro-ph.EP},
       adsurl = {https://ui.adsabs.harvard.edu/abs/2023A&A...672A..30K}
}

@ARTICLE{Ribas_2020A&A...642A.171R,
       author = {{Ribas}, {\'A}. and {Espaillat}, C.~C. and {Mac{\'\i}as}, E. and {Sarro}, L.~M.},
        title = "{Modeling protoplanetary disk SEDs with artificial neural networks. Revisiting the viscous disk model and updated disk masses}",
      journal = {\aap},
         year = 2020,
        month = oct,
       volume = {642},
          eid = {A171},
        pages = {A171},
          doi = {10.1051/0004-6361/202038352},
archivePrefix = {arXiv},
       eprint = {2009.03323},
 primaryClass = {astro-ph.SR},
       adsurl = {https://ui.adsabs.harvard.edu/abs/2020A&A...642A.171R}
}

@ARTICLE{Silva_2008PNAS..10512713S,
       author = {{Silva}, R. and {Gichuhi}, W.~K. and {Huang}, C. and {Doyle}, M.~B. and {Kislov}, V.~V. and {Mebel}, A.~M. and {Suits}, A.~G.},
        title = "{Chemical Dynamics Special Feature: H elimination and metastable lifetimes in the UV photoexcitation of diacetylene}",
      journal = {Proceedings of the National Academy of Science},
         year = 2008,
        month = sep,
       volume = {105},
       number = {35},
        pages = {12713-12718},
          doi = {10.1073/pnas.0801180105},
       adsurl = {https://ui.adsabs.harvard.edu/abs/2008PNAS..10512713S}
}

@ARTICLE{Weingartner&Draine2001ApJ...548..296W,
       author = {{Weingartner}, Joseph C. and {Draine}, B.~T.},
        title = "{Dust Grain-Size Distributions and Extinction in the Milky Way, Large Magellanic Cloud, and Small Magellanic Cloud}",
      journal = {\apj},
         year = 2001,
        month = feb,
       volume = {548},
       number = {1},
        pages = {296-309},
          doi = {10.1086/318651},
archivePrefix = {arXiv},
       eprint = {astro-ph/0008146},
 primaryClass = {astro-ph},
       adsurl = {https://ui.adsabs.harvard.edu/abs/2001ApJ...548..296W}
}

\begin{appendix}
    \section{UV shielding}
    \label{appendix:UV shielding}
\subsection{New implementation}
\begin{figure}[ht]
\centering
\includegraphics[scale=0.22, trim={0 0 0 0}, clip]{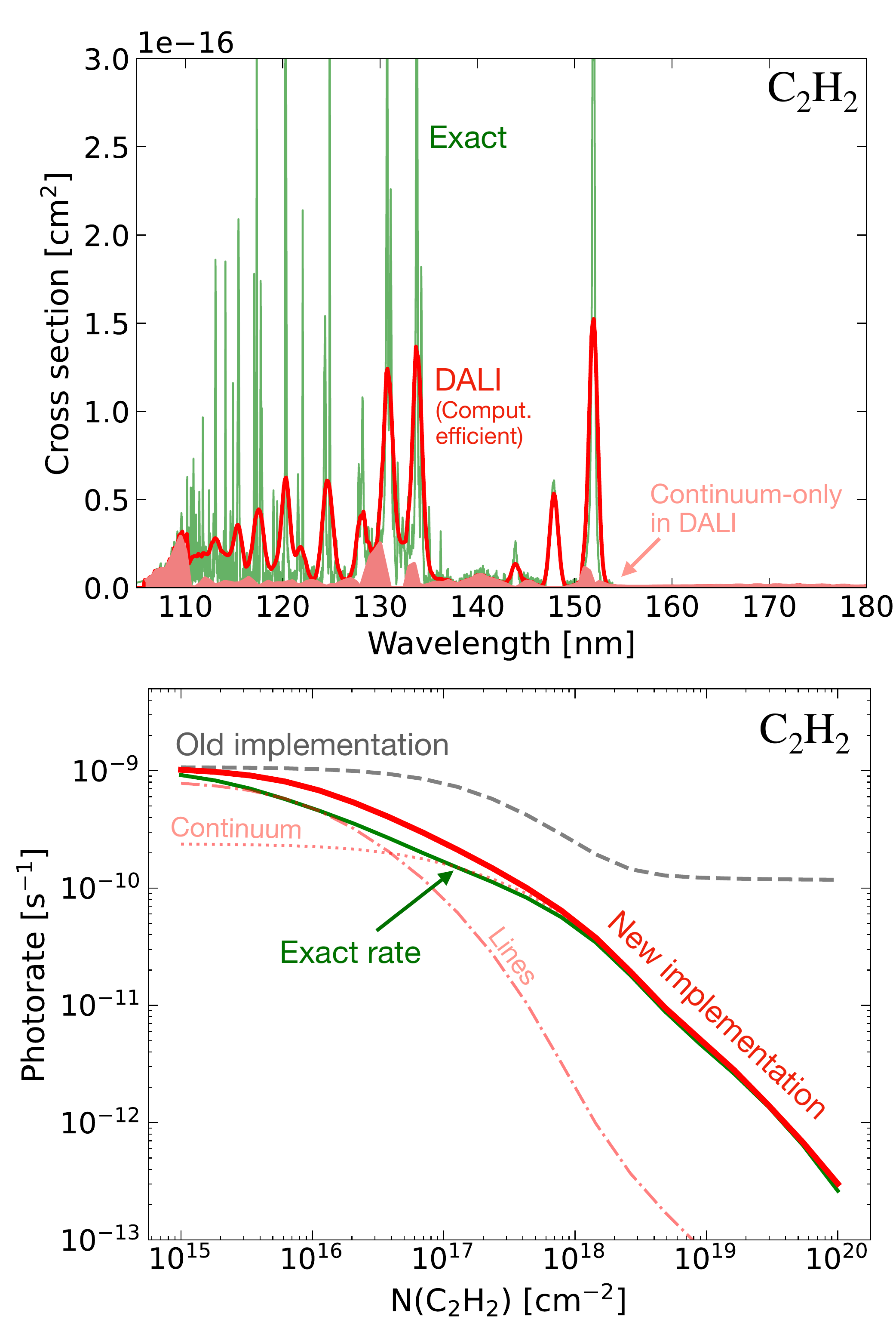}
\caption{\label{fig:C2H2test}Top: Difference between the exact cross section of C$_2$H$_2$ (in green) and used in DALI (red, corresponding to the computational efficiency format in the Leiden Database). The light red curve shows only the continuum part of the cross section in the computational efficiency format. Bottom: Improvement of the UV shielding with the new implementation. The green line shows the exact photodissociation rate (computed with the exact cross section). The continuum (dashed red line) and lines (dash-dotted line) correspond, respectively, to the contribution of the dissociation in the continuum (i.e light red curve in the top panel) and in the lines.  }
\end{figure}
The results from the new implementation of the UV shielding are shown in Fig.  \ref{fig:C2H2test}. The exact cross section of C$_2$H$_2$ is shown in green in the top panel \citep{heays_photodissociation_2017}. Since it is highly wavelength dependent with many narrow lines, DALI uses the computational efficiency format in the Leiden Database (red in the top panel of Fig. \ref{fig:C2H2test}). This format splits the cross section in two parts: lines (if  FWHM < 1 nm) and continuum. In the top panel of Fig. \ref{fig:C2H2test}, the continuum cross section is highlighted by the light red curve, while the rest of the DALI cross section corresponds to the contribution of the lines. This approach significantly reduces the number of wavelength points (< 500 for the continuum) which avoids very fine sampling of the radiative field. 
The bottom panel of Fig. \ref{fig:C2H2test} shows how well this approximation agrees with the exact photodissociation rate. As the column density of C$_2$H$_2$ increases, the photodissociation rate drops because the molecule protects itself. The previous implementation, which considered only the continuum cross section (light red in the top panel), is shown as a dashed gray line. The new implementation is shown as a solid red line, for which both the “continuum” and the “lines” cross sections (see the top panel) are taken into account:
\begin{equation}
    k = \int \sigma^{cont}(\lambda)I(\lambda)d\lambda\ + \sum_{lines} \sigma^{int}I(\lambda_{0})\times \alpha(\sigma^{int}, N)
.\end{equation}
The first term corresponds to the dissociation in the continuum, with $\sigma^{cont}$ the continuum cross section (in light red the top panel, and illustrated with the dotted line “continuum” in the bottom panel). The second term corresponds to the dissociation in the lines, with the integrated cross section $\sigma^{int}$ (illustrated in the bottom panel with the dash-dotted line “lines”). The attenuation factor in the lines, $\alpha(\sigma^{int}, N)$, can be expressed as
\begin{equation}
    \alpha = \int \frac{1}{\sqrt{\pi}}e^{-y^2}e^{-\frac{\kappa}{\sqrt{\pi}} e^{-y^2}}dy
,\end{equation}
where $\kappa = \sigma^{int} N ~/~\Delta\lambda\sqrt{2}$. The FWHM of the Gaussian is set to 1 nm ($\Delta\lambda = \rm{FWHM}~/~2.355$) and $N$ is the column density of the absorbing species.

The difference observed between the old and the new implementation at low column densities ($N_{C_2H_2} < 10^{17}$ cm$^{-2}$) is due to the photodissociation in the lines: as mentioned in Sect. \ref{sec: UV_shielding, method}, self-shielding appears first in the lines since cross sections peak $\sim$10-100 times above the continuum. This effect is highlighted by the dash-dotted light red line that drops first, very quickly. This also reflects the importance of photodissociation in lines, by a drop of $\sim$1 order of magnitude at column densities around $N=5\times10^{17}$ cm$^{-2}$. At higher column densities, the continuum begins to attenuate the UV field (dotted red line). The old implementation reaches a plateau and strongly overestimates the photodissociation rate compared to the exact rate (green). Indeed, C$_2$H$_2$ could still be photodissociated in the lines since their shielding was not taken into account. The new implementation gives a much better estimation at high column densities, almost perfectly matching the exact rate. In the case of C$_2$H$_2$, the slight overestimation of the new implementation at $N\sim 10^{16}$ cm$^{-2}$ reflects the limit of approximating narrow lines (< 1 nm) to much wider lines (FWHM = 1 nm), the latter peaking at much smaller values to conserve the total energy absorbed. 

\subsection{Mutual shielding}
\label{Appendix: mutual shielding}
\begin{figure*}[ht]
\centering
\resizebox{\hsize}{!}{ 
\includegraphics[scale=0.30, trim={0 0 0 0}, clip]{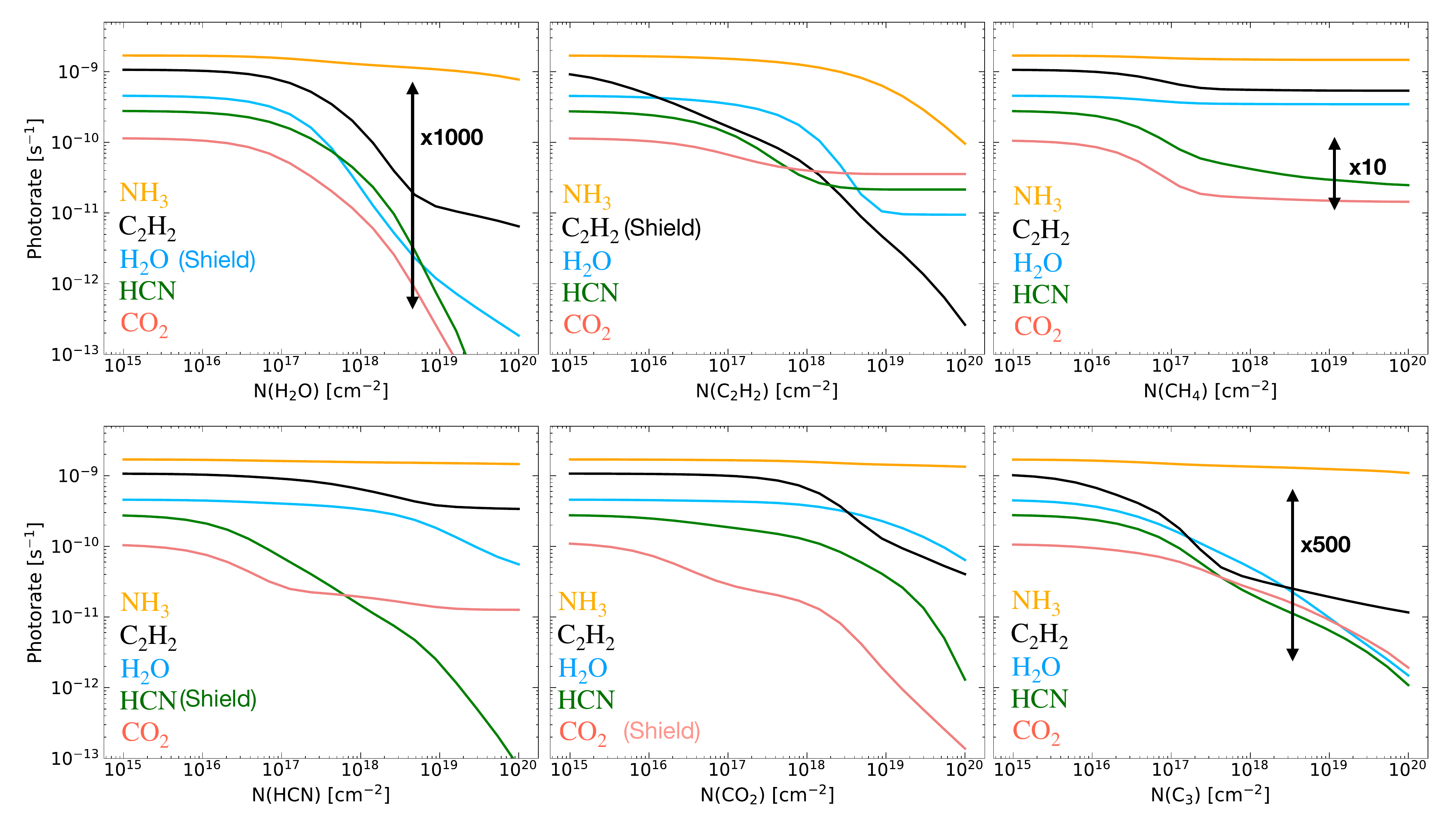}}
\caption{UV shielding efficiency of six abundant molecules: H$_2$O, C$_2$H$_2$, CH$_4$ (top) and HCN, CO$_2$, C$_3$ (bottom). H$_2$O and C$_3$ are powerful molecules to shield the gas{, by reducing the photodissociation rate of most of the molecules by two to three orders of magnitude. On the contrary, CH$_4$ and HCN are not relevant for the mutual shielding. NH$_3$ is relatively unaffected by molecular shielding, as its cross section extends up to 225 nm.} }
\label{fig: mutualshield}
\end{figure*}
\noindent Figure \ref{fig: mutualshield} shows the ability of molecules to protect themselves (self-shielding) and protect other molecules (mutual shielding). The top panel shows the evolution of photodissociation rates of molecules with H$_2$O, C$_2$H$_2$, and CH$_4$ shielding the gas. For the bottom panel, HCN, CO$_2$, and C$_3$ are shielding. The attenuation seen in these figures is only due to gas absorption, dust being excluded. The figure highlights the special role played by water, being the most efficient molecule in protecting other species. We can note that C$_3$ is also particularly efficient, having a broad cross section as well. However, the specific width of the cross section of C$_3$ was assumed from the work of \citet[see Sect. 4.3.24. in \citealt{heays_photodissociation_2017}]{van_hemert_photodissociation_2008}, but it has never been measured experimentally. Its ability to mutually shield thus relies on this assumption. C$_2$H$_2$ can protect water only at high column densities. Regarding HCN, CO$_2$ and CH$_4$, they are not relevant in the mutual shielding. NH$_3$ is relatively unaffected, even with water shielding due to a cross section extending up to 225 nm \citep{heays_photodissociation_2017}. 

\subsection{Species with a $c-\rm C_3H_2$ cross section}
\label{appendix: water cross section}
\begin{table}[]
    \centering
    \caption{Species included in the network with the c-C$_3$H$_2$ UV cross section.}
    \begin{tabular}{c|c}
    \hline
    \hline
    \textbf{nb C} & \textbf{Species}\\
    \hline
    1 & CH$_3^+$ \\
    2 & C$_2^+$, C$_2$N \\
    3 & CH$_2$CCH$_2$, CH$_3$CCH, CH$_3$CHCH$_2$, C$_3$H$_5$\\
    4 & CH$_2$CHCCH, C$_4$H$_3$, CH$_2$CHCHCH$_2$, C$_4$H$_2$\tablefootmark{a}\\
    5 & C$_3$CCH, C$_5$, C$_5$H$_2$, C$_5$H$_5$, C$_5$H$_6$, CH$_3$C$_4$H, \\
     & H$_2$CCCHCCH, C$_3$HCCH\\
\hline
\end{tabular}
    \tablefoot{\tablefoottext{a}{C$_4$H$_2$ has the C$_4$H cross section.}}
    
    \label{tab: water UV cross}
\end{table}
\noindent Several species added to the network do not have an available UV cross section in the Leiden Database. For these species, we arbitrarily choose the {c-C$_3$H$_2$} UV cross section. The list of species for which {the c-C$_3$H$_2$} cross section is used to compute the photodissociation rate is in Table \ref{tab: water UV cross}.{The c-C$_3$H$_2$ cross section has a line at 250 nm, which means that these hydrocarbons can still be photodissociated in a region shielded by water. }

{For C$_4$H$_2$, we choose the C$_4$H cross section as it should be more appropriate. Indeed, the dissociation of C$_4$H$_2$ leads to C$_2$H and C$_4$H, which are very similar to the dissociation fragments of C$_4$H (C$_2$ and C$_4$). In addition, \cite{Silva_2008PNAS..10512713S} found that the dissociation rate of C$_4$H$_2$ is negligible at $\lambda > 180$ nm, as is that of C$_4$H.}

\subsection{Effect of UV shielding on C$_2$H$_2$}
\label{appendix: effect water uv shielding on c2H2}
\begin{figure}[ht]
\centering
\resizebox{\hsize}{!}{
\includegraphics[scale=0.50, trim={12cm 0 300 0cm}, clip]{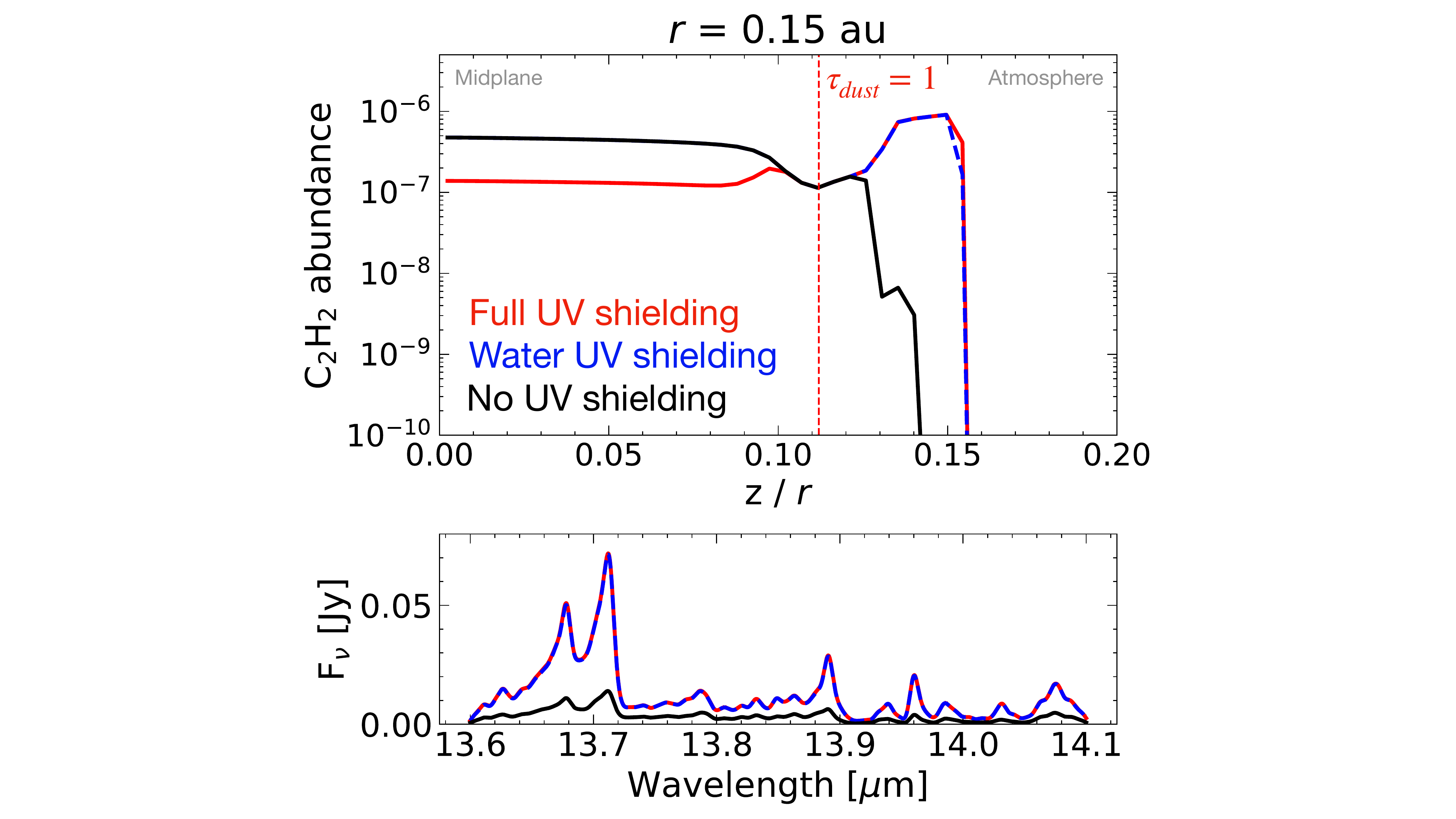}}
\caption{Top: Vertical cut at $r=0.15$ au showing the abundance of acetylene when there is no UV shielding by the gas (black), water UV shielding (blue), or the 12 species mentioned in Sect. \ref{sec: UV_shielding, method} (red). Bottom: C$_2$H$_2$ spectrum for these different UV shielding prescriptions.}
\label{fig: UVshielding_C2H2}
\end{figure}
\noindent Figure \ref{fig: UVshielding_C2H2} shows the difference between no UV shielding by the gas (in black) and water UV shielding (in dashed blue). Acetylene is 4 orders of magnitude more abundant where it emits, increasing the emission of the $Q$ branch at 13.7 $\mu$m by a factor of 5. However, the red line shows that when we include the shielding from other abundant species (S, Fe, H$_2$O, OH, CO$_2$, HCN, CN, C$_2$H$_2$, C$_3$, C$_2$H$_4$, CH$_4$, and C$_2$H$_6$), the difference is not significant for a solar C/O. It shows that it is indeed the water UV shielding that determines the position of the molecular layer in the inner disk. 

\section{Chemistry}
\subsection{Initial abundances}
\label{appendix:init_abundances}
\noindent Table \ref{tab:init_ab} indicates the initial elemental abundances for the fiducial model. For the model grid, we vary the C/O ratio by keeping O/H constant and changing C/H. {For enhanced O/H, the O/H ratio is increased by a factor of 10, and we vary the C/H ratio to match the desired C/O ratio. We follow the same procedure for the depleted grid, by reducing the O/H ratio by a factor of 10.}
\begin{table}[]
    \centering
    \caption{Initial abundances for the fiducial model. }
    \begin{tabular}{c|c}
    \hline
    \hline
    \textbf{Element}    &  \textbf{Number fraction}\\
    \hline
    H    & 1.00 \\
    He    & 7.59e-2 \\
    C    & 1.35e-4 \\
    N    & 2.14e-5 \\
    O    & 2.88e-4 \\
    Mg    & 4.17e-7 \\
    Si    & 7.94e-6 \\
    S    & 1.91e-6 \\
    Fe    & 4.27e-7\\
    \hline
    \end{tabular}
    \tablefoot{Adopted from \cite{bruderer_warm_2012}.}
    \label{tab:init_ab}
\end{table}
\subsection{New species}
\label{appendix:new_species}
\noindent Table \ref{tab:new species} lists all the species added to the chemical network. It includes all hydrocarbons (C$_x$H$_y$) available in UMIST with fewer than 6 atoms of carbon. 
\begin{table*}[]
    \centering
    \caption{New species added to the chemical network.}
    \begin{tabular}{c|c|c|c}
    \hline
    \hline
    \textbf{2C} & \textbf{3C} & \textbf{4C} & \textbf{5C}\\
    \hline
    & C$_3$, C$_3^+$ & C$_4$, C$_4^+$ & C$_5$, C$_5^+$ \\
    & C$_3$H, C$_3$H$^+$ & C$_4$H, C$_4$H$^+$ & C$_3$CCH, C$_5$H, C$_3$CCH$^+$, C$_5$H$^+$ \\
    & C$_3$H$_2$, H$_2$CCC, C$_3$H$_2^+$ & C$_4$H$_2$, C$_4$H$_2^+$ & C$_5$H$_2$, C$_3$HCCH, C$_5$H$_2^+$, C$_3$HCCH$^+$, C$_3$CCH$_2^+$ \\
    & CH$_2$CCH, H$_2$CCCH$^+$, C$_3$H$_3^+$ & C$_4$H$_3$, C$_4$H$_3^+$ & C$_3$HCCH$_2^+$, C$_5$H$_3^+$ \\
C$_2$H$_4$, C$_2$H$_4^+$ & CH$_2$CCH$_2$, CH$_3$CCH, C$_3$H$_4^+$ & CH$_2$CHCCH, C$_4$H$_4^+$ & CH$_3$C$_4$H, H$_2$CCCHCCH, CH$_3$C$_4$H$^+$, C$_5$H$_4^+$ \\
C$_2$H$_5$, C$_2$H$_5^+$ & C$_3$H$_5$, CH$_2$CCH$_3^+$, C$_3$H$_5^+$ & C$_4$H$_5^+$, CH$_2$CHCCH$_2^+$ & C$_5$H$_5$, C$_5$H$_5^+$ \\
CH$_3$CH$_3$, CH$_3$CH$_3^+$ & CH$_3$CHCH$_2$, C$_3$H$_6^+$ & CH$_2$CHCHCH$_2$ & C$_5$H$_6$, C$_5$H$_6^+$ \\
C$_2$H$_7^+$ & C$_3$H$_7^+$ & C$_4$H$_7^+$ & C$_5$H$_7^+$ \\
\hline
\end{tabular}

    \label{tab:new species}
\end{table*}
   
\subsection{Updated endothermicity for KIDA reactions}
\label{appendix:endo_tinacci}
\noindent Table \ref{tab: Tinacci corrected} indicates the list of reactions included in our chemical network with the corrected endothermicity from \cite{tinacci_gretobape_2023}. Table \ref{tab: Tinacci removed} lists all the reactions that have not been included in the chemical network because the reaction enthalpy is above the threshold of $\Delta H_r > 12,000$ K after the correction of \cite{tinacci_gretobape_2023}. These two tables show that carbon chains with three carbons and five carbons are much less reactive than expected. In particular, the correction makes C$_3$ much more abundant by suppressing its reactivity with H$_2$.
\begin{table*}[ht]
    \centering
    \caption{Reactions from KIDA corrected with the endothermicity found by \cite{tinacci_gretobape_2023}. }
    \begin{tabular}{c|c|c|c|c}
    \hline
    \hline
    \textbf{Reaction} & $\boldsymbol{\alpha}$ \textbf{[cm$\boldsymbol{^3}$.s$\boldsymbol{^{-1}}$]} & $\boldsymbol{\beta}$ & $\boldsymbol{\gamma_{KIDA}}$ \textbf{[K]} & $\boldsymbol{\gamma_{corrected}}$ \textbf{[K]}\\
    \hline
    $\rm{C_2H_3+CH_3^+ \xrightarrow{} H_2CCH^+ + H_2 + H}$ & $9.50\times10^{-11}$ & -0.5 & 0.0 & 1878 \\
    $\rm{CH_4+H_2CCC \xrightarrow{} CH_3 +CH_2CCH}$ & $5.90\times10^{-11}$ & 0.0 & 0.0 & 2210 \\
    $\rm{H_2+C_3H_2 \xrightarrow{} H +CH_2CCH}$ & $3.74\times10^{-12}$ & 2.0 & 1740 & 8766 \\
    $\rm{CH_2+C_4H_2 \xrightarrow{} CH_3 +C_4H}$ & $2.16\times10^{-11}$ & 0.0 & 2160 & 11400 \\
    $\rm{H_2+C_4H_2^+ \xrightarrow{} H +C_4H_3^+}$ & $1.00\times10^{-9}$ & 0.0 & 2000 & 5339 \\
    $\rm{H_2+C_5H \xrightarrow{} H +C_5H_2}$ & $1.14\times10^{-11}$ & 0.0 & 950 & 9578 \\
    $\rm{H_2+C_5 \xrightarrow{} H +C_5H}$ & $1.60\times10^{-10}$ & 0.0 & 1420 & 9780 \\
    
\hline
\end{tabular}
    \label{tab: Tinacci corrected}
\end{table*}

\begin{table*}[ht]
\centering
\caption{Reactions corrected with the reaction enthalpy found by \cite{tinacci_gretobape_2023}, but not included in the chemical network. }
\begin{tabular}{c|c|c|c|c}
\hline
\hline
\textbf{Reaction} & $\boldsymbol{\alpha}$ \textbf{[cm$\boldsymbol{^3}$.s$\boldsymbol{^{-1}}$]} & $\boldsymbol{\beta}$ & $\boldsymbol{\gamma_{KIDA}}$ \textbf{[K]} & $\boldsymbol{\gamma_{corrected}}$ \textbf{[K]} \\
\hline
\hline
$\rm{H_2+C_5H^+\xrightarrow{}H+C_5H_2^+}$ & 1e-17 & 0.0 & 0 & 20421\\
$\rm{H_2+C_2^+\xrightarrow{}C_2H+H^+}$ & 1.5e-09 & 0.0 & 1260 & 18358\\
$\rm{C_4H_2+C_3H^+\xrightarrow{}C_2H+C_5H_2^+}$ & 1e-09 & 0.0 & 0 & 24113\\
$\rm{CH_3+CH_3CCH\xrightarrow{}C_2H+CH_3CH_3}$ & 8.32e-13 & 0.0 & 4430 & 18587\\
$\rm{CH_3+C_4H_3\xrightarrow{}C_2H+C_3H_5}$ & 1.2e-10 & 0.0 & 0 & 12037\\
$\rm{C_2H_2+C_2H_5\xrightarrow{}C_2H+CH_3CH_3}$ & 4.5e-13 & 0.0 & 11800 & 16172\\
$\rm{H+C_5H_3^+\xrightarrow{}H_2+C_5H_2^+}$ & 7e-10 & 0.0 & 700 & 13558\\
$\rm{H_2+C_3\xrightarrow{}H+C_3H}$ & 8e-12 & 0.0 & 1420 & 14287\\
$\rm{H+C_4H_2\xrightarrow{}C_2H+C_2H_2}$ & 9.96e-10 & 0.0 & 7750 & 14194\\
$\rm{H+C_5H_2\xrightarrow{}C_2H+H_2CCC}$ & 4.98e-10 & 0.0 & 7750 & 13973\\
$\rm{H_2+C_4H_2\xrightarrow{}H+C_4H_3}$ & 4.91e-10 & -0.16 & 27900 & 29610\\
\hline
\end{tabular}
\tablefoot{The activation barrier ($\gamma_{corrected}$) is above the threshold of 12000 K.}
\label{tab: Tinacci removed}
\end{table*}

\subsection{Focus on C$_n$+H$_2$}
\label{focus_C2_H2}
\begin{figure}[ht]
\centering
\resizebox{\hsize}{!}{
\includegraphics[scale=0.50, trim={0cm 0 0 0cm}, clip]{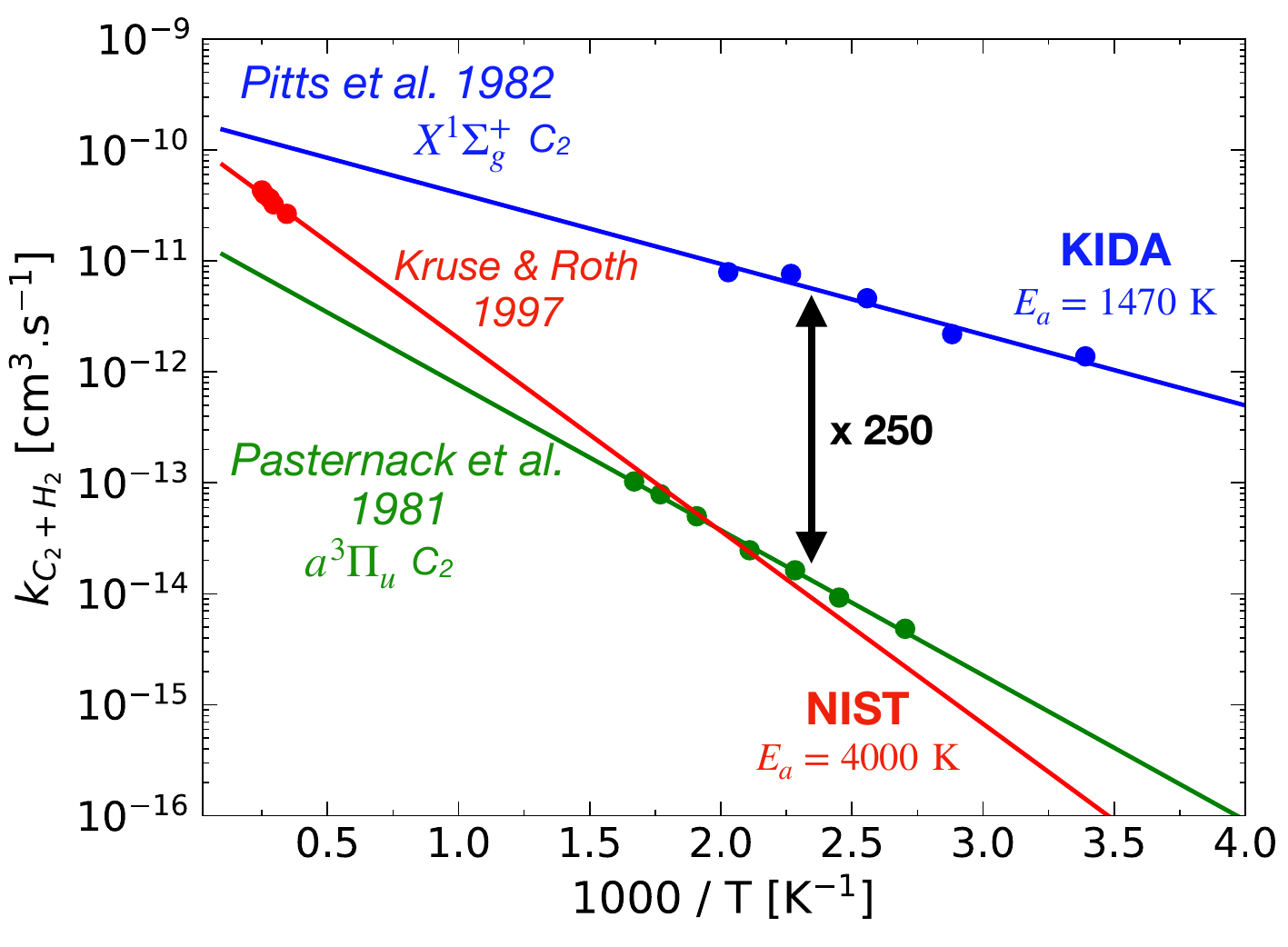}}
\caption{Inconsistency of the rate coefficient of $\rm{H_2+C_2 \xrightarrow{} C_2H +H}$ between KIDA (blue) and NIST (red) database, coming from a different initial state of C$_2$: the rate coefficient of NIST \citep{kruse_kinetics_1997} corresponds to the triplet state $a^3\Pi_u$ according to \cite{pasternack_temperature_1981}.}
\label{fig: C2_H2_react}
\end{figure}
\noindent The reaction $\rm{H_2+C_2 \xrightarrow{} C_2H +H}$ is crucial for the final abundance of C$_2$H$_2$ in the inner disk, since the reaction $\rm{H_2+C_2H \xrightarrow{} C_2H_2 +H}$ is very fast. However, NIST and KIDA databases give an inconsistent rate coefficient for this reaction. It does not exist in UMIST (RATE22 \citealt{millar_umist_2023}), possibly explaining why this reaction has been neglected in studies based on UMIST \citep{walsh_molecular_2015}. KIDA reports an activation barrier of 1420 K \citep{harada_new_2010}, while NIST places this activation energy at 4000 K. In the literature, two values of the activation barrier have been proposed. \cite{pitts_temperature_1982} found an activation energy of 1470 K, based on laser experiments at room temperature, while \cite{kruse_kinetics_1997} used shock experiments to find that this activation energy should be at 4000 K, for a temperature between 2500 and 4500 K. By extrapolating this latter rate coefficient, we found that it is consistent with the results of \cite{pasternack_temperature_1981}, but inconsistent with the values from \cite{pitts_temperature_1982} tabulated in KIDA. In fact, this disagreement comes from different initial states of C$_2$ when it reacts with H$_2$ \citep{vandishoeck_black_1982ApJ...258..533V}. The measurements done by \cite{pitts_temperature_1982} are related to the singlet $X^1\Sigma_g$ state of C$_2$ (ground state) from which the reaction proceeds very fast. In contrast, \cite{pasternack_temperature_1981} measured it for the metastable triplet state $a^3\Pi_u$ of C$_2$ and found an activation energy of $E_a$ = 3000 K, which is consistent with the
value adopted in NIST and found by \citet[see our Fig. \ref{fig: C2_H2_react}]{kruse_kinetics_1997}. Under inner disk conditions, we see no reason why C$_2$ would only be in his metastable state, so we decide to consider the rate coefficient with the ground state ($E_a=1500$ K) in the chemical network. To verify the activation energy found experimentally by \cite{pitts_temperature_1982}, M. van Hemert (2025, private communication) performed quantum calculations (multi-reference configuration interaction) and estimated an activation barrier in good agreement with the value of \cite{pitts_temperature_1982}. Therefore, we add in our network the rate coefficient found by \cite{pitts_temperature_1982}, which was used in \cite{harada_new_2010} and available in the KIDA database.

Extending these results to longer pure carbon chains (C$_n$, $n>2$), KIDA, following \cite{harada_new_2010}, uses the same rate coefficient as for C$_2$. Similarly to C$_2$, the reaction with C$_4$ is exothermic, so this assumption seems reasonable. However, only considering the reaction enthalpy with C$_3$ and C$_5$, \cite{tinacci_gretobape_2023} showed that they are much less reactive with H$_2$ (14287 K and 9780 K). To go further, this might suggest that pure carbon chains with an even number of carbon are much more reactive than those with an odd number of carbon. 

\subsection{The reaction C+H$_2$O}
\label{Appendix : C+H2O}
\begin{figure}[ht]
\centering
\resizebox{\hsize}{!}{
\includegraphics[scale=0.50, trim={0cm 0 0 0cm}, clip]{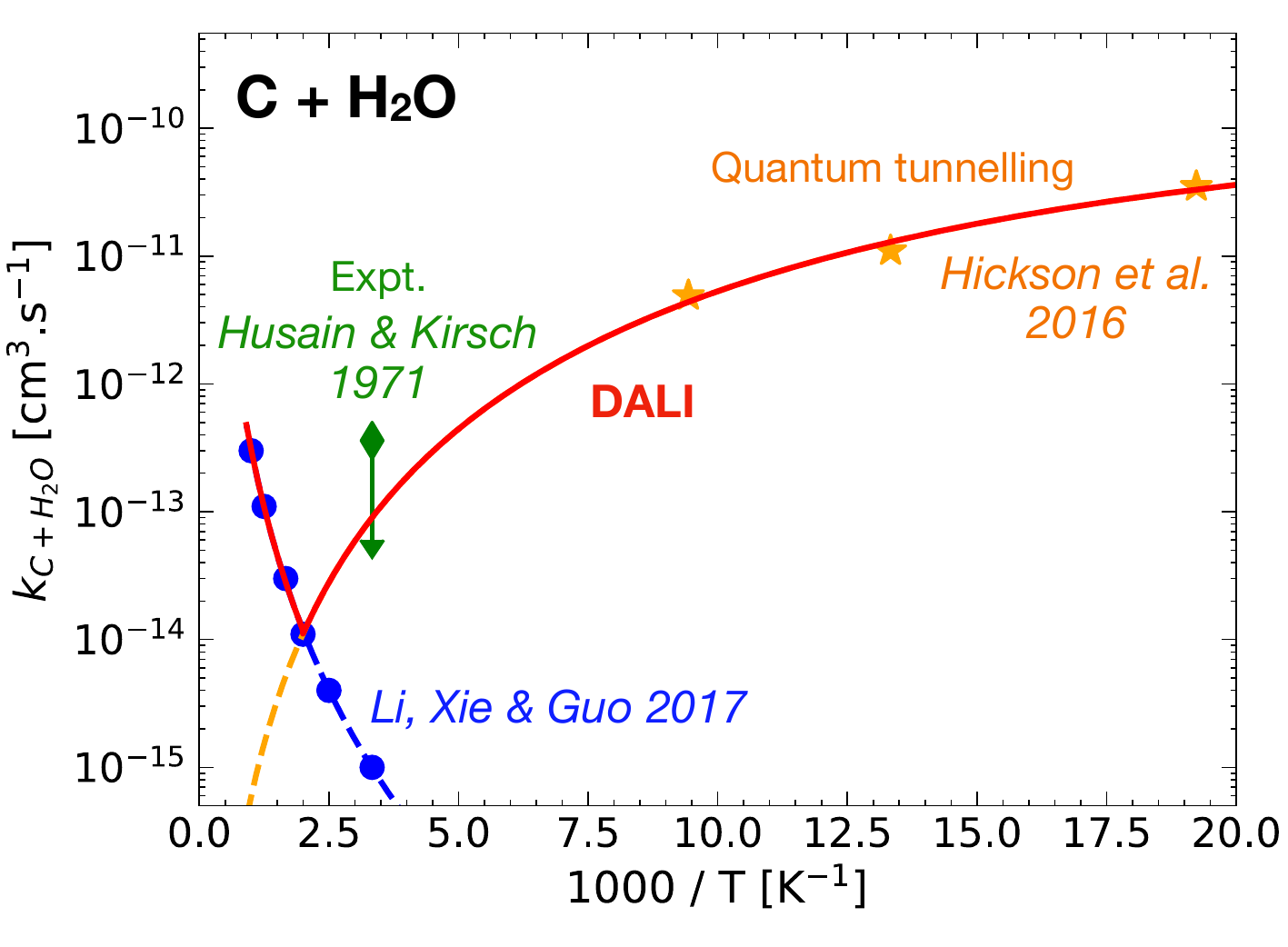}}
\caption{Rate coefficient of the reaction $\rm C+H_2O \xrightarrow{} HCO + H$ adopted in this work (red). This rate is in agreement with the upper limit of \cite{husain_reactions_1971}, the ab initio calculations of \cite{Li_CH2O_2017PCCP...1923280L} and the measurements of \cite{Hickson_2016arXiv160808877H}.}
\label{fig: C_H2O_react}
\end{figure}
{The reaction $\rm C+ H_2O \xrightarrow{} HCO + H$ can strongly reduce the abundance of hydrocarbons by putting the carbon back to CO (HCO leading to CO). Its reaction rate is known to be very low, with an upper limit of $k < 3.6\times 10^{-13}$ cm$^3$.s$^{-1}$ (\citealt{husain_reactions_1971}, NIST), and might explain why this reaction is missing in UMIST. However, the measurements of \cite{Hickson_2016arXiv160808877H} at low temperature revealed that the reaction rate is much higher than the upper limit of \citet[see our Fig. \ref{fig: C_H2O_react}]{husain_reactions_1971}. This has been interpreted as an efficient quantum tunnelling at low temperature, which allows it to cross the activation barrier. The online KIDA database includes this reaction (leading to "Products"), but as noted by \cite{woitke_2d_2024}, the rate is unreasonably high and inconsistent with \cite{husain_reactions_1971,Hickson_2016arXiv160808877H}. The ab initio calculations of \cite{Li_CH2O_2017PCCP...1923280L} confirmed that the rate is indeed low at high temperature (see Fig. \ref{fig: C_H2O_react}), consistent with \cite{husain_reactions_1971}, which means that tunnelling is not efficient in this regime. We decided to include this reaction in our chemical network by adopting the reaction rate of \cite{Hickson_2016arXiv160808877H} at low temperature and \cite{Li_CH2O_2017PCCP...1923280L} at high temperature (in red in Fig. \ref{fig: C_H2O_react}). The reaction rate is uncertain between $T\sim 200 - 500$ K, so we fitted the data points with a modified Arrhenius law for two separate temperature ranges and extrapolated the low temperature rate until 500 K:
\begin{equation}
    k(T) = \left \{
    \begin{array}{ll}
        1.314\times 10^{-13} \left ( \frac{T}{300} \right )^{-4.394} e^{-112.9/T} & \mbox{if T $\in [10,500]$ K } \\
        1.502\times 10^{-15} \left ( \frac{T}{300} \right )^{4.522} e^{-125.7/T} & \mbox{if T $\in [500,1000 ]$ K }
    \end{array}
 \right.
.\end{equation}
The emission of C$_2$H$_2$ is reduced by $\sim30$\% in the fiducial model, due to a lower abundance around 0.3-0.7 au, where the temperature is below 300 K. 
}

\subsection{Key reactions in KIDA missing in UMIST}
\label{appendix:key_reactiionkida}
\begin{table*}[ht]
\centering
\caption{Key reactions from KIDA explaining the difference between the chemical network based on UMIST+KIDA and UMIST only. }
\begin{tabular}{c|c|c|c|c}
\hline
\hline
\textbf{Reaction} & $\boldsymbol{\alpha}$ \textbf{[cm$\boldsymbol{^3}$.s$\boldsymbol{^{-1}}$]} & $\boldsymbol{\beta}$ & $\boldsymbol{\gamma}$ \textbf{[K]} & \textbf{Reference} \\
\hline
\hline
$\rm{C_2 + H_2\xrightarrow{}C_2H + H}$ & 1.60e-10 & 0.0 & 1420 & \cite{pitts_temperature_1982,harada_new_2010}\\
$\rm{C_3H + H_2\xrightarrow{}C_3H_2+H}$ & 1.14e-11 & 0.0 & 950 & \cite{harada_new_2010}\\
$\rm{H_2CCC+H_2\xrightarrow{}CH_2CCH+H}$ & 3.74e-12 & 2.0 & 1740 & \cite{harada_new_2010}\\
$\rm{C_4+H_2\xrightarrow{}C_4H+H}$ & 1.60e-10 & 0.0 & 1420 & \cite{harada_new_2010}\\
$\rm{C_4H + H_2\xrightarrow{}C_4H_2+H}$ & 1.14e-11 & 0.0 & 950 & \cite{harada_new_2010}\\
$\rm{N + CH_2CCH\xrightarrow{}C_2H_2+HCN}$ & 5.00e-11 & 0.0 & 0.0 & \cite{LOISON_2017MNRAS.470.4075L}\\
 $\rm{C_4H_3^+ + e^-\xrightarrow{}C_2H_2+C_2H}$ & 1.10e-07 & 0.0 & 0.0 & \cite{LOISON_2017MNRAS.470.4075L}\\
\hline
\end{tabular}
\tablefoot{Most of these reactions are H-abstractions, revealing the crucial role of H$_2$ in the formation of acetylene.} 
\label{tab: Reactions diff UIST and KIDA}
\end{table*}
The difference up to three orders of magnitude in C$_2$H$_2$ abundance between a chemical network based on UMIST+KIDA and based only in UMIST (Fig. \ref{fig: Umistvskida}) can be explained by seven key reactions in KIDA, listed in Table \ref{tab: Reactions diff UIST and KIDA}. Five of these seven reactions are H-abstractions revealing the crucial role of H$_2$ in the formation of C$_2$H$_2$. The rate coefficients of C$_n$ + H$_2$ and C$_n$H + H$_2$ are the same as C$_2$ + H$_2$ and C$_2$H + H \citep{harada_new_2010}.

\section{Line overlap}
\label{appendix:line_overlap}
\begin{figure}[ht]
\centering
\includegraphics[scale=0.35, trim={0cm 0 0 0cm}, clip]{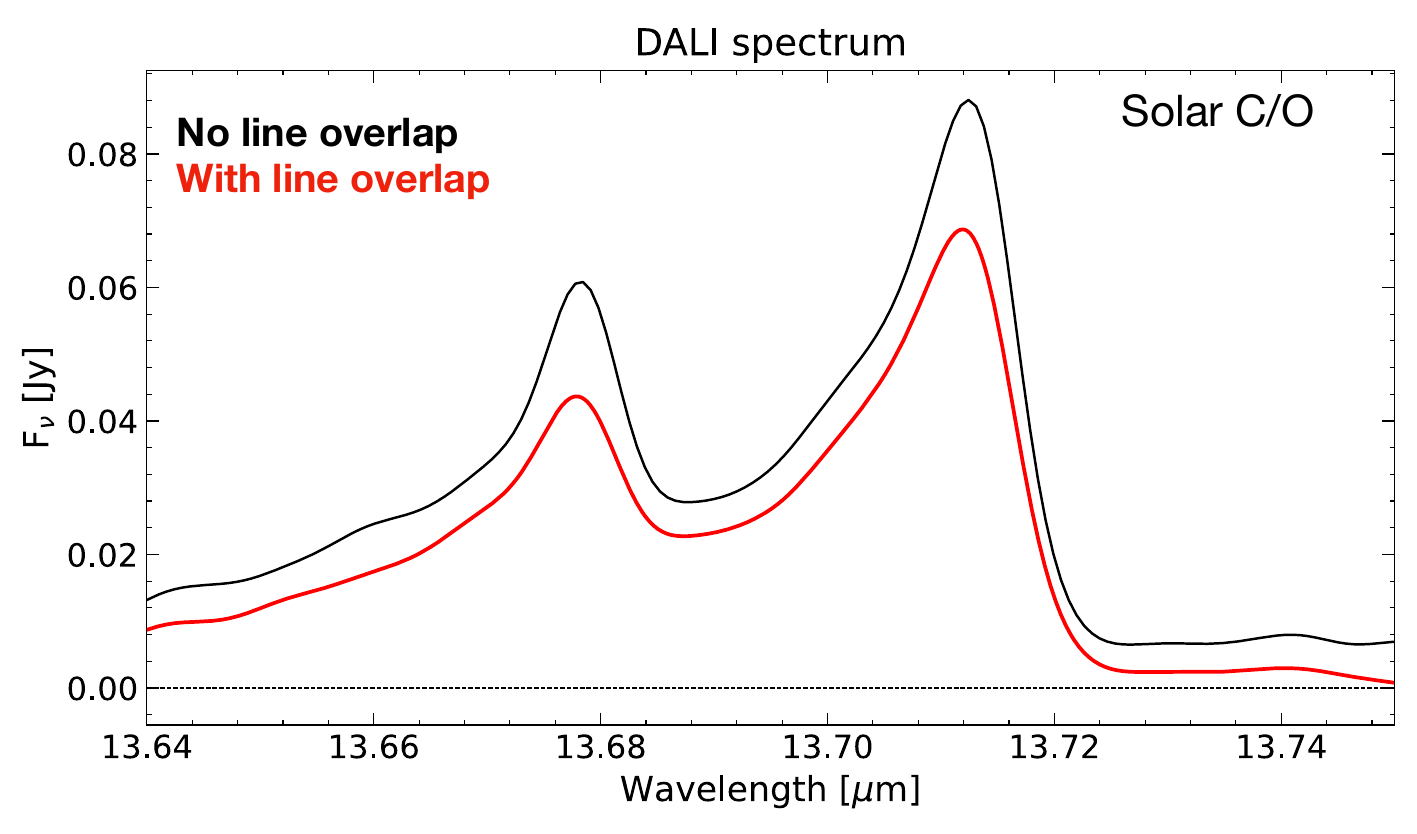}
\caption{DALI spectrum of the $Q$ branch of C$_2$H$_2$ at 13.7 $\mu$m without line overlap (in black) and with line overlap (in red). With a solar C/O, the difference is not significant in the $Q$ branch (factor 1.5) whereas it is more pronounced in the pseudo-continuum (factor $\sim 4$ near $\lambda=13.74$ $\mu$m).}
\label{fig: C2H2_overlap_spectrum}
\end{figure}
\noindent Figure \ref{fig: C2H2_overlap_spectrum} shows the result of including the line overlap for the ray-tracing of C$_2$H$_2$ (in red). It does not significantly reduce acetylene emission in the $Q$ branch. The presence of the pseudo-continuum near $\lambda=13.74$ $\mu$m reveals that there is still optically thick emission of C$_2$H$_2$, originating from the deepest layers in its emitting regions.

\section{Depleted O/H grid}

\label{appendix:subsolargrid}
\begin{figure}[ht]
\centering
\resizebox{\hsize}{!}{
\includegraphics[scale=0.50, trim={0cm 0 0 0cm}, clip]{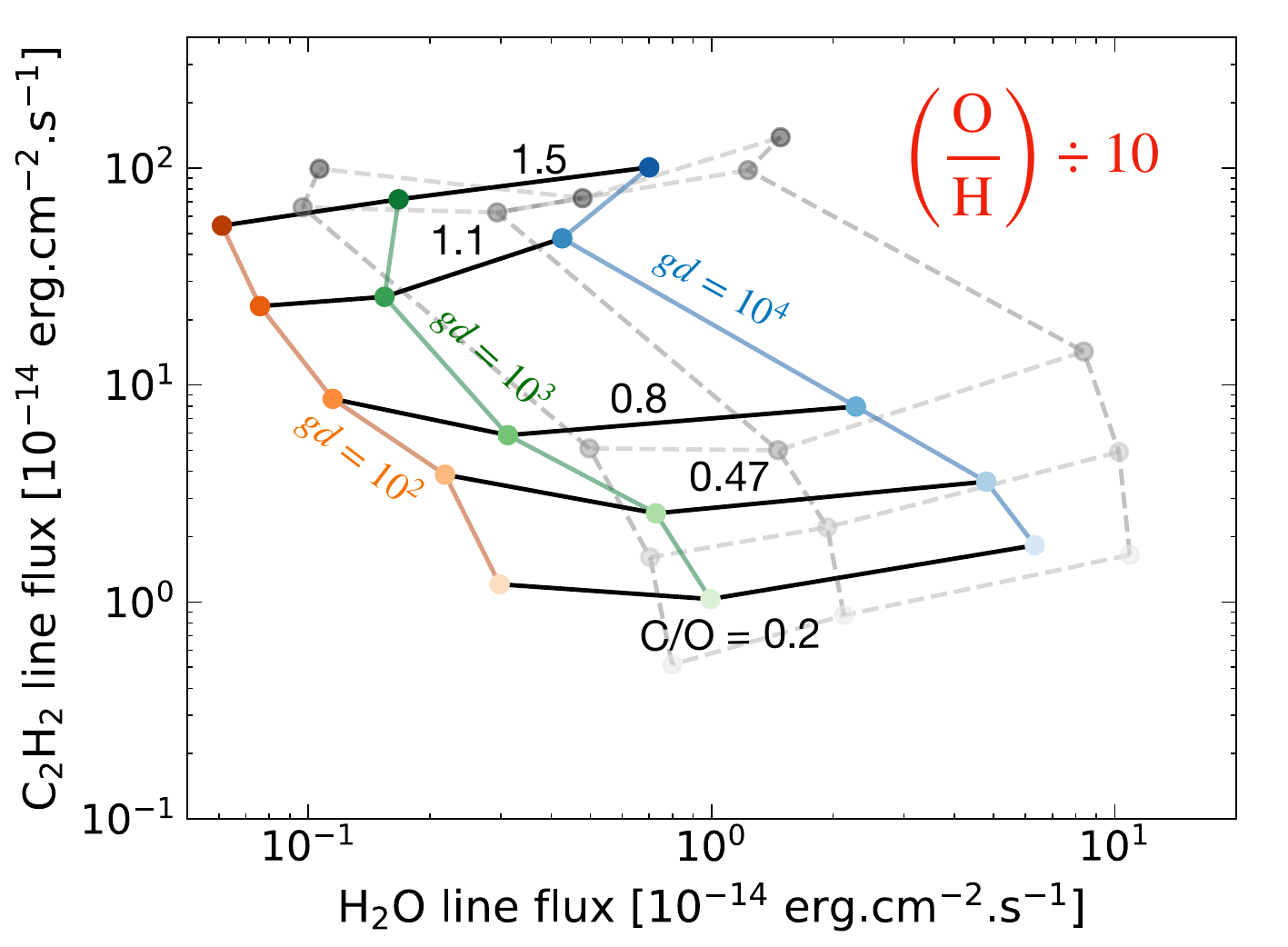}}
\caption{Same as Fig. \ref{fig: diagnostic_plot}b but showing the depleted O/H grid .}
\label{fig: diag_plot_subsolar}
\end{figure}

\noindent Figure \ref{fig: diag_plot_subsolar} presents the C$_2$H$_2$ and H$_2$O emissions obtained with the depleted O/H grid (10 times less oxygen than the fiducial grid) to mimic late times in protoplanetary disks, where the water vapor is advected onto the star and the metal-poor gas from the outer disk is reaching the inner regions \citep{mah_close-ice_2023, Sellek_2025A&A...701A.239S}. The decrease in oxygen, which enhances C$_2$H$_2$ formation, is balanced by a water shielding becoming less effective, leading to an increase of a factor of 2 in acetylene emission. The effect of the oxygen depletion is stronger for H$_2$O, shifting the grid to the left side (small H$_2$O emission) rather than the top left.

\section{Other parameters}
\label{appendix:paramwithnegligibleimpact}
\begin{figure}[ht]
\centering
\resizebox{\hsize}{!}{
\includegraphics[scale=0.50, trim={0cm 0 0 0cm}, clip]{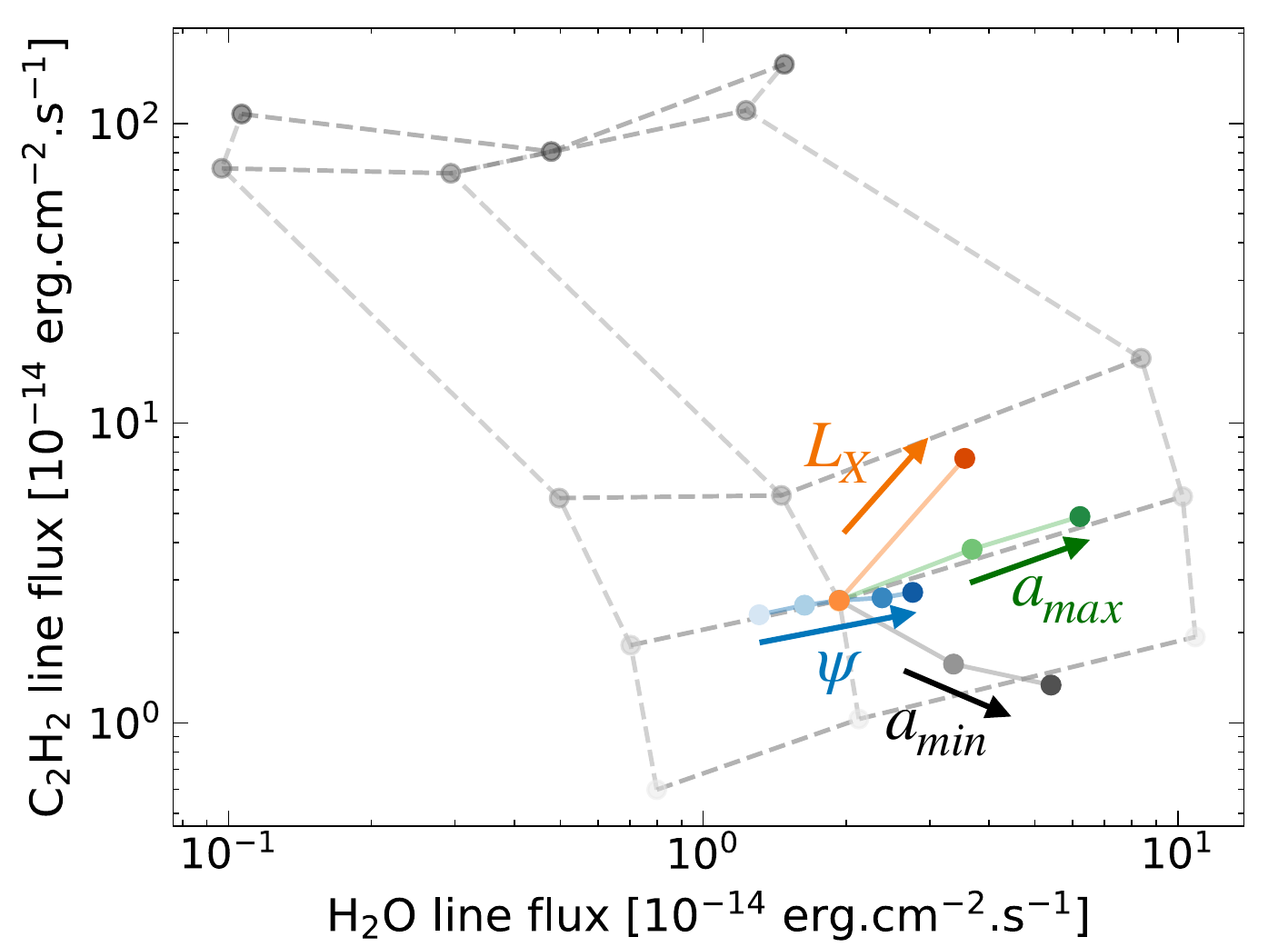}}
\caption{Same as Fig. \ref{fig: diagnostic_plot}c but showing the impact of $a_{min}$ (black), $a_{max}$ (green), $\psi$ (blue), and $L_X$ (orange). The fiducial grid is shown in gray.}
\label{fig: diag_plot_negligible_effect}
\end{figure}

\begin{figure}[ht]
\centering
\resizebox{\hsize}{!}{
\includegraphics[scale=0.50, trim={0cm 0 0 0cm}, clip]{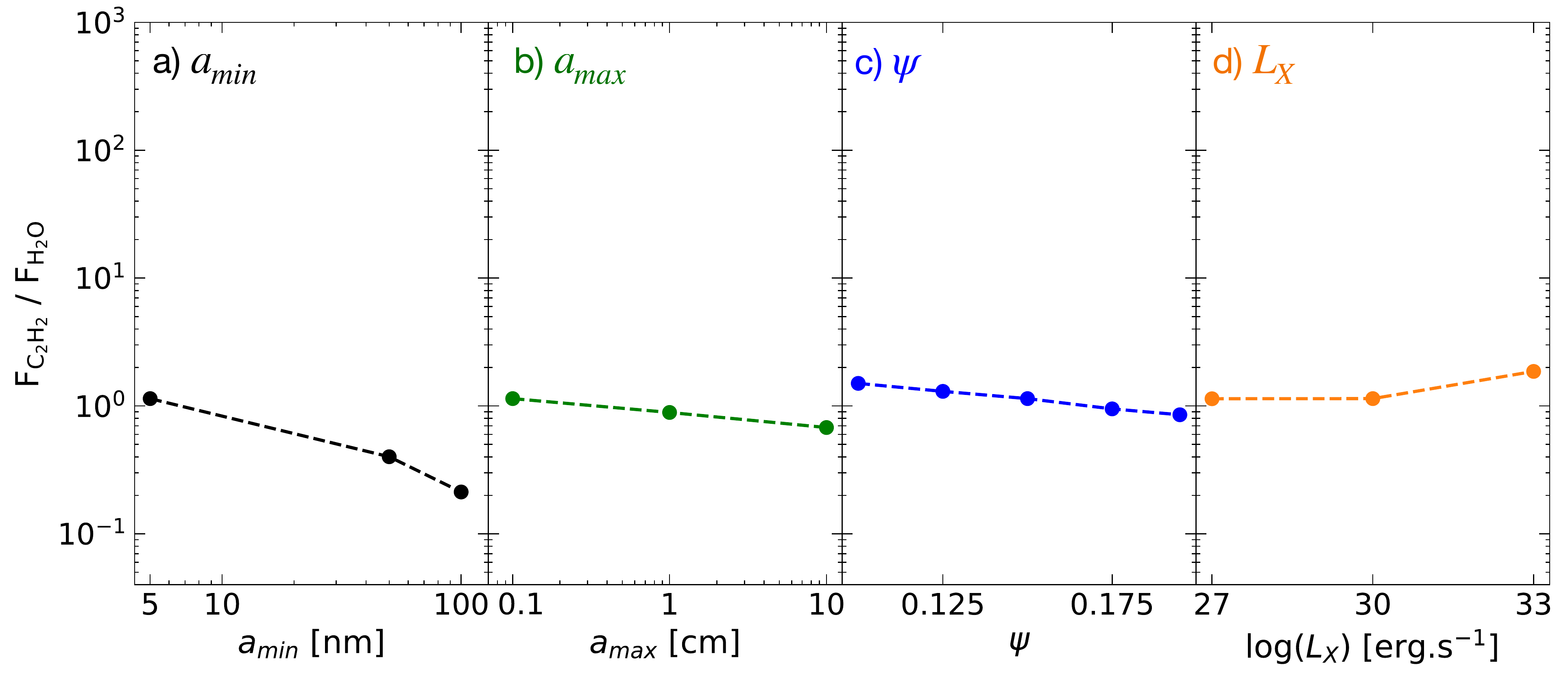}}
\caption{Evolution  of the line flux ratio C$_2$H$_2$/H$_2$O with the minimum and maximum grain size, the flaring angle, and the X-ray luminosity.}
\label{fig: flux ratio dep balec}
\end{figure}
Figure \ref{fig: diag_plot_negligible_effect} presents the results for the other parameters explored in this work: the minimum and maximum grain size $a_{min}$ and $a_{max}$, the flaring angle $\psi$ and the X-ray luminosity $L_X$. As mentioned in Sect. \ref{sec: dust_prop}, {increasing $a_{min}$ lowers the opacity in the upper atmosphere, pushing deeper the molecular emitting layers, but keeping the dust emitting layer the same. It reduces C$_2$H$_2$ emission as less acetylene is above the $\tau_{dust}=1$ surface. Water is less affected by the dust continuum as it emits higher up. Pushing its emission into denser layers increases its emission.} An increase of 6 orders of magnitude in the X-ray luminosity increases C$_2$H$_2$ and H$_2$O emission by only a factor of 3, showing that these two molecules are not sensitive to X-rays. In particular, it confirms that X-rays are important for both the formation and destruction of C$_2$H$_2$, thus canceling their effect on C$_2$H$_2$ emission.  

Figure \ref{fig: flux ratio dep balec} shows the evolution  of the line flux ratio C$_2$H$_2$/H$_2$O with these parameters ($a_{min}$, $a_{max}$, $\psi$, $L_X$). Considering the two-orders-of-magnitude variation in the observed C$_2$H$_2$/H$_2$O line flux ratio \citep{Grant_2025A&A...702A.126G}, these parameters can be considered irrelevant compared to C/O, O/H, or $q$.

\section{Model with C/O = 1.5}
\label{appendix:model_co1.5}
\noindent Figure \ref{fig: Fid_model_co1.5}e presents the result for a model with C/O > 1 (other parameters fixed to the values in the fiducial model). It reveals the dramatic difference with abundances maps of C/O < 1. C$_2$H$_2$ is abundant almost everywhere, and its emitting region is much more extended than with C/O < 1. On the contrary, water is much less abundant and emits from a smaller region, closer to the star. Figure \ref{fig: Fid_model_co1.5}d also interestingly shows that the UV field map is significantly different than when the C/O < 1: this time, C$_2$H$_2$ and C$_3$ absorb the UV photons instead of water with the footprint of C$_2$H$_2$ around $r \sim$ 0.5-5 au. 

The H/H$_2$ transition is much higher compared to C/O < 1, mainly due to the fact that H$_2$ is no longer destroyed by the warm oxygen chemistry leading to OH and H$_2$O. However, just above the surface where C$_2$H$_2$ becomes abundant, the hydrogen becomes atomic again. In this layer, it is the warm carbon chemistry that consumes H$_2$, as it was with oxygen in C/O < 1. Indeed, H$_2$ is destroyed by the reactions 
\begin{equation}
    \rm{} C_2 + H_2 \xrightarrow[]{} C_2H + H
\end{equation}
\begin{equation}
    \rm{} C_2H + H_2 \xrightarrow[]{} C_2H_2 + H
.\end{equation}
Then, C$_2$H$_2$ reacts with C to form C$_3$. This layer is still strongly irradiated ($G_0 \sim 10^6$), photodissociating C$_3$ to produce C$_2$. Thus, this layer creates a loop between C$_2$ and C$_3$ which consumes H$_2$. This layer is irradiated enough to photodissociate C$_3$ but not C$_2$, creating a sweet spot where H$_2$ would be dramatically destroyed. Indeed, in upper layers, C$_2$ is photodissociated as well, and in deeper layer, C$_3$ can survive to the UV field. Therefore, this loop is broken for layers above or below this specific region. 

\begin{figure*}[ht]
\centering
\resizebox{\hsize}{!}{
\includegraphics[scale=0.4, trim={0 0 0 0cm}, clip]{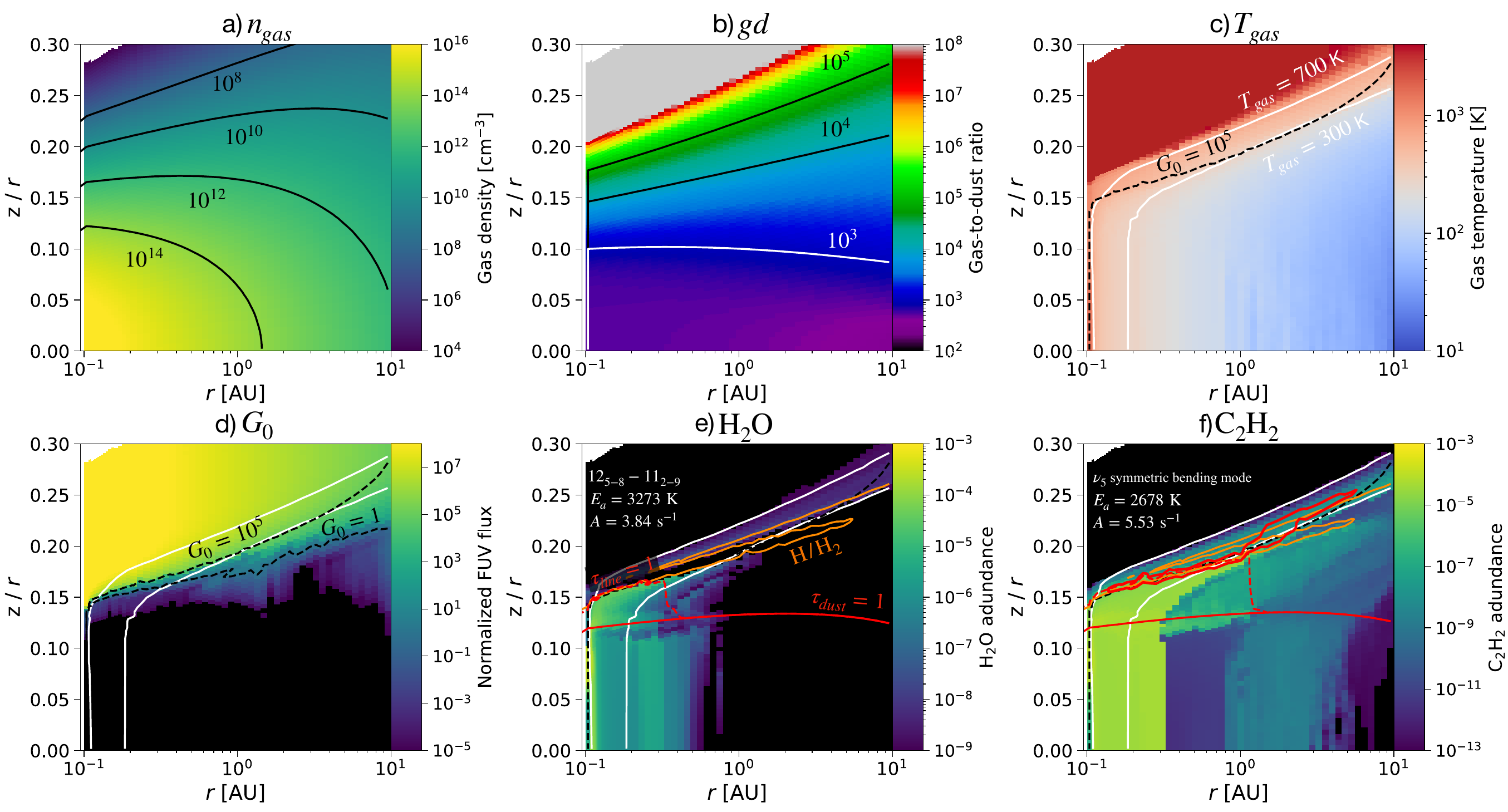}}
\caption{Disk structure of the fiducial model but with C/O = 1.5. Top panels: Gas density, gas-to-dust ratio, and gas temperature. Bottom panels: Normalized UV field $G_0$ (Habing units) and the abundance of H$_2$O and C$_2$H$_2$. The white lines indicate the 300 K and 700 K gas temperature contours. The bottom solid red line shows the dust optically thick surface ($\tau_{\rm{dust}}=1$ at 14 $\mu$m), while the dashed red line represents the surface where $\tau_{\rm{line}}=1$. The red contours correspond to 80\% of the total emitting flux. The dashed orange line in the bottom panels indicates the H/H$_2$ transition, sometimes difficult to distinguish from the dashed black line $G_0=10^5$. }
\label{fig: Fid_model_co1.5}
\end{figure*}
\end{appendix}
\end{document}